\documentclass[opre,nonblindrev]{informs3_MOD} 
\usepackage{booktabs} 

\SingleSpacedXI

\usepackage{natbib}
 \bibpunct[, ]{(}{)}{,}{a}{}{,}%
 \def\bibfont{\small}%

 \TheoremsNumberedThrough
 \ECRepeatTheorems  
 \EquationsNumberedThrough

\newenvironment{xproof}{\proof{\textit{Proof.}}}{\halmos\endproof\medskip}

\usepackage{tikz}
\usepackage{multicol}
\usepackage{multirow}
\usepackage{siunitx}
\usepackage{nicefrac}

\usepackage{thmtools}
\usepackage{mleftright}
\usepackage{mleftright}
\usepackage{bm}
\usepackage{xcolor}
\usepackage{hyperref} 
\usepackage[noabbrev,capitalise]{cleveref}
\usepackage{enumitem}
\usepackage{float}

\usepackage{import}
\usepackage{xifthen}
\usepackage{pdfpages}
\usepackage{mathtools}
\usepackage{transparent}

\usepackage{xparse}
\RenewDocumentCommand{\div}{O{}mm}{%
    D_{#1}\mleft(#2\,\mright\|\mleft.\,#3\mright)
}

\newcommand{\emptyseq}{\varnothing}

\renewcommand{\vec}[1]{\bm{#1}}

\newcommand{\cA}{\mathcal{A}}

\newcommand{\va}{\vec{a}}
\newcommand{\vx}{\vec{x}}
\newcommand{\vz}{\vec{z}}
\newcommand{\vw}{\vec{w}}
\newcommand{\vA}{\vec{A}}
\newcommand{\vy}{\vec{y}}
\newcommand{\vc}{\vec{c}}
\newcommand{\vg}{\vec{g}}

\newcommand{\vu}{\vec{u}}
\newcommand{\vv}{\vec{v}}
\newcommand{\vm}{\vec{m}}
\newcommand{\intQ}{\relint Q}

\DeclareMathOperator*{\relint}{relint}
\DeclareMathOperator{\diag}{diag}

\newcommand{\regu}{\varphi}
\newcommand{\depth}{\mathfrak{D}}
\newcommand{\bbR}{\mathbb{R}}
\newcommand{\Rp}{\mathbb{R}_{\ge 0}}

\newcommand{\defeq}{\coloneqq}
\newcommand{\cJ}{\mathcal{J}}
\newcommand{\children}[1]{\mathcal{C}_{#1}}

\newcommand{\bbN}{\mathbb{N}}
\newcommand{\cX}{\mathcal{X}}
\newcommand{\cY}{\mathcal{Y}}
\newcommand{\cZ}{\mathcal{Z}}
\newcommand{\cU}{\mathcal{U}}
\newcommand{\cV}{\mathcal{V}}
\newcommand{\prox}[2]{\text{\normalfont prox}_{#1}\mleft(#2\mright)}

\DeclareMathOperator*{\ext}{\,\triangleleft\,}
\DeclareMathOperator{\aff}{aff}

\newtheorem{setup}{Setup}

\newcommand{\numberthis}[1]{\refstepcounter{equation}\tag{\theequation}\label{#1}}

\usepackage{etoolbox}
\usepackage[linesnumbered,ruled,lined,noend]{algorithm2e}

\SetAlFnt{\small}
\SetAlCapFnt{\small}
\SetAlCapNameFnt{\small}
\SetAlCapHSkip{0pt}
\IncMargin{-\parindent}
\makeatletter
    \patchcmd\algocf@Vline{\vrule}{\vrule \kern-0.4pt}{}{}
    \patchcmd\algocf@Vsline{\vrule}{\vrule \kern-0.4pt}{}{}
\makeatother
\SetKwComment{Hline}{}{\vspace{-3mm}\textcolor{gray}{\hrule}\vspace{1mm}}
\definecolor{darkgrey}{gray}{0.3}
\definecolor{commentcolor}{gray}{0.3}
\SetKwComment{Comment}{\color{commentcolor}$\triangleright$\ }{}
\SetCommentSty{}
\SetNlSty{}{\color{darkgrey}}{}
\crefname{algocf}{Algorithm}{Algorithms}
\SetKwProg{Fn}{function}{}{}
\SetKwProg{Subr}{subroutine}{}{}

\crefalias{AlgoLine}{line}%
\crefname{setup}{Setup}{Setups}

\makeatletter
\let\cref@old@stepcounter\stepcounter
\def\stepcounter#1{%
  \cref@old@stepcounter{#1}%
  \cref@constructprefix{#1}{\cref@result}%
  \@ifundefined{cref@#1@alias}%
    {\def\@tempa{#1}}%
    {\def\@tempa{\csname cref@#1@alias\endcsname}}%
  \protected@edef\cref@currentlabel{%
    [\@tempa][\arabic{#1}][\cref@result]%
    \csname p@#1\endcsname\csname the#1\endcsname}}
\makeatother
\newcommand{\cfrp}[1]{CFR$^+$}

\begin{document}

\RUNAUTHOR{Farina, Kroer, and Sandholm}

\RUNTITLE{Better Regularization for Sequential Decision Spaces}

\ARTICLEAUTHORS{%
\AUTHOR{Gabriele Farina}
\AFF{Computer Science Department, Carnegie Mellon University, Pittsburgh, PA 15213, \EMAIL{gfarina@cs.cmu.edu}} 
\AUTHOR{Christian Kroer}
\AFF{Industrial Engineering and Operations Research Department, Columbia University, New York City, NY 10027, \EMAIL{christian.kroer@columbia.edu}}
\AUTHOR{Tuomas Sandholm}
\AFF{Computer Science Department, Carnegie Mellon University, Pittsburgh, PA 15213, \EMAIL{sandholm@cs.cmu.edu}} 
} 

\TITLE{Better Regularization for Sequential Decision Spaces: Fast Convergence Rates for Nash, Correlated, and Team Equilibria}
\ABSTRACT{
    We study the application of iterative first-order methods to the problem of computing equilibria of large-scale two-player extensive-form games. First-order methods must typically be instantiated with a regularizer that serves as a distance-generating function for the decision sets of the players. For the case of two-player zero-sum games, the state-of-the-art theoretical convergence rate for Nash equilibrium is achieved by using the dilated entropy function. In this paper, we introduce a new entropy-based distance-generating function for two-player zero-sum games, and show that this function achieves significantly better strong convexity properties than the dilated entropy, while maintaining the same easily-implemented closed-form proximal mapping. Extensive numerical simulations show that these superior theoretical properties translate into better numerical performance as well.

    We then generalize our new entropy distance function, as well as general dilated distance functions, to the scaled extension operator. The scaled extension operator is a way to recursively construct convex sets, which generalizes the decision polytope of extensive-form games, as well as the convex polytopes corresponding to correlated and team equilibria. By instantiating first-order methods with our regularizers, we develop the first accelerated first-order methods for computing correlated equilibra and ex-ante coordinated team equilibria. Our methods have a guaranteed $1/T$ rate of convergence, along with linear-time proximal updates.
}
\maketitle






    \section{Introduction}

Large-scale \textit{extensive-form game (EFG)} models have been used in several recent AI milestones, where equilibrium approximation was used as the approach for building AI agents~\cite{Brown17:Superhuman,Brown19:Superhuman,Moravvcik17:DeepStack,Bowling15:Heads}. A crucial component for constructing these agents is a fast method for computing approximate Nash equilibria in large and very-large game models. For the two-player zero-sum setting, an EFG can be solved in polynomial time using a linear program (LP) whose size is linear in the size of the game tree~\cite{Stengel96:Efficient}.
However, this LP-based approach was not used in any of these AI milestones. Instead, fast iterative methods are preferred~\cite{Zinkevich07:Regret,Hoda10:Smoothing,Tammelin15:Solving,Brown19:Solving,Farina19:Online,Kroer20:Faster} as well as sampling-based variants~\cite{Lanctot09:Monte,Gibson12:Generalized,Kroer15:Faster,Brown17:Superhuman,Brown19:Superhuman,Schmid19:Variance,Farina20:Stochastic}.
The reason for this is that constructing the LP, and running simplex or interior-point methods on it, is too expensive for these large-scale models. In contrast, iterative methods only require oracle access to one or two gradient computations, or even estimates thereof, in order to perform an iteration.

From a theoretical standpoint, the fastest iterative methods for solving
    two-player zero-sum games are \textit{first-order methods (FOMs)} such as the
    excessive gap technique~\cite{Nesterov05:Excessive} or mirror
    prox~\cite{Nemirovski04:Prox}, which converge at a rate of $1/T$, where $T$
    is the number of iterations.
In order to apply these methods to EFGs, they must be instantiated with a
    \emph{distance-generating function} (DGF), which yields an appropriate
    notion of how to measure distances between strategies in the game.
In this framework, the convex set of all strategies belonging to a player is
    referred to as the sequence-form polytope, and alternatively as a \emph{treeplex}~\cite{Hoda10:Smoothing}, which is a tree-like structure of scaled
    simplexes.
Essentially the only sequence-form poltyope DGFs that are known are based on the dilated DGF
    framework introduced by \citet{Hoda10:Smoothing} (apart from using the
    standard $\ell_2$ distance, which is unsuitable due to projection
    requirements at each iteration).
For example, the dilated entropy distance yields the best current rate of
    convergence for $1/T$ methods~\cite{Kroer20:Faster}.
One drawback of the dilated entropy DGF, as well as other dilated DGFs, is that
    current analyses incur a dependence of the form $2^{\depth}$, where $\depth$ is the
    depth of the decision
    space~\cite{Hoda10:Smoothing,Kroer20:Faster,Kroer18:Solving}.
In some cases this is reasonable, since the decision space itself may have
    exponential size in the depth of the game tree.
However, in other cases the decision space may have substantial structure, such
    that this exponential complexity in depth makes the bounds exponential in the size of the
    game tree.

In this paper, we introduce the first DGF for sequence-form polytopes whose strong convexity
    is not derived from its structure as  a dilated distance function (again,
    the standard Euclidean distance also satisfies this, but it requires
    difficult projections).
In particular, we show that a weighted version of the negative entropy for the
    nonnegative unit cube is a superior DGF for sequence-form polytopes.
First, we show that this DGF can achieve strong convexity modulus $1/M_Q$ (where $M_Q$ is the maximum value of the $\ell_1$ norm on $Q$), with the largest weights at individual decision points being on the order of $M_Q\log n$ (where $\log n$ is the largest number of actions at any decision point), which improves upon that of the dilated entropy DGF by a
    factor of $2^\depth$. This also translates into an improvement to the theoretical convergence rate of FOMs by a factor $2^{\depth+2}$.
A particularly appealing part of this result is that our analysis depends only
    on the $\ell_1$ norm of the sequence-form polytope, and has no exponential dependence on
    the depth.
At the same time, we must also ensure that the new DGF allows fast computation
    of the associated proximal steps required by, for example, mirror prox or EGT.
We show that this is indeed the case: the weights in our new DGF are chosen in
    a way that allows us to show that this new DGF corresponds to a particular
    dilated entropy DGF on the sequence-form polytope (while being different outside the
    sequence-form polytope).
This allows us to use existing results on fast proximal-step computation for
dilated entropy.
We call our new DGF the \emph{dilatable global entropy} (DGE).

After introducing DGE for sequence-form polytopes, we switch our focus to studying DGFs for
    the more general \emph{scaled extension}
    operator~\citep{Farina19:Efficient}.
The scaled extension operator is a
    method for iteratively constructing a convex
set as a sequence of convexity-preserving compositions of convex sets. This
operator can be used to construct the sequence-form polytope, but more importantly for our
purposes it can also be used to construct more general sets such as the polytope of correlation plans needed for computing optimal extensive-form correlated equilibria and ex-ante coordinated team strategies in certain classes of games where it is known that those solution concepts can be computed in polynomial time. First, we show
how to extend the class of dilated DGFs to polytopes constructed via scaled
extension, thereby generalizing the framework of \citet{Hoda10:Smoothing} beyond
sequence-form polytopes, while also giving a simpler proof of strong convexity. This enables
DGFs such as the dilated entropy or dilated Euclidean distance to be applied to
a much broader class of polytopes. Then, we show that our DGE construction can
also be extended to scaled extension. Taken together, we generalize the entire
class of known ``nice'' DGFs for sequence-form polytopes to the set of all polytopes which
can be constructed via scaled extension. Applying these results to the problems
of computing optimal correlated solution concepts and ex-ante coordinated team strategies yields the \emph{first} method for iteratively solving these
problems at a rate of $1/T$, while enjoying fast closed-form solutions at each
iteration. In contrast, the only prior result of this form required using the
standard Euclidean distance, and thus had to perform expensive projections at
every iteration of the algorithm~\cite{Farina19:Correlation}.

Extensive experiments validate the efficacy of our new DGFs. We find that these new
DGFs lead to much smaller amounts of smoothing, while still ensuring correctness of
the algorithms. Intuitively, this means that we can safely take much larger steps
at each iteration.

\subsubsection*{Paper Outline}
The paper is structured as follows. \cref{sec:foms} presents background on first-order methods, which includes the description of the DGFs needed for setting up these methods. That section can be skimmed for notation, if the reader is already familiar with FOMs. 
\cref{sec:efg prelims} gives an introduction to extensive-form games. That section can be skimmed for notation, if the reader is already familiar with EFGs. 
\cref{sec:treeplexes} introduces the sequence-form polytope and dilated DGFs.
\cref{sec:global dgf} presents our new DGF for the sequence-form polytope, along with the convergence rate obtained when combined with a FOM.
\cref{sec:scaled extension} develops DGFs for the scaled extension operator, and shows how this leads to efficient FOMs with a $1/T$ convergence rate for correlated and team equilibria.
\cref{sec:experiments} provides an extensive set of computational evaluations of our new DGFs for various games and types of equilibrium.

\section{Preliminaries on Extensive-Form Games}
\label{sec:efg prelims}

An extensive form game (EFG) is a game played on a tree. Every node in the tree belongs to some player, whose turn it is to act, and the set of branches at the node correspond to the set of actions available to the player. In general, a strategy for a player may consist of choosing a probability distribution over the actions at each node in the game. Additionally, there may be special nodes called \emph{chance nodes}, which have a fixed distribution over actions associated with them. These nodes model stochastic outcomes, for example, the dealing out of cards in a card game or the valuation signals sent to buyers in a sequential auction. At leaf nodes the game ends, and each leaf node is associated with a vector of payoffs, one payoff per player. The goal of each player in the game is to maximize the expected value of their leaf-node payoffs. Finally, an EFG can model imperfect information: an \emph{information set} is a group of nodes belonging to a player such that the player cannot distinguish among those nodes, and is therefore required to have the same probability distribution over actions at each node in the information set. An example of an information set would be in a poker game, where the information set represents all the cards seen by the player, as well as all bets (which are public). Each node in the information set would correspond to different possible hands held by the other player(s).

\begin{figure}\centering
\scalebox{0.75}{
\tikzstyle{level 1}=[level distance=3cm, sibling distance=3.5cm]
\tikzstyle{level 2}=[level distance=3cm, sibling distance=2cm]
\tikzstyle{level 3}=[level distance=3cm, sibling distance=1cm]
\tikzstyle{player} = [text width=4em, draw, text centered, rectangle, fill=blue!20, inner sep=3pt]
\tikzstyle{nature} = [minimum width=3pt,circle,  draw, fill=red!20, inner sep=3pt]
\tikzstyle{end} = [circle, minimum width=3pt, fill, inner sep=0pt, right]
\begin{tikzpicture}[grow=right, sloped]
    \node[nature]{Card}
        child { node[player] {P1}
            child{ node[player] (checkA) {P2} node[right,xshift=22] {...}  edge from parent node[below] {Check}}
            child{ node[player] (raiseA) {P2}
                child{ node[end] {} node[right] {$(1,-1)$} edge from parent node[below] {Fold}}
                child{ node[end] {} node[right] {$(2,-2)$} edge from parent node[below] {Call}}
            edge from parent node[below] {Raise}}
        edge from parent node[above] {$1/6$} node[below] {K, Q}}
        child { node[player] {P1}
            child{ node[player] (checkB) {P2} node[right,xshift=22] {...}  edge from parent node[below] {Check}}
            child{ node[player] (raiseB) {P2}
                child{ node[end] {} node[right] {$(1,-1)$} edge from parent node[below] {Fold}}
                child{ node[end] {} node[right] {$(-2,2)$} edge from parent node[below] {Call}}
            edge from parent node[below] {Raise}}
        edge from parent node[above] {$1/6$} node[below] {J, Q}}
        ;
        \draw[dashed] (raiseA) to[out=25,in=-25] (raiseB);
        \draw[dashed] (checkA) to[out=25,in=-25] (checkB);
\end{tikzpicture}
}
  \caption{Example poker game. The red ``Card'' node is a chance node. Only a subset of the possible cards dealt out by chance are shown (``J,Q'' and ``K,Q''). Dotted lines denote nodes that belong to the same information set.}
  \vspace{-4mm}
\label{fig:poker example}
\end{figure}
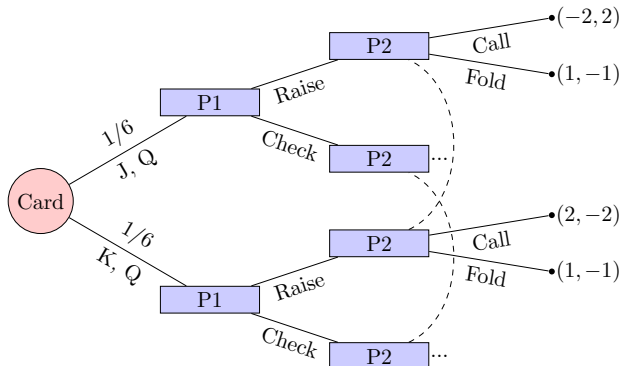

\cref{fig:poker example} illustrates a part of a poker game tree. At each node, either a card is dealt at random to each player (for space reasons, we only show two branches of cards dealt, even though there would be more in a real poker game) or some player acts. The payoffs are at the leaves. They are zero sum in this example.

A \emph{solution concept} provides a definition of rationality. For a given EFG, the application of a solution concept yields a set of equilibria, where each equilibrium has one strategy per player. A strategy describes how a player acts at every one of her information sets. For example, a \emph{Nash equilibrium} is a set of strategies such that each player cannot improve their expected utility by switching to another strategy, when the strategies of other players are held fixed. We will introduce various solution concepts in the sections where we give algorithms for them.


    \section{Preliminaries on First-Order Methods}
\label{sec:foms}

The types of FOMs that we will consider rely on access to a function $d$ which is used to construct a notion of distance between pairs of points in the decision space. The soundness of the algorithm requires such a function to satisfy a number of properties: 

\begin{definition}[Distance-generating function]\label{def:dgf}
    A \emph{distance-generating function (DGF)} $d$ for a compact and convex set $\cX \subseteq \bbR^n$ is a function $d: \cX \to \bbR$ such that:
    \begin{itemize}
        \item it is continuous on $\cX$ and differentiable in the relative interior of $\cX$;
        \item it is strongly convex in the relative interior of $\cX$ with respect to some norm $\|\cdot\|$, that is, there exists a constant $\mu > 0$ such that
        \[
                    (\nabla d(\vx) - \nabla d(\vx'))^\top( \vx-\vx') \geq \mu \| \vx-\vx' \|^2\qquad
                    \forall\, \vx,\vx' \in \relint\cX.
        \]
        For twice-differentiable $d$, the strong convexity condition is automatically verified as long as
        \begin{align}
            \label{eq:second order sc}
            \vec{m}^\top \nabla^2 d(\vx)\vec{m}  \geq \mu\|\vec{m}\|^2,\
            \forall \vx\in \relint\cX, \vec{m} \in \bbR^{n}.
        \end{align}
    \end{itemize}
    Furthermore, we make the common assumption that
    \[
        \min_{\vx \in \cX} d(\vx) = 0.
    \]
    This can always be assumed without loss of generality, as $d$ can always be
    shifted by a constant amount without losing the other properties.
\end{definition}

Given a convex set $\cX\subseteq\bbR^n$ and a distance-generating function $d$ for it, several important tools can be defined, which collectively form a \emph{proximal setup} for $\cX$:
\begin{itemize}
\item The \emph{Bregman divergence} $D_d : \cX \times \relint\cX \to \Rp$
    associated with $d$ yields a notion of distance between points defined as%
    \footnote{A Bregman divergence need not be symmetric and thus might not be a metric in the technical sense.}
    \[
        \div[d]{\vx}{\vx'} \defeq d(\vx) - d(\vx') - \nabla d(\vx')^\top (\vx-\vx')\qquad \forall\, \vx\in\cX,\vx'\in\relint\cX.
    \]
\item The \emph{$d$-diameter of $\cX$} is
    \[
        \Omega_{d,\cX} \defeq \max_{\vx,\vx'\in \cX} \div[d]{\vx}{\vx'} \leq \max_{\vx\in \cX} d(\vx) - \min_{\vx\in \cX} d(\vx)
    \]
\item Finally, we denote the largest possible value of the $\ell_1$ norm on a $\cX$ with the symbol $M_\cX \defeq \max_{\vx\in \cX} \|\vx\|_1$.
\end{itemize}

\subsection{``Nice'' distance-generating functions}
\label{sec:nice dgf}

While not a part of the assumptions on the DGF $d$, it is typically assumed that $d$ allows one to efficiently compute the following two quantities, which come up at every iteration of most FOMS:
\begin{itemize}
    \item the \emph{gradient} $\nabla d(\vx)$ of $d$ at any point $\vx \in \relint\cX$;
    \item the gradient of the convex conjugate $d^*$ of $d$ at any point $\vg \in \bbR^n$:
    \[
        \nabla d^*(\vec{g}) = \argmax_{\vx\in \cX} \big\{  \vec{g}^\top \vx - d(\vx)\big\}.
    \]
\end{itemize}
The gradient of the convex conjugate can be intuitively thought of as a linear
maximization problem over $\cX$ (i.e., the \emph{support function} of $\cX$, which is a non-smooth convex optimization problem),
\emph{smoothed} by the regularizer $d$.
For that reason, in this paper we shall refer to $\nabla d^*(\vg)$ either symbolically, or occasionally as the \emph{smoothed support function}.

Because the above two quantities arise so frequently in optimization methods, it
is important that the chosen distance-generating function allow for efficient
computation of them. In particular, in this paper we are concerned with ``nice''
DGFs that enable linear-time (in the dimension $n$) exact computation of those
two quantities.

\begin{definition}
    A distance-generating function $d$ is said to be ``nice'' if $d(\vx)$, $\nabla d(\vx)$ and $\nabla d^*(\vg)$ can be computed exactly in linear time in the dimension of the domain of $d$.
\end{definition}

\citet{Hoda10:Smoothing} also introduce a notion of a ``nice'' DGF. Their definition is similar to ours, but only states that $\nabla d^*(\vg)$ should be ``easily computable''. In contrast, we attach a concrete meaning to that statement: we take it to mean linear time in the dimension of the domain.

Finally, we mention a closely related operation that comes up often in 
optimization methods: the \emph{proximal operator} (or \emph{prox operator}),
defined as
\begin{align*}
    \prox{\tilde\vx}{\vg}
    &\defeq \argmin_{\vx\in \cX} \left\{ \vg^\top \vx - \div[d]{\vx}{\tilde\vx}\right\}
    = \nabla d^*(-\vg + \nabla d(\tilde\vx)) \in \cX\numberthis{eq:prox as argconjugate}
\end{align*}
for any $\tilde\vx \in \cX$ and $\vg \in \bbR^n$.
In light of~\eqref{eq:prox as argconjugate}, the prox operator can be implemented
efficiently provided that $\nabla d$ and $\nabla d^*$ can. So, prox operators can be computed exactly in linear time in the dimension $n$ for ``nice'' DGFs.

\subsection{Bilinear saddle-point problems}
We will be interested in solving \emph{bilinear saddle-point problems} (BSPPs), whose general form is 
\begin{align}
    \min_{\vx \in \cX} \max_{\vy \in \cY} \vx^\top \vA\vy,
    \label{eq:bspp}
\end{align}
where $\vA\in \mathbb{R}^{n\times m}$ and $\cX,\cY$ are convex and compact sets.
We will now present the EGT and mirror prox algorithms for solving BSPPs. 
These algorithms depend on two proximal setups: one for $\cX$ and one for $\cY$, denoted $d_x$ and $d_y$, respectively.
Let $\|\cdot\|_x$ and $\|\cdot\|_y$ be the norms associated with the strong convexity of $d_x$ and $d_y$ in the given proximal setup.
The convergence rate then depends on the following \emph{operator norm} of the payoff matrix $\vA$:
\[
    \|\vA\| \defeq \max\{\vx^\top \vA\vy : \|\vx\|_x\leq 1, \|\vy\|_y\leq 1\}.
\]
We will primarily be concerned with DGFs that are strongly convex with respect
to either the $\ell_1$ or $\ell_2$ norms. The magnitude of $\|\vA\|$ is the
primary way in which the norm matters: if both $d_x$ and $d_y$ are strongly
convex with respect to the $\ell_2$ norm, then $\|\vA\|$ can be on the order of
$\sqrt{nm}$, whereas if both are with respect to the $\ell_1$ norm, then
$\|\vA\|$ is simply equal to its largest entry.


\subsection{The Excessive Gap Technique (EGT)}

The \emph{excessive gap technique (EGT)} is a first-order method introduced by
\citet{Nesterov05:Smooth}, and one of the primary applications is to solve
BSPPs such as \cref{eq:bspp}. EGT assumes access to a proximal setup for $\cX$
and $\cY$, with one-strongly-convex DGFs $d_x,d_y$, and
constructs smoothed approximations of the optimization problems faced by the
$x$ and $y$ players.
%
Based on this setup, we formally state the
EGT of \cite{Nesterov05:Excessive} in \cref{algo:egt}.
EGT alternatingly takes steps focused on decreasing one or the other smoothing parameter. These steps are called \textsc{ShrinkX} and \textsc{ShrinkY} in \cref{algo:egt}.

\begin{figure}[H]
    \refstepcounter{algocf}\label{algo:egt}%
    \def\scale{0.94}%
    \makeatletter%
    \setlength{\algocf@ruledwidth}{\linewidth}%
    \@algocf@pre@ruled%
    \global\sbox\algocf@capbox{\hskip\AlCapHSkip
    \setlength{\algocf@lcaptionbox}{\hsize}\addtolength{\algocf@lcaptionbox}{-\AlCapHSkip}%
    \parbox[t]{\algocf@lcaptionbox}{\algocf@captiontext{\fnum@algocf}{Excessive Gap Technique (EGT) algorithm.}}
    }%
    \box\algocf@capbox\kern\interspacetitleruled\hrule%
      width\algocf@ruledwidth height\algotitleheightrule depth0pt\kern\interspacealgoruled%
    \renewcommand{\@ResetCounterIfNeeded}{}%
    \makeatother%
    \scalebox{\scale}{\begin{minipage}[t]{.360\linewidth}%
        \RestyleAlgo{plain}%
        \begin{algorithm}[H]%
            \DontPrintSemicolon
            \Fn{\normalfont\textsc{Initialize}()}{
                $t \gets 0$\;
                $\mu_x^0 \gets \|\vA\|,~~\mu_y^0 \gets \| \vA \|$\;
                $\tilde{\vx} \gets \argmin_{\hat{x}\in \cX} d_x\mleft( \hat{x} \mright) $\;
                $\vy^0 \gets \nabla d_y^*( \vA^\top \tilde{\vx} / \mu^0_y )$\;
                $\vx^0 \gets \prox{\tilde{\vx}}{\vA\vy^0 / \mu^0_x}$\;
            }
            \Hline{}
            \Fn{\normalfont\textsc{Iterate}()}{
                $t \gets t + 1,~~\tau \gets 2 / (t + 2)$\;
                \textbf{if }\emph{$t$ is even} \textbf{then} \textsc{ShrinkX()}\!\!\!\;
                \textbf{else} \textsc{ShrinkY()}\;
            }
        \end{algorithm}
    \end{minipage}}
    \scalebox{\scale}{\begin{minipage}[t]{.350\linewidth}\small
        \RestyleAlgo{plain}
        \begin{algorithm}[H]
            \DontPrintSemicolon
            \Fn{\normalfont\textsc{ShrinkX}()}{
                    $\bar{\vx} \gets -\nabla d_x^*(-\vA\vy^{t-1}/\mu_x^{t-1})$\;
                    $\hat{\vx}\gets (1 - \tau)\vx^{t-1} + \tau\bar{\vx} $\;
                    $\bar{\vy}\gets \nabla d_y^*(\vA^\top\hat{\vx} / \mu_y^{t-1})$\;
                    $\tilde{\vx}\gets \prox{\bar{\vx}}{\frac{\tau}{(1-\tau)\mu_x^{t-1}}\vA\bar{\vy}}$\;
                    $\vx^{t} \gets (1-\tau)\vx^{t-1} + \tau \tilde{\vx}$\;
                    $\vy^{t} \gets (1-\tau)\vy^{t-1} + \tau \bar{\vy}$\;
                    $\mu_x^t \gets (1-\tau)\mu_x^{t-1}$\;
            }
        \end{algorithm}
    \end{minipage}}
    \scalebox{\scale}{\begin{minipage}[t]{.360\linewidth}\small
        \RestyleAlgo{plain}
        \begin{algorithm}[H]
            \DontPrintSemicolon
            \Fn{\normalfont\textsc{ShrinkY}()}{
                    $\bar{\vy} \gets \nabla d_y^*(\vA^\top\vx^{t-1}/\mu_y^{t-1})$\;
                    $\hat{\vy}\gets (1 - \tau)\vy^{t-1} + \tau\bar{\vy} $\;
                    $\bar{\vx}\gets -\nabla d_x^*(-\vA\hat{\vy} / \mu_x^{t-1})$\;
                    $\tilde{\vy}\gets \prox{\bar{\vx}}{\frac{-\tau}{(1-\tau)\mu_y^{t-1}}\vA^\top\bar{\vx}}$\;
                    $\vy^{t} \gets (1-\tau)\vy^{t-1} + \tau \tilde{\vy}$\;
                    $\vx^{t} \gets (1-\tau)\vx^{t-1} + \tau \bar{\vx}$\;
                    $\mu_y^t \gets (1-\tau)\mu_y^{t-1}$\;
            }
        \end{algorithm}
    \end{minipage}}
    \makeatletter
    \@algocf@post@ruled
    \renewcommand{\@ResetCounterIfNeeded}{\setcounter{AlgoLine}{0}}
    \@ResetCounterIfNeeded{}
    \makeatother
\end{figure}

\cref{algo:egt} shows how initial points are selected and the alternating steps
and stepsizes are computed. \citet{Nesterov05:Excessive} proves that the EGT
algorithm converges at a rate
of $O(1/T)$:

\begin{theorem}[\citet{Nesterov05:Excessive} Theorem 6.3]\label{thm:nesterovEGT}
At every iteration $t \geq 1$ of the EGT algorithm, the solution $(\vx^t, \vy^t)$ satisfies
$\vx^t\in \cX$, $\vy^t\in \cY$, and
\[
    \max_{\vy\in \cY} (\vx^t)^\top \vA\vy - \min_{\vx\in \cX} \vx^\top \vA\vy^t  \leq  \frac{4\|\vA\| \sqrt{\Omega_{d_x,\cX}\Omega_{d_y,\cY}} }{t+1}.
\]
\end{theorem}

\subsection{Mirror Prox (MP)}
Next we consider the \textit{Mirror Prox (MP)} algorithm~\cite{Nemirovski04:Prox}. Rather than construct smoothed approximations, mirror prox directly uses the DGFs to take first-order steps. Hence, the MP algorithm is best understood as an algorithm that operates on the product space $\cX\times \cY$ directly. As such, in most analyses of the MP algorithm, a single $1$-strongly convex DGF for the product space $\cX \times \cY$ is required. To better align with the setup used for EGT, we will define the DGF for the product space $\cX \times \cY$ starting from proximal setups for both $\cX$ and $\cY$, with $1$-strongly convex DGFs $d_x,d_y$ with respect to norms $\|\cdot\|_x$ and $\|\cdot\|_y$, respectively. With this setup, it is immediate to see that the function
\[
    d : \cX\times\cY \ni (\vec{x},\vec{y}) \mapsto d_x(\vx) + d_y(\vy)
\]
is a DGF for the product space $\cX\times\cY$, which is strongly convex with modulus one with respect to the norm $\|(\vx,\vy)\| \defeq \sqrt{\|\vx\|^2_x + \|\vy\|^2_y}$. Furthermore, each proximal step taken with respect to $d$ can be expressed as two independent proximal steps with respect to $d_x$ and $d_y$:
\newcommand{\stack}[2]{\begin{pmatrix}#1\\#2\end{pmatrix}}
\begin{align*}
    \prox{(\tilde{\vx}, \tilde{\vy})}{\begin{array}{c}\vg_x\\\vg_y\end{array}} &= \argmin_{(\vx,\vy)\in\cX\times\cY} \mleft\{ \stack{\vg_x}{\vg_y}^\top\stack{\vx}{\vy} - \div[d]{\stack{\vx}{\vy}}{\stack{\tilde{\vx}}{\tilde{\vy}}}\mright\}\\
    &=\stack{
        \argmin_{\vx\in\cX} \mleft\{ \vg_x^\top \vx - d_x(\vx) + \nabla d_x(\tilde{\vx})^\top \vx\mright\}
    }{
        \argmin_{\vy\in\cY} \mleft\{ \vg_y^\top \vy - d_y(\vy) + \nabla d_y(\tilde{\vy})^\top \vy\mright\}
    } = \stack{\prox{\tilde{\vx}}{\vg_x}}{\prox{\tilde{\vy}}{\vg_y}}.
\end{align*}
Similarly, the $d$-diameter of the product space $\cX\times\cY$ is equal to the sum of diameters of $\cX$ and $\cY$ in their respective proximal setups. Finally, we note that the function
\[
    F: \cX\times\cY \ni (\vx,\vy) \mapsto \stack{\vec{A}\vy}{-\vec{A}^\top \vx},
\]
critical in the analysis of MP~\citep{Ben01:Lectures}, satisfies
\begin{align*}
    \mleft\|F\stack{\vx}{\vy} - F\stack{\vx'}{\vy'}\mright\|_* &= \sqrt{\|\vec{A}(\vy-\vy')\|^2_{x*} + \|\vec{A}^\top(\vx-\vx')\|^2_{y*}\}}\\
    &\le \sqrt{\mleft[\max_{\|\tilde{\vx}\|_x \le 1} \tilde{\vx}^\top \vec{A} (\vy-\vy')\mright]^2 +
    \mleft[\max_{\|\tilde{\vy}\|_y \le 1} (\vx-\vx')^\top \vec{A} \tilde{\vy}\mright]^2}\\
    &\le \sqrt{\|\vec{A}\|^2\cdot\|\vy-\vy'\|_y^2 + \|\vec{A}\|^2\cdot\|\vx-\vx'\|_x^2} = \|\vec{A}\|\cdot\mleft\|\stack{\vx-\vx'}{\vy-\vy'}\mright\|,
\end{align*}
that is, it is $\|\vec{A}\|$-Lipschitz with respect to the norm $\|\cdot\|$ on $\cX\times\cY$.

\cref{algo:mirror prox} shows the sequence of steps taken in every iteration of the MP algorithm.
Compared to EGT, mirror prox has a somewhat simpler structure: it simply takes repeated extrapolated proximal steps. First, a proximal step in the descent direction is taken for both $x$ and $y$. Then, the gradient at those new points is used to take a proximal step starting from the previous iterate (this is the extrapolation part: a step is taken starting from the previous iterate, but with the extrapolated gradient). Finally, the \textit{average} strategy is output.

\begin{figure}[H]\raggedright
    \refstepcounter{algocf}\label{algo:mirror prox}%
    \def\scale{0.94}%
    \makeatletter%
    \setlength{\algocf@ruledwidth}{\linewidth}%
    \@algocf@pre@ruled%
    \global\sbox\algocf@capbox{\hskip\AlCapHSkip
    \setlength{\algocf@lcaptionbox}{\hsize}\addtolength{\algocf@lcaptionbox}{-\AlCapHSkip}%
    \parbox[t]{\algocf@lcaptionbox}{\algocf@captiontext{\fnum@algocf}{Mirror Prox (MP) algorithm.}}
    }%
    \box\algocf@capbox\kern\interspacetitleruled\hrule%
      width\algocf@ruledwidth height\algotitleheightrule depth0pt\kern\interspacealgoruled%
    \renewcommand{\@ResetCounterIfNeeded}{}%
    \makeatother%
    \scalebox{\scale}{\begin{minipage}[t]{.33\linewidth}%
        \RestyleAlgo{plain}%
        \begin{algorithm}[H]%
            \DontPrintSemicolon
            \Fn{\normalfont\textsc{Initialize}()}{
                $t \gets 0$\;
                $\vz_x^0 \gets \argmin_{\hat{\vx} \in \cX} d_x(\hat{\vx})$\;
                $\vz_y^0 \gets \argmin_{\hat{\vy} \in \cY} d_y(\hat{\vy})$\;
            }
        \end{algorithm}
    \end{minipage}}
    \scalebox{\scale}{\begin{minipage}[t]{.38\linewidth}\small
        \RestyleAlgo{plain}
        \begin{algorithm}[H]
            \DontPrintSemicolon
            \Fn{\normalfont\textsc{Iterate}()}{
                $t \gets t + 1$\;
                $\vw_x^t \gets \prox{\vz_x^{t}}{\eta^t \vA\vz_y^{t-1}}$ \;
                $\vw_y^t \gets \prox{\vz_y^t}{-\eta^t \vA^\top \vz_x^{t-1}}$ \;
                $\vz_x^{t+1} \gets \prox{\vz_x^t}{\eta^{t} \vA\vw_y^{t}}$\;
                $\vz_y^{t+1} \gets \prox{\vz_y^t}{-\eta^{t} \vA^\top\vw_x^{t}}$\;
                $\vx^t \gets [\sum_{\tau=1}^t \eta^{\tau}]^{-1}\sum_{\tau=1}^t \eta^{\tau} \vw_x^\tau$\;
                $\vy^t \gets [\sum_{\tau=1}^t \eta^{\tau}]^{-1}\sum_{\tau=1}^t \eta^{\tau} \vw_y^\tau$\;
            }
        \end{algorithm}
    \end{minipage}}
    \raisebox{-3mm}{\scalebox{\scale}{\begin{minipage}[t]{.33\linewidth}\small
                \textbf{Note}: $\{\eta^t\}$ is a sequence of stepsize parameters. A well-known and
    theoretically-sound choice for $\eta^t$ is $\eta^t \defeq \frac{1}{\|A\|}$ 
    for all $t = 0, 1, \dots$ (see also \cref{thm:mirror prox}).
    \end{minipage}}}
    \makeatletter
    \@algocf@post@ruled
    \renewcommand{\@ResetCounterIfNeeded}{\setcounter{AlgoLine}{0}}
    \@ResetCounterIfNeeded{}
    \makeatother

\end{figure}

As we recall in the next theorem, like EGT the mirror prox algorithm converges at rate $O(1/T)$.
\begin{theorem}[\citet{Ben01:Lectures}, Theorem 5.5.1]\label{thm:mirror prox}
    Suppose the stepsize in \cref{algo:mirror prox} is set as $\eta_t = 1/\|A\|$. Then we have
    \[
        \max_{\vy\in \cY} (\vx^t)^\top \vA\vy - \min_{\vx\in \cX} \vx^\top \vA\vy^t  \leq  \frac{\|\vA\| (\Omega_{d_x,\cX} + \Omega_{d_y,\cY}) }{2t}.
    \]
\end{theorem}



    \section{The Sequence-Form Polytope}
\label{sec:treeplexes}


We now describe how the set of Nash equilibria of a two-player zero-sum EFG can
be represented as a bilinear saddle-point problem. The sequential nature of the
decision spaces is represented via the \emph{sequence form}, where each
strategy space $\cX$ and $\cY$ has the form of a convex polytope.

For each player (a.k.a. agent), we assume that we
have a set of decision points $\cJ$, and each decision point $j \in \cJ$ has a
set of actions $A_j$, with $|A_j| = n_j$ actions in total. If the agent takes a
given action $a\in A_j$ at decision point $j$, then $\children{ja} \subset \cJ$
denotes the set of next potential decision points that the agent may face
(which may be empty if no more decisions can occur after taking action $a$ at
$j$). We assume that the decision points form a tree, meaning that
$\children{ja} \cap \children{j'a'} = \emptyset$ for any two pairs $ja$ and
$j'a'$ such that $j\ne j'$ or $a\ne a'$. This is equivalent to assuming that
the corresponding EFG has \emph{perfect recall}, meaning that no agent ever
forgets any past information. 

For an EFG, the decision points $\cJ$ for a given player correspond to the set
of information sets in the game belonging to that player, and a pair $ja$
consisting of a decision point $j$ and action $a$ is referred to as a
\emph{sequence}. The special element $\emptyseq$ is called the \emph{empty} sequence. We use the symbol $\Sigma \defeq \{\emptyseq\} \cup \{ja: j\in \cJ, a\in A_j\}$ to denote the set of all sequences.  Given a decision point $j\in \cJ$ we denote its parent sequence, defined as the last sequence (decision point-action pair) encountered on the path from the root to $j$ with the symbol $p_j$. If the player does not act before $j$, then we conventionally let $p_j$ be set to the empty sequence $\emptyseq$.

Conceptually, a strategy for a player is an assignment of probability distributions over actions $A_j$ at each decision point $j \in \cJ$. To enable expressing the expected utility in the game as a linear function, in this paper we make use of the \emph{sequence-form} representation of strategies~\citep{Stengel96:Efficient,Koller96:Efficient,Romanovskii62:Reduction}. In the sequence-form representation, a strategy is a vector $\vec{x} \in \bbR^{|\Sigma|}_{\ge 0}$ whose entries are indexed over $\Sigma$. The entry corresponding to sequence $ja\in\Sigma$ contains the \emph{product} of the probabilities of all actions at all decision points on the path from the root down to action $a$ at decision point $j$ included. In order to be a valid sequence form strategy, the vector $\vec{x} \in \bbR_{\ge 0}^{|\Sigma|}$ must satisfy the constraints
\begin{equation}\label{eq:sf constraints}
    x_\emptyseq = 1, \qquad\text{and}\quad \sum_{a\in A_j} x_{ja} = x_{p_j} \quad \forall\ j\in\cJ.
\end{equation}
We call the set of all valid sequence-form strategies (that is, all vectors $\vec{x}\in\bbR_{\ge 0}^{|\Sigma|}$ that satisfy the constraints in~\eqref{eq:sf constraints}) the \emph{sequence-form polytope} of the player.

For a two-player zero-sum EFG with perfect recall, the problem of computing a
Nash equilibrium can be cast as a BSPP in the form of \cref{eq:bspp}. In this
formulation, $\cX$ and $\cY$ are the sequence-form polytopes for Player~$1$ and Player~$2$, respectively. The
\emph{payoff matrix} $\vA$ is such that for a pair of sequence-form strategies
$\vx,\vy$, the objective $\vx^\top \vA\vy$ is equal to the expected value achieved by the
second player under those strategies. Thus, the second player wishes to
maximize this objective, while the first player wishes to minimize it. Each
cell in $\vA\in \mathbb{R}^{|\Sigma_1|\times |\Sigma_2|}$, where $\Sigma_1$ and $\Sigma_2$ denote the sets of sequences of Player~$1$ and Player~$2$ respectively, corresponds to a pair
of sequences, one for each player. The matrix is often sparse: each non-zero
entry corresponds to a pair of sequences such that they are the last sequences
on the path to some leaf node (and thus we have zeroes for all cells such that
the corresponding sequences are never the last pair of sequences before the
game ends). The value at a cell is the payoff to the second player at that
leaf, times the product of all chance probabilities on the path to the leaf.

 \subsection{Preliminaries on Dilated Distance-Generating Functions}
\label{sec:dilated dgfs} 

Dilated distance-generating functions are a general framework for constructing
``nice'' DGFs (in the
sense of \cref{sec:nice dgf}) for (the relative interior of the) sequence-form polytopes~\citep{Hoda10:Smoothing}.
Specifically, a dilated DGF for a sequence-form polytope is constructed by taking a weighted
sum over suitable \emph{local} regularizers $d_\emptyseq$ and $d_j$ ($j\in \cJ$), and is of the form
\begin{equation}\label{eq:dilated dgf}
    d: Q \ni \vx \mapsto \alpha_\emptyseq d_\emptyseq(x_\emptyseq) + \sum_{j\in\cJ} \alpha_j d_j^\square(x_{p_j}, (x_{ja})_{a\in A_j}),
\end{equation}
where
\begin{equation}\label{eq:djsquare}
d_j^\square(x_{p_j}, \vec{x}) \defeq \begin{cases}
        0 & \text{if } x_{p_j} = 0\\
        x_{p_j} d_j\mleft(\frac{(x_{ja})_{a\in A_j}}{x_{p_j}}\mright) & \text{otherwise}.
    \end{cases}
\end{equation}

Each local  function $d_j: \Delta^{|A_j|}\to\bbR$ is assumed to be continuously differentiable and strongly convex modulus one on the relative interior of the probability simplex $\Delta^{|A_j|}$. By dividing $(x_{ja})_{a\in A_j}$ by $x_{p_j}$ in~\eqref{eq:djsquare}, we renormalize $(x_{ja})_{a\in A_j}$ to the simplex, measure the DGF there, and then scale that value by $x_{p_j}$ to make it proportional to the ``size'' of $x_{p_j}\cdot \Delta^{|A_j|}$. Finally, the weight $\alpha_j$ is a flexible weight term that can be chosen to ensure good properties. \cite{Hoda10:Smoothing} showed that if each local DGF $d_j$ is strongly convex, then the dilated DGF $d$ is also strongly convex (although they do not give an explicit modulus), and they show that the associated smoothed support function can easily be computed, provided that the smoothed support function for each $d_j$ can easily be computed. 

%

The gradient of a dilated DGF and of its convex conjugate can be computed exactly
in closed form by combining the gradients of each $d_j$ and their convex conjugates, as
shown in \cref{algo:dilated dgf stub}. 

\begin{figure}[ht]\raggedright
    \refstepcounter{algocf}\label{algo:dilated dgf stub}%
    \def\scale{0.96}%
    \makeatletter%
    \setlength{\algocf@ruledwidth}{\linewidth}%
    \@algocf@pre@ruled%
    \global\sbox\algocf@capbox{\hskip\AlCapHSkip
    \setlength{\algocf@lcaptionbox}{\hsize}\addtolength{\algocf@lcaptionbox}{-\AlCapHSkip}%
    \parbox[t]{\algocf@lcaptionbox}{\algocf@captiontext{\fnum@algocf}{Gradient and smoothed support function implementation for general dilated DGFs.}}
    }%
    \box\algocf@capbox\kern\interspacetitleruled\hrule%
      width\algocf@ruledwidth height\algotitleheightrule depth0pt\kern\interspacealgoruled%
    \makeatother%
    \scalebox{\scale}{\begin{minipage}[t]{.52\linewidth}%
        \RestyleAlgo{plain}%
        \begin{algorithm}[H]%
            \DontPrintSemicolon
            \Fn{\normalfont\textsc{Gradient}($\vx \in \relint Q$)}{
                $\vg \gets \vec{0} \in \bbR^{|\Sigma|}$\;
                \For{$j \in \cJ$ in bottom-up order}{
                    $(g_{ja})_{a\in A_j} \gets (g_{ja})_{a\in A_j} + \alpha_j\, \nabla d_j\mleft(\frac{(x_{ja})_{a\in A_j}}{x_{p_j}}\mright)$\;
                    $g_{p_j}\gets g_{p_j} + \alpha_j \,d_j\mleft(\frac{(x_{ja})_{a\in A_j}}{x_{p_j}}\mright)$\;
                    $g_{p_j}\gets g_{p_j} - \alpha_j\,\nabla d_j\mleft(\frac{(x_{ja})_{a\in A_j}}{x_{p_j}}\mright)^\top\mleft(\frac{(x_{ja})_{a\in A_j}}{x_{p_j}}\mright)$\hspace*{-1cm}
                }
                $g_\emptyseq \gets g_\emptyseq + \alpha_\emptyseq \nabla d_\emptyseq(x_\emptyseq)$\;
                \textbf{return} $\vg$\;
            }
        \end{algorithm}
    \end{minipage}}
    \scalebox{\scale}{\begin{minipage}[t]{.50\linewidth}\small
        \RestyleAlgo{plain}
        \begin{algorithm}[H]
            \DontPrintSemicolon
            \Fn{\normalfont\textsc{ConjugateGradient}($\vg \in \bbR^{|\Sigma|}$)}{
                $\vz \gets \vec{0} \in \bbR^{|\Sigma|}$\;
                $z_\emptyseq \gets 1$\;
                \For{$j\in\cJ$ in bottom-up order}{
                    $(z_{ja})_{a\in A_j} \gets \nabla d_j^*((g_{ja})_{a \in A_j})$\;
                    $g_{p_j} \gets g_{p_j} - d_j((z_{ja})_{a\in A_j}) + \sum_{a\in A_j} g_{ja} z_{ja}$\;
                }
                \For{$j\in\cJ$ in top-down order}{
                    \For{$a\in A_j$}{
                        $z_{ja} \gets z_{p_j} \cdot z_{ja}$\;
                    }
                }
                \textbf{return} $\vz$\;
            }
        \end{algorithm}
    \end{minipage}}
    \makeatletter
    \@algocf@post@ruled
    \makeatother
\end{figure}


The local DGFs must be chosen so that they are compatible with the relative interior of the simplex. For a given simplex $\Delta^k$, these are usually chosen either as the \emph{entropy DGF} $d(\vy) = \log k + \sum_{i} y_i\log y_i $ (we let $y_i \log y_i = 0$ whenever $y_i = 0$) or \emph{Euclidean DGF} $d(\vy) = \frac{1}{2}\sum_i(y_i - 1/k)^2 $. These are both 1-strongly convex on $\relint\Delta^k$ (for entropy wrt. the $\ell_1$ norm and for Euclidean wrt. the $\ell_2$ norm), and their associated smoothed support functions can be computed in $O(k)$ time (see e.g.~\citet{Ben01:Lectures,Condat16:Fast}).

One of the most important properties of dilated DGFs is that
they lead to a ``nice'' DGF as long as each local convex conjugate gradient $\nabla d_j^*$ can be computed in time linear in $|A_j|$. In particular, this makes the dilated entropy and dilated Euclidean DGFs ``nice'' DGFs. 
In contrast, the standard Euclidean DGF applied to the overall polytope $Q$ is not ``nice'': It requires $|\Sigma|\log|\Sigma|$ time to resolve its convex conjugate gradient.

For dilated DGFs, the strongest general result on the strong-convexity modulus comes from \citet{Farina19:Optimistic}, where the authors show that if each local DGF $d_j$ is strongly convex modulus one with respect to the $\ell_2$ norm, and we set $\alpha_j =  2 + 2\max_{a\in A_j} \sum_{j' \in \children{ja}} \alpha_{j'}$ for each $j\in \cJ$, then $d$ is strongly convex modulus one with respect to the $\ell_2$ norm on $Q$.

 \subsection{Preliminaries on the Dilated Entropy Distance-Generating Function}
The dilated \emph{entropy} DGF is the instantiation of the general dilated DGF framework of \cref{sec:dilated dgfs} with the particular choice of using the (negative) entropy function at each decision node. 
In particular, for any choice of weights $\alpha_\emptyseq,\alpha_j > 0$ it is the regularizer of the form%
\footnote{In this paper, we let $0\log(0) = 0\log(0/0)=0$. Since the dilated entropy DGF is a Legendre function, it is guaranteed that all iterates and prox-steps will remain the relative interior of the optimization domain at all times, thus avoiding the non-differentiability issue of the entropy function at the boundary of $Q$.}
\begin{equation}\label{eq:regu}
    Q \ni \vx \mapsto \alpha_\emptyseq x_\emptyseq \log x_\emptyseq + \sum_{j\in\cJ}\alpha_j\mleft(x_{p_j}\log |A_j| + \sum_{a \in A_j} x_{ja}\log\mleft(\frac{x_{ja}}{x_{p_j}}\mright)\mright).
\end{equation}

We will now briefly review existing results specific to the dilated entropy DGF, for which stronger results are known than for the general class of dilated DGFs.
A central result in the present paper is to show that there exist DGFs for sequence-form polytopes which are better than the dilated entropy DGF, but that these DGFs can be partially recast in a dilated form, in order to enable efficient computation of the smoothed support function.

First, as a direct consequence of the more general discussion in \cref{sec:dilated dgfs} and \cref{algo:dilated dgf stub}, the dilated entropy DGF is a ``nice'' DGF (in the precise sense of \cref{sec:nice dgf}) no matter the choice of weights $\alpha$. In particular, in the case of the negative entropy functions $d_j(\vx) = \log |A_j| + \sum_{a\in A_j} x_{ja} \log x_{ja}$, one has
\[
\big(\nabla d_j(\vx)\big)_{a} = 1 + \log x_{a}, \quad \big(\nabla d_j^*(\vg)\big)_a = \frac{e^{g_a}}{\sum_{a' \in A_j} e^{g_{a'}}} \qquad\forall a \in A_j, \vx \in \relint \Delta^{|A_j|}, \vg \in \bbR^{|A_j|}.
\]
By plugging the above expression in the template of \cref{algo:dilated dgf stub} we obtain linear-time exact algorithms to compute $\nabla \regu$ and $\nabla \regu^*$.

\citet{Kroer20:Faster} show that the dilated entropy DGF is strongly convex modulus $1 / M_Q$ with respect to the $\ell_1$ norm, when the weights $\alpha$ are chosen as in the following definition.
\begin{definition}[Kroer et al. dilated entropy DGF, $\psi$]
    Define the DGF weights $\beta_j$ recursively as
\[
    \beta_\emptyseq \defeq 2 + 2\sum_{j \in \children{\emptyseq}} \beta_{j},\qquad\quad
    \beta_j \defeq 2 + 2\max_{a\in A_j} \sum_{j' \in \children{ja}} \beta_{j'}\qquad\forall j\in\cJ.
\]
The resulting instantiation of the dilated entropy DGF is called the \emph{Kroer et al. dilated entropy DGF} and denoted $\psi$:
\[
\psi : Q \ni \vx \mapsto \beta_\emptyseq x_\emptyseq \log x_\emptyseq + \sum_{j\in\cJ}\beta_j\mleft(x_{p_j}\log |A_j| + \sum_{a \in A_j} x_{ja}\log\mleft(\frac{x_{ja}}{x_{p_j}}\mright)\mright).
\]
\label{def:kroer dgf}
\end{definition}

\begin{remark}
    On the surface, the strong convexity modulus of $\frac{1}{M_Q}$ with respect to the
    $\ell_1$ norm might appear less appealing than the modulus $1$ obtained by using
    the $\ell_2$ norm. However, recall that the norm that is used
    to measure strong convexity affects the value of the operator norm of $\vA$, which
    is significantly smaller under the $\ell_1-\ell_\infty$ operator norm (where it is equal to $\max_{ij} |A_{ij}|$) than the $\ell_2-\ell_2$ operator norm for strong convexity with respect to the $\ell_2$ norm.
\end{remark}

One drawback of both the general and entropy-specific dilated DGFs developed in the past is that they have an exponential dependence on the depth of the sequence-form polytope. In particular, note that the factor of $2$ in the recursive definition of the weights means that the factor $\beta_j$ for some root decision point is growing at least on the order of $2^{\depth_Q}$, where $\depth_Q$ is the depth of the tree-form sequential decision process. For many sequence-form polytopes this might be acceptable: if the tree-form sequential decision process is reasonably balanced, then the number of decision points is also exponential in depth. However, for other sequence-form polytopes this would be unacceptable: the most extreme case would be a single line of decision points, where the number of decision points is linear in $\depth_Q$, but the $\beta_j$ at the root is exponentially large. This exponential dependence on depth also enters the convergence rate of the FOMs, since it effectively acts as a scalar on the polytope diameter $\Omega$ induced by $\depth_Q$.
The present paper was motivated by the need to soundly resolve that drawback, thus introducing the first ``nice'' DGF (in the sense of \cref{sec:nice dgf}) with guaranteed polynomially-small diameter for any decision point.

 \section{The Dilatable Global Entropy Distance-Generating Function}\label{sec:global dgf}
We now develop our new DGF for sequence-form polytopes. The DGF is based on a scaled variant of the standard entropy DGF (which we will also refer to as the ``global entropy'' to distinguish it from the dilated entropy), which is strongly convex modulus one on the hypercube $[0,1]^{|\Sigma|}$. 

\begin{definition}[Dilatable global entropy]
    The \emph{dilatable global entropy distance generating function} $\tilde\regu$ 
    is the function $\tilde\regu : Q \to \Rp$ defined as
    \begin{align*}
        \tilde\regu: Q \ni \vx\mapsto w_\emptyseq x_\emptyseq \log(x_\emptyseq) + \sum_{j \in\cJ}\sum_{a \in A_j} w_{ja} x_{ja} \log x_{ja} + \sum_{j\in\cJ} \gamma_j x_{p_j} \log |A_j|
    ,\end{align*}
    where each $\gamma_j \ge 1$ ($j\in\cJ$) is defined resursively as
    \begin{align}
        \gamma_\emptyseq = 1 + \sum_{j\in\children{\emptyseq}} \gamma_{j},\qquad\quad
    \gamma_j \defeq 1 + \max_{a \in A_j} \mleft\{ \sum_{j' \in \children{ja}} \gamma_{j'}
        \mright\} \qquad \forall\ j\in\cJ
        \label{eq:dilated weights}
    ,\end{align}
    and each $w_\sigma \ge 1$ ($\sigma \in \Sigma$) is defined recursively as
    \begin{align*}
        w_\emptyseq &\defeq \gamma_\emptyseq - \sum_{j \in \children{\emptyseq}} \gamma_j,\\ w_{ja} &\defeq \gamma_j - \sum_{j' \in \children{ja}} \gamma_{j'} = 1 + \max_{a' \in A_j}\mleft\{\sum_{j' \in \children{ja'}} \gamma_{j'}\mright\} - \sum_{j' \in \children{ja}} \gamma_{j'} \qquad\forall ja\in\Sigma.
    \end{align*}
    \label{def:dilatable global entropy}
\end{definition}

The weights $\gamma_j$ defined in~\eqref{eq:dilated weights} are very similar to the ones given for the dilated DGFs in the previous section, except that the whole expression is smaller by a factor of two. Avoiding this factor of two is crucial, because it allows us to avoid the exponential dependence on depth.
Here, it is easy to see that $\gamma_j$ is upper bounded by the number of decision points in the subtree
rooted at $j$, so $\gamma_j$ is at most polynomial in the size of the sequential
decision problem. In fact, it is not hard to show that if $j$ is the sole root decision point, then $\gamma_j$ is equal to $\max_{\vx\in Q} \|\vx\|_1$.
%


\paragraph{Dilatability.}\quad The adjective \emph{dilatable} comes from the key property that the dilatable global entropy is equal to a specific dilated entropy regularizer $\regu$, \emph{on the sequence-from strategy space} $Q$. That equality does not hold outside the sequence-form polytope, and this means that the gradient of $\tilde\regu(\vx)$, which is used in the FOMs that we consider, may differ as well. More precisely, consider the dilated entropy DGF defined as
\[
    \regu: Q \ni \vx \mapsto \gamma_\emptyseq x_\emptyseq\log x_\emptyseq + \sum_{j\in\cJ} \gamma_j \mleft(x_{p_j} \log |A_j| + \sum_{a\in A_j} x_{ja} \log\mleft(\frac{x_{ja}}{x_{p_j}}\mright)\mright).  
\]
Then, we have the following.

\begin{theorem}\label{thm:dilatability}
    The dilatable global entropy DGF and the dilated entropy DGF coincide on the
    polytope of sequence-form strategies $Q$, that is, $\tilde\regu(\vx) = \regu(\vx)$ for all $\vx \in Q$.
\end{theorem}
\begin{xproof}
We start by expanding the definition of $\regu(\vx)$:
\begin{align*}
    \regu(\vx) &\defeq \sum_{j \in \cJ}\sum_{a \in A_j} \gamma_j x_{ja} \log\mleft( \frac{x_{ja}}{x_{p_j}} \mright) + \sum_{j\in\cJ} \gamma_j x_{p_j} \log |A_j|\\
             &= \sum_{j\in\cJ} \sum_{a\in A_j} \gamma_j x_{ja} \log x_{ja}  - \sum_{j \in \cJ}\sum_{a \in A_j} \gamma_j x_{ja} \log x_{p_j}+ \sum_{j\in\cJ} \gamma_j x_{p_j} \log |A_j|.\numberthis{eq:xxx}
\end{align*}
Given the assumption $\vx \in Q$, it holds that $\sum_{a \in A_j} x_{ja} = x_{p_j}$ for all $j \in \cJ$ and so we can simplify the middle summation in \eqref{eq:xxx} and obtain
\begin{align*} 
  \regu(\vx) &= \sum_{j\in\cJ} \sum_{a\in A_j} \gamma_j x_{ja} \log x_{ja}  - \sum_{j \in \cJ} \gamma_j x_{p_j} \log x_{p_j} + \sum_{j\in\cJ} \gamma_j x_{p_j} \log |A_j|\\
             &= \sum_{j\in\cJ} \sum_{a\in A_j} \gamma_j x_{ja} \log x_{ja}  - \sum_{j \in \cJ} \sum_{a \in A_j} \sum_{j' \in \children{ja}} \gamma_{j'} x_{ja} \log x_{ja} + \sum_{j\in\cJ} \gamma_j x_{p_j} \log |A_j|\\
             &= \sum_{j\in\cJ} \sum_{a\in A_j} \gamma_j x_{ja} \log x_{ja}  - \sum_{j \in \cJ} \sum_{a \in A_j}\mleft(\sum_{j' \in \children{ja}} \gamma_{j'}\mright) x_{ja} \log x_{ja} + \sum_{j\in\cJ} \gamma_j x_{p_j} \log |A_j|\\
			 &= \sum_{j\in\cJ}\sum_{a\in A_j} w_{ja} x_{ja} \log x_{ja} + \sum_{j\in\cJ} \gamma_j x_{p_j} \log |A_j| = \tilde\regu(\vx)
,\end{align*}
as we wanted to show.
\end{xproof}

\paragraph{``Nice''ness.}\quad We now show that our dilatable global entropy regularizer is ``nice'' in the sense of \cref{sec:nice dgf}, that is, its gradient and the gradient of its convex conjugate can be computed exactly in linear time in $|\Sigma|$.
\begin{itemize}
    \item The gradient of $\tilde\regu$ can be trivially computed in closed form and linear time in $|\Sigma|$ starting from \cref{def:dilatable global entropy} as
    \[
        (\nabla\tilde\regu(\vx))_{\sigma} = (1 + \log x_\sigma)w_\sigma + \sum_{j\in\children{\sigma}} \gamma_j \log |A_j|\qquad\forall \sigma\in\Sigma, \vx \in \intQ.
    \]
    \item Using the dilatability property, we have that the gradient of the convex conjugate satisfies
        \begin{align*}
        \nabla\tilde\regu^*(\vg) = \argmax_{\vx \in Q} \mleft\{ \vg^\top \vx - \tilde\regu(\vx) \mright\}
            = \argmax_{\vx \in Q} \mleft\{ \vg^\top \vx - \regu(\vx) \mright\}
            = \nabla\regu^*(\vg)
        ,\end{align*}
        where we used the dilatability property (\cref{thm:dilatability}) in the second equality. Therefore, since $\regu$ is a dilated DGF and its smoothed support function can be computed in linear time, the smoothed support function of $\tilde\regu$ can be computed in linear time in $|\Sigma|$. Similarly, $\prox{\vc}{\vg} = \nabla\regu^*(-\vg + \nabla\tilde\regu(\vc))$, where the internal gradient is with respect to $\tilde\regu$, as opposed to $\regu$ as in the general reduction in~\cref{eq:prox as argconjugate}. 
\end{itemize}

\paragraph{Strong convexity.}\quad
On the other hand, we now show that $\tilde\regu$ has the advantage of a better strong convex modulus, compared to the class of dilated entropy DGFs.

\begin{theorem}
    The dilatable global entropy function $\tilde\regu: Q\to\Rp$ is a DGF for the sequence-form polytope $Q$, $1$-strongly convex on $\relint Q$ with respect to the $\ell_2$ norm.
    \label{thm:DGE sc l2}
\end{theorem}
\begin{xproof}
The function $\tilde\regu$ is twice-differentiable on $(0,1)^{|\Sigma|}\supseteq \intQ$. Using~\eqref{eq:second order sc} we conclude that $\tilde\regu$ is $1$-strongly convex, since the Hessian is
\begin{align*}
    \nabla^2 \tilde\regu(\vx) = \textrm{diag}\mleft(\mleft\{\frac{w_{ja}}{x_{ja}}\mright\}_{ja\in\Sigma}\mright)
                            \succeq I,
\end{align*}    
where we used the inequalities $0\le x_{ja} \le 1$ and $w_{ja} \ge 1$. 
Next, we verify that the minium of $\tilde\regu$ is $0$. Because of dilatability, $\argmin_{\vx\in Q}\tilde\regu(\vx)=\argmin_{\vx\in Q}\regu(\vx) = \nabla\regu^*(\vec{0}) = \vec{m}$ is the uniform strategy, that is, the strategy that assigns probability $1/|A_j|$ to each action at each decision point $j\in\cJ$ (this corresponds, in the sequence-form representation, to $x_{ja} = x_{p_j} / |A_j|$ for all $ja\in\Sigma$ and $x_\emptyseq=1$). Substituting into the definition of $\tilde\regu$, we obtain
\begin{align*}
    \tilde\regu(\vec{m}) &= \sum_{j\in\cJ}\gamma_j \mleft(x_{p_j}\log|A_j| + \sum_{a\in A_j} \frac{x_{p_j}}{|A_j|}\log\frac{1}{|A_j|}\mright) = \sum_{j\in\cJ} \gamma_j (x_{p_j}\log |A_j| - x_{p_j}\log |A_j|) = 0.
\end{align*}
\end{xproof}

\begin{theorem}
    The dilatable global entropy function $\tilde\regu$ is strongly convex modulus $1 / M_Q$ with respect to the $\ell_1$ norm on $\intQ$.
    \label{thm:DGE sc}
\end{theorem}
\begin{xproof}
    Using the second-order definition of strong convexity, we wish to show that
    the inequality $\vec{m}^\top\nabla^2 \tilde\regu(\vx) \vec{m} \geq
    \frac{1}{M_Q}\|\vec{m}\|_1^2$ holds for any $\vec{m}\in \mathbb{R}^{|\Sigma|}$.
    Expanding the Hessian matrix and using the fact that $w_{ja} \ge 1$ for all $ja\in\Sigma$ gives
\begin{align*}
    \vec{m}^\top \nabla^2 \tilde\regu(\vx) \vec{m} = \vec{m}^\top\diag\mleft(\mleft\{\frac{w_{ja}}{x_{ja}}\mright\}_{ja\in\Sigma}\mright) \vec{m}
    \geq \sum_{ja \in \Sigma} \frac{m_{ja}^2}{x_{ja}}.\numberthis{eq:bound hessian}
\end{align*}    
On the other hand, by expanding the definition of $\|\vec{m}\|_1^2$ and applying the Cauchy-Schwarz inequality, we have
\begin{align*}
    \|\vec{m}\|_1^2 &= \left(\sum_{ja\in\Sigma} |m_{ja}|\right)^2 
    = \left(\sum_{ja\in\Sigma} \frac{|m_{ja}|}{\sqrt{x_{ja}}} \sqrt{x_{ja}}\right)^2 
    \leq \left( \sum_{ja\in\Sigma} \frac{m_{ja}^2 }{ x_{ja}} \right) \left(\sum_{ja\in\Sigma} x_{ja}\right)
    \leq \left( \sum_{ja\in\Sigma} \frac{m_{ja}^2 }{ x_{ja}} \right) M_Q.
\end{align*}
Substituting~\eqref{eq:bound hessian} into the last inequality yields a proof of the desired
strong convexity modulus $1 / M_Q$.
\end{xproof}

\paragraph{Diameter.}\quad
The properties above immediately imply that our dilatable global entropy DGF satisfies all the requirements for a prox setup on the polytope of sequence-form strategies $Q$. Here we complete the analysis by giving bounds on the diameter induced by $\tilde\regu$.
\begin{theorem}
    The $\tilde\regu$-diameter $\Omega_{\tilde\regu,Q}$ of $Q$ is at most $M_Q^2 \max_{j'\in \cJ} \log |A_{j'}| $.
    \label{thm:DGE treeplex diameter}
\end{theorem}
\begin{xproof}
    By the definition of the polytope diameter and the fact that we chose our DGFs such that $\min_{\vx \in \cA}\tilde\regu(\vx) = 0$, we have
    \begin{align*}
        \Omega_{\tilde\regu,Q} &\leq  \max_{\vx\in Q} \tilde\regu(\vx)
        \leq \max_{\vx\in Q}\sum_{j\in\cJ} \gamma_j x_{p_j} \log |A_j|
        \leq \max_{\vx\in Q}\max_{j' \in \cJ}\;\log |A_{j'}| \sum_{j\in\cJ} \gamma_j x_{p_j} \\
        &\leq M_Q \max_{j' \in\cJ}\;\log |A_{j'}| \sum_{j\in\cJ}  x_{p_j} 
        \leq M_Q^2 \max_{j'\in \cJ}\;\log |A_{j'}|,
    \end{align*}
    where the second inequality is by noting that $\log x_{ja} \leq 0$ since $x_{ja} \leq 1$ for all $ja\in \Sigma$,
    the fourth inequality is by noting that $\gamma_j$ is largest at root decision points, where it is at most $M_Q$, and 
    the fifth inequality upper bounds $\sum_{j\in\cJ}  x_{p_j}$ by $M_Q$.
\end{xproof}

\citet{Kroer20:Faster} show that the dilated entropy DGF with weights $\beta$ leads to a polytope diameter
\[
   2^{\depth_Q+2}M_Q^2 \max_{j'\in \cJ} \log |A_{j'}|.
\]
Our DGF improves that polytope diameter by a factor of $2^{\depth_Q+2}$. Thus, we are the first to achieve a polytope diameter with no exponential dependence on the depth $\depth_Q$ of the sequence-form polytope.

Summing up our results on the dilatable global entropy, we have shown that it enjoys the same fast smoothed support function computation as the dilated entropy DGF while having a better way to achieve strong convexity modulus $1 / M_Q$. In particular, the existing dilated entropy setup requires the weight parameters $\beta$ to grow exponentially in the depth of the sequence-form polytope, whereas we have only a linear growth in those weights. 
More concretely, this means that the largest weights $\max_{j\in \cJ}\beta_j$ in the dilated entropy DGF are larger than the largest weights $\max_{j\in \cJ}\gamma_j$ in the dilatable global entropy DGF by a factor of more than $2^{\depth_Q}$. 
This in turn allowed us to achieve a better polytope diameter by a factor of $2^{\depth_Q+2}$ while retaining the same strong convexity modulus.


\section{Scaled Extension and Correlated Decision Spaces}
In this section we extend and generalize both the framework of dilated DGFs and
the dilatable global entropy DGF to more complex combinatorial domains
than sequence-form polytopes. In particular, we show that dilated DGFs and the
dilatable global entropy apply to sets that can be constructed through
composition of \emph{scaled extension}, a convexity-preserving operation that
was recently proposed as a general way of constructing sequential decision spaces
in the presence of correlation between the strategies of two or more players~\citep{Farina19:Efficient}.

Our generalization begets the first ``nice'' regularizers (in the sense of \cref{sec:nice dgf}) for correlated strategy spaces, which in turn enables us to construct the first FOMs that guarantee convergence to optimal correlated equilibria and optimal ex-ante team coordinated equilibria at a rate of $1/T$ in certain classes of games where these equilibria can be found in polynomial time.

\label{sec:scaled extension}

We start by recalling the definition of scaled extension.

\begin{definition}[Scaled extension~\citep{Farina19:Efficient}]\label{def:scaled extension}
    Let $\cU$ and $\cV$ be nonempty convex sets, and let $h : \cU \to\Rp$ be a
        nonnegative affine real function. The \emph{scaled extension} of $\cU$
        with $\cV$ via $h$ is defined as the convex set
    \[
        \cU \ext^h \cV \defeq \{(\vec{u}, h(\vu)\,\vec{v}) : \vec{u} \in \cU,\ \vec{v} \in \cV\}
    .\]
\end{definition}

In the appendix we prove the following property, which will be useful for the construction of our DGF for sets obtained through scaled extension.

\begin{restatable}{lemma}{lemrelintscext}\label{lem:relint scext}
    Let $\cU \subset \bbR^m, \cV \subset \bbR^n$ be bounded sets, and $h : \cU \to \bbR$ be an affine function $h : \vu \mapsto \vec{a}^\top \vu + b$, nonnegative on $\cU$ and strictly positive on $\relint \cU$. Then,
    \[
        \relint(\cU \ext^h \cV) = (\relint \cU) \ext^h (\relint\cV).
    \]
\end{restatable}



\subsection{Preliminaries on Correlation and Triangle-Freeness}
\label{sec:correlation polytope}
As reviewed earlier in the paper, Nash equilibria in two-player zero-sum EFGs can be expressed as BSPPs. It turns out that several other solution concepts can also be formulated as BSPPs via more intricate convex-polytope constructions.
In this section we briefly describe two important solution concepts that can be expressed as BSPPs: several variants of optimal correlated equilibria and ex-ante team coordinated equilibria.

In correlated equilibria, the rationality assumption of Nash equilibrium is relaxed in order to allow for coordination between the players. It is assumed that a \emph{mediator} will recommend actions to be taken. 
In \emph{correlated equilibria}, each player 
sees the recommended action before deciding whether to take it.
In \emph{coarse correlated equilibria}, the players must commit to acting according to the recommended strategy before the recommendation is revealed. 
In all these solution concepts, the recommended strategy is sampled by a mediator, from some correlated distribution which is known to the players.


We will consider three types of correlated equilibria in two-player EFGs. 
\begin{enumerate}
    \item   
    \textbf{Extensive-form correlated equilibrium (EFCE)}:
    The mediator incrementally recommends individual moves to the players. Every time a player faces a decision point, the mediator privately reveals a recommended move for that decision point to that player. If a player chooses to disregard a recommendation, then the mediator immediately stops issuing recommendations to that player forever~\citep{Stengel08:Extensive}.
    \item   
    \textbf{Extensive-form coarse correlated equilibrium (EFCCE)}:
    The mediator incrementally recommends individual moves to the players, but at each decision point, the player must decide whether to follow the recommendation \emph{before} seeing the recommendation~\citep{Farina20:Polynomial}.
    \item 
    \textbf{Normal-form coarse correlated equilibrium (NFCCE)}: each player will be recommended a strategy from the normal-form representation of the EFG, but they must decide whether to commit to playing the recommended strategy before seeing the recommendation~\citep{Moulin78:Strategically}. 
\end{enumerate}

\citet{Farina20:Polynomial} show that EFCE are a subset of EFCCE, and that EFCCE are a subset of NFCCE.
They also show that for triangle-free decision problems (or EFGs),\footnote{The triangle-free condition is rather technical, and so we omit its exact definition here as it is beyond the scope of the paper. The most natural class of games that it captures is the set of EFGs where all chance moves are public, that is, observed by all players.} the set $\Xi$ of all possible correlated plans between two players can be represented via recursive applications of the scaled extension operator. Thus, we can write BSPPs of the form
\begin{align}
    \argmin_{\vx\in \Xi} \max_{\vy \in \cY} \vx^\top \cA\vy,
    \label{eq:correlation bspp}
\end{align}
where the minimization over $\vx$ represents the choice of a correlated plan for the two players, and the maximization over $\cY$, intuitively, represents different ways of rejecting the mediator's recommendation.
By carefully choosing $\cY$, we can enforce different types of correlated equilibrium behavior. 
Now, as long as we have a ``nice'' DGF for scaled extensions and a ``nice'' DGF for the polytope $\cY$,  we can apply fast FOMs to the computation of the corresponding type of correlated equilibrium. Appropriate characterizations for $\cY$ exist in the case of EFCE~\citep{Farina19:Correlation}, EFCCE~\citep{Farina20:Coarse}, and NFCCE~\citep{Farina20:Coarse}. In each case, $\cY$ is itself a  polytope that can be constructed via scaled extension (though simpler than $\Xi$).

Reviewing the details of the scaled-extension-based construction of the polytope of correlation plans $\Xi$
is beyond the scope of this paper. Here, we take the decomposition as a given, and we are concerned with the task of constructing a suitable proximal setup for sets that, like $\cY$ and $\Xi$, can be expressed through composition of scaled extension operations. 

In addition to two-player correlated equilibrium problems discussed above, one can also capture \emph{adversarial team games} with the scaled extension DGFs that we will construct. In the adversarial team game that we consider, two players on a team (meaning that they share the same payoffs) are trying to correlate their strategies so as to maximize utility against an opponent whose utility is exactly the opposite of theirs (i.e., it is a zero-sum game between the team and the opponent). This solution concept is called \emph{team-maxmin equilibrium with coordination (TMECor)}~\citep{Celli18:Computational}.
The set of correlated plans for the two players on the team can again be expressed with $\Xi$ if their two decision spaces are triangle free~\citep{farina2020faster}. Then, the strategies of the opponent are simply a sequence-form polytope $Q$, so $\cY=Q$. It follows that DGFs for both polytopes can be chosen as the DGE from our paper.

For EFCE, EFCCE, NFCCE, and TMECor, it will follow from our results below that it is possible to construct a ``nice'' DGF for the polytope of correlation plans when the game is triangle free. By applying this DGF to a method such as EGT or mirror prox, we get the first $1/T$ iterative method for converging to each of these solution concepts, with only a linear cost per iteration. In contrast, prior iterative approaches  converge at a rate of $1/\sqrt{T}$, and sometimes still require significantly more expensive projections at every iteration (e.g.,~\citet{Farina19:Correlation,Farina19:Efficient}).

\subsection{Dilated Distance-Generating Functions for Scaled Extension}
\label{sec:scext dilated}

In this section we show that the construction of dilated DGFs can be generalized to sets obtained through scaled extension. Let $\cZ \defeq \cU \ext^h \cV$ be a set constructed by scaled extension of $\cU$ with $\cV$ using a linear function $h$, and assume that ``nice'' DGFs $d_u, d_v$ for $\cU$ and $\cV$ respectively have been chosen. In \cref{prop:dilated scaled ext} we show that $d_u$ and $d_v$ can be combined to give a composite ``nice'' DGF for $\cZ$. 

\begin{proposition}\label{prop:dilated scaled ext}
    Let
    \begin{itemize}
        \item $\displaystyle\cZ \defeq \cU \ext^h \cV$, where $\cU\subseteq\bbR^m, \cV\subseteq\bbR^n$ are compact convex sets, and $h$ is a linear function $\vu \mapsto \va^\top \vu$ such that $h(\vu)\ge 0$ for all $\vu \in \cU$, and $h(\vu) > 0$ for all $\vu\in \relint\cU$;
        \item $d_u: \cU \to \bbR$ and $d_v: \cV \to \bbR$ be DGFs for $\cU$ and $\cV$, strongly convex with respect to norms $\|\cdot\|_u$ and $\|\cdot\|_v$, respectively;
        \item $\alpha_v > 0$ be a positive scalar.
    \end{itemize}
    Then, the function
    \[
        d_z: \cZ \ni (\vu, \vw) \mapsto d_u(\vu) + \alpha_v\, d^\square_v(\vu,\vw),
        \qquad d^\square_v(\vu, \vw) \defeq \begin{cases}
            0 & \text{if } h(\vu) = 0\\
            h(\vu) d_v\mleft(\frac{\vw}{h(\vu)}\mright) & \text{if } h(\vu) > 0
        \end{cases}
        \numberthis{eq:dz}
    \]
    is a DGF for $\cZ$. Furthermore, if $d_u$ and $d_v$ are ``nice'' DGFs, so is $d_z$. 
\end{proposition}
\begin{xproof}
    We verify that all conditions stated in \cref{def:dgf} are satisfied.
    \begin{description}
    \item[Continuity] 
    In order to show that $d_z$ is continuous, it suffices to show that $d_v^\square$ is. Let $(\vu',\vw')\in\cZ$. If $h(\vu') > 0$, then by continuity of $h$ there exists a neighborhood of $(\vu',\vw')$ where $d_v^\square(\vu,\vw)$ coincides with the function $h(\vu)d_v(\vw/h(\vu))$, which is clearly continuous since $d_v$ is continuous by hypothesis and $h$ is a linear function. Hence, we are only left with checking continuity in the case where $h(\vu') = 0$. First, note that $d^\square_v$ is a nonnegative function given the hypotheses, and that $M \defeq \max_{\vv\in\cV} d_v(\vv)$ is a finite constant by Weierstrass' theorem since $d_v$ is a continuous function defined over a compact domain. For any $\epsilon > 0$, consider the neighborhood $N_{r(\epsilon)}$ of radius $r(\epsilon) \defeq \epsilon/(2\|\vec{a}\|_2 M)$ centered in $(\vu',\vw')$. Then, for any point $(\vu,\vw)$ in the neighborhood, we have that either $h(\vu) = 0$, yielding $d_v^\square(\vu,\vw) = 0$, or $h(\vu) > 0$, yielding
    \begin{align*}
        d_v^\square(\vu,\vw) &= h(\vu) d_v\mleft(\frac{\vw}{h(\vu)}\mright) = \vec{a}^\top (\vu-\vu') d_v\mleft(\frac{\vw}{h(\vu)}\mright)
        \le \|\vec{a}\|_2 \|\vu-\vu'\|_2 M \le \frac{\epsilon}{2} < \epsilon.
    \end{align*}
    So, in either case we have that $0 \le d^\square(\vu,\vw) < \epsilon$ for all $(\vu,\vw) \in N_{r(\epsilon)}$, that is, $d_v^\square$ is continuous at $(\vu',\vw')$ as we wanted to show.

    \item[Differentiability]
    Since $h(\vu) > 0$ for all $\vu \in \relint\cU$  by hypothesis, \cref{lem:relint scext} guarantees that $\relint\cZ = (\relint \cU) \ext^h (\relint \cV)$. Furthermore, using the definition of $d_v^\square$, we have that
    \[
        d_z(\vu,\vw) = d_u(\vu) + \alpha_v h(\vu) d_v\mleft(\frac{\vw}{h(\vu)}\mright)\qquad\forall\,(\vu,\vw) \in \relint \cZ.
    \]
    Hence, direct computation shows that at any $(\vu, \vw) \in \relint\cZ$,
    \begin{equation}\label{eq:gradient dz}
        \nabla d_z(\vu,\vw) = \begin{pmatrix} \nabla d_u(\vu)  + \alpha_v \mleft(d_v\mleft(\frac{\vw}{h(\vu)}\mright)- \nabla d_v\mleft(\frac{\vw}{h(\vu)}\mright)^\top \frac{\vw}{h(\vu)}\mright)\vec{a} \\
        \alpha_v \nabla d_v\mleft(\frac{\vw}{h(\vu)}\mright)
        \end{pmatrix}.
    \end{equation}
    (Note that~\eqref{eq:gradient dz} only needs the evaluation of $\nabla d_u$ and $\nabla d_v$ in the relative interior of $\cU$ and $\cV$ respectively, since $\relint\cZ = (\relint\cU)\ext^h(\relint\cV) $ by \cref{lem:relint scext}.)

    \item[Strong convexity] 
    We now verify that $d_z$ is strongly convex on $\relint\cZ$. We start from the following result, which follows easily from algebraic manipulation (see the appendix for details).
    \begin{restatable}{lemma}{lemhandysc}\label{lem:handy sc}
        The function $d_z$ as defined in \cref{prop:dilated scaled ext} satisfies
        \begin{align*}
            (\nabla d_z(\vu,\vw) - \nabla d_z(\vu',\vw'))^\top \begin{pmatrix}\vu - \vu'\\\vw - \vw'\end{pmatrix}
           &\geq \big(\nabla d_u(\vu) - \nabla d_u(\vu')\big)^\top(\vu-\vu')\\&\hspace{2.5cm} + \alpha_v h\mleft(\frac{\vu+\vu'}{2}\mright)\mleft\|\frac{\vw}{h(\vu)} - \frac{\vw'}{h(\vu)}\mright\|_v^2
           \numberthis{eq:handy sc}
        \end{align*}
        for all $(\vu,\vw), (\vu',\vw') \in \relint\cZ = (\relint\cU)\ext^h(\relint\cV)$.
    \end{restatable}
    Since $d_u$ is strongly convex, and since $\|\cdot\|_u,\|\cdot\|_v$ are norms for finite-dimensional spaces, there exist constants $\mu_u,\mu_v > 0$ such that
    \[
        \big(\nabla d_u(\vu) - \nabla d_u(\vu')\big)^\top(\vu-\vu') \ge \mu_u\|\vu - \vu'\|_2^2
    \]
    and
    \[
        \mleft\|\frac{\vw}{h(\vu)} - \frac{\vw'}{h(\vu)}\mright\|_v^2 \ge \mu_v\,\mleft\|\frac{\vw}{h(\vu)} - \frac{\vw'}{h(\vu)}\mright\|_2^2.
    \]
    and thus we can write
    \[
        (\nabla d_z(\vu,\vw) \!-\! \nabla d_z(\vu',\vw'))^\top\!\! \begin{pmatrix}\vu - \vu'\\\vw - \vw'\end{pmatrix} \ge \mu_u\|\vu-\vu'\|_2^2 + \mu_v \alpha_v h\mleft(\frac{\vu+\vu'}{2}\mright)\mleft\|\frac{\vw}{h(\vu)} - \frac{\vw'}{h(\vu)}\mright\|_v^2\!\!.\numberthis{eq:bound ell2}
    \]
    Now,
    \begin{align*}
        \|\vw-\vw'\|_2
        &= \mleft\|\frac{h(\vu)+h(\vu')}{2}\mleft(\frac{\vw}{h(\vu)} - \frac{\vw'}{h(\vu')}\mright) + (h(\vu) - h(\vu'))\cdot\frac{1}{2}\mleft(\frac{\vw}{h(\vu)} + \frac{\vw'}{h(\vu')}\mright)\mright\|_2\\
            &\le \frac{h(\vu)+h(\vu')}{2}\mleft\|\frac{\vw}{h(\vu)} - \frac{\vw'}{h(\vu')}\mright\|_2 + |h(\vu) - h(\vu')|\cdot \mleft\|\frac{1}{2}\mleft(\frac{\vw}{h(\vu)} + \frac{\vw'}{h(\vu')}\mright)\mright\|_2\\
            &= h\mleft(\frac{\vu+\vu'}{2}\mright)\mleft\|\frac{\vw}{h(\vu)} - \frac{\vw'}{h(\vu')}\mright\|_2 + |h(\vu) - h(\vu')|\cdot \mleft\|\frac{1}{2}\mleft(\frac{\vw}{h(\vu)} + \frac{\vw'}{h(\vu')}\mright)\mright\|_2.
    \end{align*}
    Since $\vw/h(\vu), \vw'/h(\vu')\in\relint\cV \subseteq \cV$ and $\cV$ is a convex set, then the argument of the last norm on the right is a point in $\cV$. Furthermore, because $\cU$ and $\cV$ are compact, there exist constants $\Omega_u,\Omega_v > 0$ such that $\max_{\vu\in\cV} \|\vu\|_2 \le \Omega_u$ and $\max_{\vv\in\cV}\|\vv\|_2 \le \Omega_v$.  Hence, we can write
    \begin{align*}
        \|\vw-\vw'\|_2 \le h\mleft(\frac{\vu+\vu'}{2}\mright) \mleft\|\frac{\vw}{h(\vu)} - \frac{\vw'}{h(\vu')}\mright\|_2 + (\|\vec{a}\|_2\,\Omega_v)\|\vu-\vu'\|_2,
    \end{align*}
    which in turn implies that
    \begin{align*}
        \mleft\|\stack{\vu}{\vw} - \stack{\vu'}{\vw'}\mright\|_2 &\le \|\vu-\vu'\|_2 + \|\vw-\vw'\|_2 \\
        &\le  (1+\|\vec{a}\|_2\,\Omega_v)\|\vu-\vu'\|_2 + h\mleft(\frac{\vu+\vu'}{2}\mright) \mleft\|\frac{\vw}{h(\vu)} - \frac{\vw'}{h(\vu')}\mright\|_2.
    \end{align*}
    Squaring an using the Cauchy-Schwarz inequality yields
    \begin{align*}
        \mleft\|\stack{\vu}{\vw} - \stack{\vu'}{\vw'}\mright\|_2^2
        &\le \mleft((1+\|\vec{a}\|_2\,\Omega_v)\|\vu-\vu'\|_2 + h\mleft(\frac{\vu+\vu'}{2}\mright) \mleft\|\frac{\vw}{h(\vu)} - \frac{\vw'}{h(\vu')}\mright\|_2\mright)^2 \\
        &\le \mleft(\frac{(1+\|\vec{a}\|_2\Omega_u)^2}{\mu_u} + \frac{h\mleft(\frac{\vu+\vu'}{2}\mright)}{\mu_v\alpha_v}\mright)\mleft(\mu_u\|\vu-\vu'\|_2^2 + \mu_v\alpha_v h\mleft(\frac{\vu+\vu'}{2}\mright) \mleft\|\frac{\vw}{h(\vu)} - \frac{\vw'}{h(\vu')}\mright\|_2^2\mright) \\
        &\le \mleft(\frac{(1+\|\vec{a}\|_2\Omega_u)^2}{\mu_u} + \frac{\|\vec{a}\|_2\Omega_u}{\mu_v\alpha_v}\mright)\mleft(\mu_u\|\vu-\vu'\|_2^2 + \mu_v \alpha_v h\mleft(\frac{\vu+\vu'}{2}\mright) \mleft\|\frac{\vw}{h(\vu)} - \frac{\vw'}{h(\vu')}\mright\|_2^2\mright),
    \end{align*}
    that is,
    \[
        \mu_u\|\vu-\vu'\|_2^2 + \mu_v\alpha_v h\mleft(\frac{\vu+\vu'}{2}\mright) \mleft\|\frac{\vw}{h(\vu)} - \frac{\vw'}{h(\vu')}\mright\|_2^2 \ge \frac{1}{\frac{(1+\|\vec{a}\|_2\Omega_u)^2}{\mu_u} + \frac{\|\vec{a}\|_2\Omega_u}{\mu_v\alpha_v}}\mleft\|\stack{\vu}{\vw} - \stack{\vu'}{\vw'}\mright\|_2^2.
    \]
    Finally, plugging the above inequality into~\eqref{eq:bound ell2} yields 
    \[
        (\nabla d_z(\vu,\vw) - \nabla d_z(\vu',\vw'))^\top \begin{pmatrix}\vu - \vu'\\\vw - \vw'\end{pmatrix} \ge \frac{1}{\frac{(1+\|\vec{a}\|_2\Omega_u)^2}{\mu_u} + \frac{\|\vec{a}\|_2\Omega_u}{\mu_v\alpha_v}}\mleft\|\stack{\vu}{\vw} - \stack{\vu'}{\vw'}\mright\|_2^2.
    \]
    Hence, $d_z$ is strongly convex (with respect to the Euclidean norm) with strong convexity modulus $1/\mleft(\frac{(1+\|\vec{a}\|_2\Omega_u)^2}{\mu_u} + \frac{\|\vec{a}\|_2\Omega_u}{\mu_v\alpha_v}\mright) > 0$.

    \item[Minimum of $d_z$]
    Since by hypothesis $\min_{\vu\in\cU} d_u(\vu) = \min_{\vv\in\cV}d_v(\vv) = 0$, $\alpha_v > 0$, and $h(\vu) \ge 0$ for all $\vu\in\cU$, from the definition~\eqref{eq:dz} follows that $d_z(\vu,\vw) \ge 0$ for all $(\vu,\vw) \in \cZ$. So, in order to conclude that $\min_{(\vu,\vw)\in\cZ} d_z(\vu,\vw) = 0$, it is enough to show that the value $0$ can be attained by $d_z$. That can be checked directly by considering the point $(\vu^*,h(\vu^*)\vv^*)\in\cZ$ where $\vu* \defeq \argmin_{\vu\in\cU} d_u(\vu) = \nabla d_u^*(\vec{0})$ and $\vv^*\defeq \argmin_{\vv\in\cV} d_v(\vv) = \nabla d_v^*(\vec{0})$.
    \end{description}

    \noindent So, $d_z$ is a DGF. We now argue that it is a ``nice'' DGF provided that $d_u$ and $d_v$ are.
    The key observation is that the gradient of the convex conjugate at any point $\vg = (\vg_u,\vg_v)\in\bbR^{m}\times\bbR^{n}$ satisfies
    \begin{align*}
        &\max_{\substack{\vu\in\cU\\\vv\in\cV}} \mleft\{\vg_u^\top \vu + h(\vu) \vg_v^\top \vv - d_u(\vu) - \alpha_v h(\vu) d_v(\vv)\mright\}\\
            &\qquad=
            \max_{\substack{\vu\in\cU\\\vv\in\cV}} \mleft\{\vg_u^\top \vu  - d_u(\vu) + \alpha_v\, h(\vu)\mleft[ \mleft(\frac{\vg_v}{\alpha_v}\mright)^\top \vv - d_v(\vv)\mright]\mright\} \\
            &\qquad=
            \max_{\vu\in\cU} \mleft\{\vg_u^\top \vu  - d_u(\vu) + \alpha_v\, h(\vu)\max_{\vv\in\cV}\mleft\{ \mleft(\frac{\vg_v}{\alpha_v}\mright)^\top \vv - d_v(\vv)\mright\}\mright\}\\
            &\qquad=
            \max_{\vu\in\cU} \mleft\{\mleft(\vg_u + \alpha_v\, d_v^*\mleft(\frac{\vg_v}{\alpha_v}\mright)\vec{a}\mright)^\top  \vu  - d_u(\vu)\mright\}
            ,
    \end{align*}
    where the second equality follows since $\alpha_v>0$ and $h(\vu) \ge 0$ by hypothesis, and the third equality follows from the definition of $h : \vu\mapsto \va^\top\vu$. Hence,
    \begin{equation}\label{eq:argconj}
        \nabla d_z^*(\vg_u, \vg_v) = \begin{pmatrix}
            \nabla d_u^*\mleft(\vg_u + \alpha_v\, d_v^*\mleft(\frac{\vg_v}{\alpha_v}\mright)\vec{a}\mright)\\
            h\mleft(\nabla d_u^*\mleft(\vg_u + \alpha_v\, d_v^*\mleft(\frac{\vg_v}{\alpha_v}\mright)\vec{a}\mright)\mright)\, \nabla d_v^*\mleft(\frac{\vg_v}{\alpha_v}\mright) 
        \end{pmatrix}\in \cU \ext^h \cV = \cZ.
    \end{equation}
    
    We now turn to the second part of the statement, and assume that $d_u$ and $d_v$ are ``nice'' DGFs.  It is clear that the gradient~\eqref{eq:gradient dz} can be computed in linear time: we only need to take inner products which takes time linear in the dimension of $\cV$, compute the value of $h(\vu)$ which is linear in the dimension of $\cU$, and compute corresponding values of $d_u$ and $d_v$, which takes linear time by assumption. Furthermore,
    since 
    \begin{align*}
        d_v^*\mleft(\frac{\vg_v}{\alpha_v}\mright) = \max_{\vv \in \cV}\mleft\{\mleft(\frac{\vg_v}{\alpha_v}\mright)^\top \vv - d_v(\vv)\mright\}
    \end{align*}
    can be computed in linear time starting from $\nabla d_v^*({\vg_v}/{\alpha_v})$, then the gradient of the conjugate of $d_z$, given in~\eqref{eq:argconj}, can be evaluated in linear time provided that $\nabla d_u^*$ and $\nabla d_v^*$ can, as is the case when $d_u$ and $d_v$ are ``nice''.
\end{xproof}


The construction of \cref{prop:dilated scaled ext} can be composed repeatedly to give rise to a ``nice'' DGF for any polytope that can be expressed as a composition of scaled extension operations, such as the sequence-form polytope $Q$ and the polytope of correlation plans $\Xi$.

To make things formal, we introduce the following setup, which is satisfied by both $Q$ and $\Xi$~\citep{Farina20:Polynomial}.
\begin{setup}\label{setup:xi}
Let $\cX$ be a set expressible in the form
\[
    \cX = \cX_1 \ext^{h_1} \cX_2 \ext^{h_2} \dots \ext^{h_{n-1}} \cX_{n},
    \numberthis{eq:cX}
\]
where:
\begin{itemize}
    \item Each $\cX_k \subseteq \bbR^{s_k}$ is a compact and convex set such that $\max_{\vv \in \cX_k} \|\vv\|_2 \le 1$. (In the case of $\cX = Q$ and $\cX = \Xi$, each $\cX_k$ is a probability simplex $\cX_k = \Delta^{s_k}$.)
    \item Each $h_k$ is a linear function nonnegative on $\cU_k \defeq \cX_1 \ext^{h_1} \dots \ext^{h_{k-1}} \cX_k$, strictly positive in the relative interior of $\cU_k$. Furthermore, we assume that each $h_k$ can be written in the form
    \[
        h_k : \cU_k \ni \vu \defeq (\vu_1,\dots,\vu_{k}) \mapsto \vec{a}_k^\top\vu = \vec{a}_{k,1}^\top \vu_1 + \dots + \vec{a}_{k,k}^\top\vu_k,
        \numberthis{eq:h expansion}
    \]
    for an appropriate vector $\vec{a}_k \defeq (\vec{a}_{k,1} ,\dots,\vec{a}_{k,k}) \in [0,1]^{s_1}\times\dots\times[0,1]^{s_k}$, and that $h_k(\vu) \le 1$ for all $\vu\in\cU_k$;
    \item For each $k=1,\dots,n$, a $1$-strongly convex DGF $d_k$ for $\cX_k$ has been chosen, where strong convexity is measured with respect to the Euclidean norm.
\end{itemize}
\end{setup}
By repeatedly composing the construction of \cref{prop:dilated scaled ext} using DGFs $d_k$ for each $\cX_k$ in \cref{setup:xi}, we obtain the DGF
\[
    d: \cX_1 \ext^{h_1} \cX_2 \ext^{h_2} \dots \ext^{h_{n-1}} \cX_{n} \ni \vec{x} \defeq (\vec{x}_1, \dots,\vec{x}_n) \mapsto \alpha_1\,d_1(\vec{x}_1) + \sum_{k=2}^{n} \alpha_k d^\square_k(\vec{x}_1, \dots, \vec{x}_k),
    \numberthis{eq:dz spell out}
\]
where, for all $k=2,\dots,n$,
\[
    d_k^\square(\vec{x}_1,\dots, \vec{x}_k) \defeq \begin{cases}
        0 & \text{if } h_{k-1}(\vec{x}_1,\dots,\vec{x}_{k-1}) = 0\\
        h_{k-1}(\vec{x}_1,\dots,\vec{x}_{k-1})\, d_k\!\mleft(\frac{\vec{x}_{k}}{h_{k-1}(\vec{x}_1,\dots,\vec{x}_{k-1})}\mright) & \text{if } h_{k-1}(\vec{x}_1,\dots,\vec{x}_{k-1}) > 0.
    \end{cases}    
\]

In the case of the sequence-form polytope $\cX = Q$, where $\cX_k = \Delta^{n_k}$ and each $h_k$ maps a sequence-form vector to the parent sequence of the $k$-th decision point, this yields exactly a dilated DGF in the precise sense of~\eqref{eq:dilated dgf}. So, the construction in \cref{prop:dilated scaled ext} subsumes that of \cref{sec:dilated dgfs} which only applied to sequence-form strategy spaces.
Additionally, in the case of the polytope of correlation plans $\cX = \Xi$, to our knowledge our construction yields the first ``nice'' DGF.

However, before the DGF~\eqref{eq:dz spell out} can be put to use in an optimization method such as EGT or mirror prox, the weights $\alpha_k$ need to be specified so that the DGF is $1$-strongly convex with respect to some norm. The next proposition provides a general way of doing so, with respect to the Euclidean norm, under conditions met for both $\cX = Q$ and $\cX = \Xi$. From now on, we will the $i$-th coordinate of a vector $\vv$ with the symbol $v[i]$.

\begin{proposition}\label{prop:bound}
    Consider the composite DGF defined in~\eqref{eq:dz spell out} for the set $\cX$~\eqref{eq:cX} under \cref{setup:xi}.
    Then, for all $\vx \defeq (\vx_1,\dots,\vx_n)$ and $\vx' \defeq (\vx_1,\dots,\vx_n)$ such that $\vx,\vx'\in\relint\cX = (\relint\cX_1)\ext^{h_1}\dots\ext^{h_{n-1}}(\relint\cX_n)$, the composite DGF $d$ satisfies
    \[
        \langle\nabla d(\vx) - \nabla d(\vx'), \vx-\vx'\rangle \ge \sum_{k=1}^n \sum_{i=1}^{s_k}\mleft(\frac{\alpha_k}{2} - \sum_{p=k}^{n-1} \alpha_{p+1} \|\vec{a}_{p}\|_0\, a_{p,k}[i]\mright)(x_k[i]-x'_k[i])^2.   
    \]
\end{proposition}
\begin{xproof}
    By induction on the sequence of scaled extension operations.
    \begin{description}
        \item[Base case] The base case corresponds to the case where $n=1$ and $\cX= \cX_1 \subseteq \bbR^{s_1}$, that is, no scaled extension is performed. In that case, $d: \cX \ni \vec{x} \mapsto \alpha_1 d_1(\vx)$, where $d_1$ is a $1$-strongly convex DGF for $\cX_1$ with respect to the Euclidean norm by hypothesis. Hence,
        \begin{align*}
            \langle \nabla d(\vx) - \nabla d(\vx'), \vx-\vx'\rangle &\ge \alpha_1\|\vx-\vx'\|_2^2 = \sum_{i=1}^{s_1} \alpha_1 (x[i]-x'[i])^2 \ge \sum_{i=1}^{s_1} \frac{\alpha_1}{2} (x[i]-x'[i])^2,
        \end{align*}
        which satisfies the statement.
        \item[Inductive step] Let $\cU \defeq \cX_1 \ext^{h_1} \dots \ext^{h_{n-2}} \cX_{n-1}$, and let $\tilde d$ be the dilated DGF constructed for $\cU$. Assume by induction that the statement of this proposition holds for $\tilde d$. We will show that the statement continues to hold after one further application of the construction, that is, for the dilated DGF
        \[
            d: ( \cU \ext^{h_{n-1}} \cX_n ) \ni (\vu, \vw) \mapsto \tilde d(\vu) + \alpha_n d_n^\square(\vu, \vw),
        \]
        where
        \[ 
            d_n^\square(\vu,\vw) \defeq \begin{cases} 0 & \text{if } h_{n-1}(\vu,\vw) = 0\\
                h_{n-1}(\vu,\vw) d_n\mleft(\frac{\vw}{h(\vu)}\mright) & \text{if } h_{n-1}(\vu,\vw) > 0.
                \end{cases}
        \]
        To lighten notation, from now on in this proof we will let $h$ be a shorthand for $h_{n-1}$.
        From \cref{lem:handy sc}, it follows that for any $(\vu,\vw),(\vu',\vw')\in\relint(\cU \ext^{h} \cX_n) = (\relint \cU) \ext^{h} (\relint\cX_n)$,
        \begin{align*}
            &\mleft\langle \nabla d(\vu,\vw)-\nabla d(\vu',\vw'), \stack{\vu-\vu'}{\vw-\vw'}\mright\rangle \ge \mleft\langle\nabla \tilde d(\vu) - \nabla\tilde d(\vu'), \vu-\vu'\mright\rangle \\
            &\hspace{6cm}+ \alpha_n h\mleft(\frac{\vu+\vu'}{2}\mright)\mleft\|\frac{\vw}{h(\vu)} - \frac{\vw'}{h(\vu')}\mright\|_2^2.\numberthis{eq:here}
        \end{align*}
        Now, 
        \begin{align*}
            \|\vw - \vw'\|_2 &= \mleft\|h(\vu) \frac{\vw}{h(\vu)} - h(\vu')\frac{\vw'}{h(\vu')}\mright\|_2\\
                &= \mleft\|h\mleft(\frac{\vu + \vu'}{2}\mright) \mleft(\frac{\vw}{h(\vu)} - \frac{\vw'}{h(\vu')}\mright) + (h(\vu) - h(\vu'))\mleft[\frac{1}{2}\mleft(\frac{\vw}{h(\vu)} + \frac{\vw'}{h(\vu')}\mright)\mright]\mright\|_2 \\
                &\le h\mleft(\frac{\vu + \vu'}{2}\mright) \mleft\|\frac{\vw}{h(\vu)} - \frac{\vw'}{h(\vu')}\mright\|_2 + (h(\vu) - h(\vu')),
        \end{align*}
        where in the last step we used the triangle inequality and the hypothesis that $\max_{\vv\in\cX_n}\|\vv\| \le 1$ together with the fact that $\frac{1}{2}\mleft(\frac{\vw}{h(\vu)} + \frac{\vw'}{h(\vu')}\mright) \in \relint\cX_n$ by convexity.
        By taking squares, rearranging, and using the fact that $h(\vu) \in [0,1]$ for all $u\in\cU$ we have
        \begin{align*}
            h\mleft(\frac{\vu + \vu'}{2}\mright) \mleft\|\frac{\vw}{h(\vu)} - \frac{\vw'}{h(\vu')}\mright\|_2^2 &\ge
            h\mleft(\frac{\vu + \vu'}{2}\mright)^2 \mleft\|\frac{\vw}{h(\vu)} - \frac{\vw'}{h(\vu')}\mright\|^2\\
            &\ge \frac{1}{2}\|\vw - \vw'\|^2 - (h(\vu) - h(\vu'))^2\\
            &=\frac{1}{2}\|\vw - \vw'\|^2 - (\va_{n-1}^\top (\vu - \vu'))^2\numberthis{eq:subtraction term}
        \end{align*}
        Plugging the inductive hypothesis and~\eqref{eq:subtraction term} into~\eqref{eq:here}, defining $\vu = (\vx_1,\dots,\vx_{n-1}), \vu' = (\vx'_1,\dots,\vx'_{n-1})\in\relint\cU, \vw = \vx_n\in\relint\cX_n$ and $\vx \defeq (\vu, \vx_{n}), \vx' \defeq (\vu',\vx'_{n})$  yields
        \begin{align*}
            \mleft\langle \nabla d(\vx)-\nabla d(\vx'), \vx-\vx'\mright\rangle &\ge \mleft[\sum_{k=1}^{n-1} \sum_{i=1}^{s_k}\mleft(\frac{\alpha_k}{2} - \sum_{p=k}^{n-2} \alpha_{p+1} \|\vec{a}_{p}\|_0\, a_{p,k}[i]\mright)(x_k[i]-x'_k[i])^2\mright] \\&\hspace{4cm}+ \frac{\alpha_n}{2}\|\vx_n - \vx'_n\|_2^2 - \alpha_n \Big(\va_{n-1}^\top (\vu-\vu')\Big)^2\\
            &\ge \mleft[\sum_{k=1}^{n-1} \sum_{i=1}^{s_k}\mleft(\frac{\alpha_k}{2} - \sum_{p=k}^{n-2} \alpha_{p+1} \|\vec{a}_{p}\|_0\, a_{p,k}[i]\mright)(x_k[i]-x'_k[i])^2\mright] \\&\hspace{1cm}+ \frac{\alpha_n}{2}\|\vx_n - \vx'_n\|_2^2 - \alpha_n \|\va_{n-1}\|_1\mleft(\sum_{q=1}^{n-1}\sum_{i=1}^{s_q} a_{n-1,q}[i] (x_q[i]-x'_q[i])^2\mright)\\
            &=\sum_{k=1}^n \sum_{i=1}^{s_k}\mleft(\frac{\alpha_k}{2} - \sum_{p=k}^{n-1} \alpha_{p+1} \|\vec{a}_{p}\|_0\, a_{p,k}[i]\mright)(x_k[i]-x'_k[i])^2.   
        \end{align*}
        where the second inequality follows from applying the Cauchy-Schwarz inequality and the fact that $\|\vec{a}_p\|_0 \ge \|\vec{a}_p\|_1$ since $\vec{a}_p\in[0,1]^{s_1+\dots+s_{p-1}}$ by the hypotheses in \cref{setup:xi}.  
    \end{description}
\end{xproof}

\cref{prop:bound} implies that when the $\alpha_k$'s are such that 
\[
    \frac{\alpha_k}{2} - \sum_{p=k}^{n-1}\alpha_{p+1} \|\vec{a}_{p}\|_0 a_{p_k}[i] \ge 1 \qquad \text{for all } k = 1,\dots,n, i = 1,\dots, s_k,    
\]
then $d$ is $1$-strongly convex with respect to the Euclidean norm. So, we have the following.
\begin{corollary}\label{cor:alpha scext dilated}
    Consider the composite DGF defined in~\eqref{eq:dz spell out} for the set $\cX$~\eqref{eq:cX} under \cref{setup:xi}, where
    the coefficients $\alpha_k$ are defined recursively as
    \[
        \alpha_n = 2,\quad\text{and}\quad \alpha_k = 2 + 2\,\mleft\| \sum_{p=k}^{n-1} \alpha_{p+1}\|\vec{a}_{p}\|_0\, \vec{a}_{p,k} \mright\|_\infty \quad\forall\, k= n-1,\dots, 1.
    \] 
    Then $d$ is $1$-strongly convex with respect to the Euclidean distance.
\end{corollary} 

Since sets $\cX$ obtained via scaled extension of \emph{simplex} domains are prevalent in game theory, we show a stronger result for the case where 
each $d_k$ is set to the negative entropy function.

\begin{definition}\label{def:dilent scext}
    Consider a set $\cX$ obtained via scaled extension of \emph{simplex} domains $\cX_k = \Delta^{s_k}$ in accordance with \cref{setup:xi}, and consider the dilated DGF for $\cX$ obtained as described in~\eqref{eq:dz spell out}, when  $d_k$ is set to the negative entropy function
\[
    d_k : \Delta^{s_k} \ni \vec{x}_k \mapsto \log(s_k) + \sum_{i=1}^{s_k} x_k[i]\log x_k[i] \ge 0\numberthis{eq:negative entropy}
\]
    for all $k$. 
    Then, the resulting function $\psi : (\cX_1 \ext^{h_1} \dots \ext^{h_{n-1}} \cX_n) \to \bbR_{\ge 0}$ is 
    \[
        (\vec{x}_1,\dots,\vx_n) \mapsto \alpha_1\mleft(\log(s_1) + \sum_{i=1}^{s_1} x_1[i]\log x_1[i]\mright) + \sum_{k=2}^n \alpha_k\mleft(\log(s_k) + \sum_{i=1}^{s_k} x_k[i]\log \frac{x_k[i]}{h_{k-1}(\vx_1,\dots,\vx_{k-1})}\mright)
    \]
    for the particular choice of weights $\alpha_k$ defined in \cref{cor:alpha scext dilated} is called the \emph{dilated entropy DGF}.
\end{definition}

Note that \cref{def:dilent scext} extends the name \emph{dilated entropy DGF}, already used for the DGF in \cref{def:kroer dgf} in the case of the sequence-form strategy $\cX = Q$, to any scaled extension of simplexes. The overload is sound, in the sense that when $\cX = Q$, \cref{def:dilent scext} recovers exactly the function in \cref{def:kroer dgf}, together with the state-of-the-art coefficients defined by \citet{Kroer17:Theoretical}. A consequence of this observation is that the coefficients defined by \cref{cor:alpha scext dilated} grow exponentially fast in the dimension of $\cX$, showing that the composite DGF constructed by means of \cref{prop:dilated scaled ext} suffers from the same problem as the Kroer et al. dilated entropy DGF discussed in \cref{sec:dilated dgfs}.
We show that the strong convexity result of \citet{Kroer20:Faster} for treeplexes generalizes as well, the proof can be found in \cref{app:details}.
\begin{restatable}{proposition}{propellonesc}\label{prop:ell1 sc}
    Let $\cX$ be obtained via scaled extension of simplex domains $\cX_k = \Delta^{s_k}$. Then, the dilated entropy DGF (\cref{def:dilent scext}) for $\cX$ is $1$-strongly convex with respect to the Euclidean norm, and $(1/M_\cX)$-strongly convex with respect to the $\ell_1$ norm.
\end{restatable}

In the next section we show how the ideas already used in \cref{sec:global dgf} generalize to scaled-extension-based sets, yielding the first ``nice'' DGF with polynomially small range on $\cX$.

\subsection{Dilatable Global Entropy for Scaled Extension}
    In this section we instantiate the generic framework of dilated DGFs, as defined in the previous section, to the chains of scaled extensions with \emph{simplex domains}. More specifically, we consider the same setting as \cref{setup:xi}, under the further assumption that each convex and compact set $\cX_k$ in the decomposition of $\cX$~\eqref{eq:cX} is a probability simplex, that is, for all $k=1,\dots,n$, $\cX_k = \Delta^{s_k}$ where $s_k \in \bbN_{\ge 1}$. This setup encompasses both sequence-form strategy spaces and the polytope of correlation plans.
    
    The \emph{dilated entropy DGF} $\psi$ for such a set is the dilated DGF~\eqref{eq:dz spell out} obtained by recursively applying the general construction of \cref{prop:dilated scaled ext} with the (negative) entropy function at each $\Delta^{s_k}$. Specifically, $\psi$ can be written as\footnote{We recall that we use the convention $0\log(0) = 0\log(0/0) = 0$.}
    \begin{align*}
        &\psi: \cX \ni (\vx_1,\dots,\vx_n) \mapsto \alpha_1 \mleft(\log s_1 + \sum_{i=1}^{s_1} x_{1}[i]\log x_1[i]\mright) \\
        &\hspace{2cm}+ \sum_{k=2}^n \alpha_k \mleft(h_{k-1}(\vx_1,\dots,\vx_{k-1}) \log s_k + \sum_{i=1}^{s_k} x_{k}[i] \log\frac{x_{k}[i]}{h_{k-1}(\vx_1,\dots,\vx_{k-1})}\mright).
    \end{align*}
    
    By using the same manipulations of the logarithms that we used in \cref{thm:dilatability}, $\psi$ coincides, on $\cX$ with the function
    \begin{align*}
        &\tilde\regu: \cX \ni (\vx_1,\dots,\vx_n) \mapsto \sum_{k=1}^n \mleft(\alpha_k\sum_{i=1}^{s_k} x_k[i] \log x_k[i]\mright) - \sum_{k=2}^n \alpha_k h_{k-1}(\vx_1,\dots,\vx_{k-1}) \log h_{k-1}(\vx_1,\dots,\vx_{k-1})\\
        &\hspace{7.5cm} + \alpha_1 \log s_1 + \sum_{k=2}^n \alpha_k h_{k-1}(\vx_1,\dots,\vx_{k-1})\log s_k.
        \numberthis{eq:dgo scext}
    \end{align*}
    For this reason, similarly to what we did for extensive-form strategy spaces, we coin $\tilde\regu_k$ the \emph{dilatable global entropy DGF}. It is immediate to see by induction that $\nabla\psi$ can be computed exactly in linear time. Furthermore, because $\psi$ is a ``nice'' DGF by virtue of \cref{prop:dilated scaled ext}, and $\psi = \tilde\regu$ on $\cX$, we immediately obtain that $\tilde\regu$ is a ``nice'' DGF.
    
    We conclude this section by showing that there exists polynomially small (in the dimension of $\cX$) DGF weights $\alpha_k$ such that $\tilde\regu$ defined in \eqref{eq:dgo scext} is $1$-strongly convex with respect to the Euclidean norm. For any $\vec{m} \defeq (\vec{m}_1,\dots,\vec{m}_n)\in \bbR^{s_1}\times\dots\times\bbR^{s_n}$ and $\vec{x} \defeq (\vx_1,\dots,\vx_n)\in\cX$, the Hessian matrix of $\tilde\regu$ satisfies 
    \begin{align*}
        \vec{m}^\top \nabla^2 \tilde\regu(\vec{x})\vec{m} &= \sum_{k=1}^n \mleft( \alpha_k \sum_{i=1}^{s_k} \frac{m_k[i]^2}{x_k[i]}\mright) - \sum_{k=2}^n \mleft(\alpha_k \frac{\mleft(\sum_{p=1}^{k-1} \vec{a}_{k-1,p}^\top \vec{m}_p\mright)^2}{h_{k-1}(\vx_1,\dots,\vx_{k-1})}\mright)\\
            &= \sum_{k=1}^n \mleft( \alpha_k \sum_{i=1}^{s_k} \frac{m_k[i]^2}{x_k[i]}\mright) - \sum_{k=2}^n \mleft(\alpha_k \frac{\mleft(\sum_{p=1}^{k-1} \vec{a}_{k-1,p}^\top \vec{m}_p\mright)^2}{\sum_{p=1}^{k-1} \vec{a}_{k-1,p}^\top \vec{x}_{p}}\mright),
            \numberthis{eq:hessian bound 1}
    \end{align*}
    where in the second equality we expanded the definition of $h_{k-1}$ according to~\eqref{eq:h expansion}. Now, expanding the following product, we find that
    \begin{align*}
        \|\vec{a}_{k-1}\|_0\,\mleft(\sum_{p=1}^{k-1} \vec{a}_{k-1,p}^\top \vec{x}_p\mright)\mleft(\sum_{p=1}^{k-1}\sum_{i=1}^{s_p} \frac{a_{k-1,p}[i] m_p[i]^2}{x_p[i]}\mright)
        &\ge \|\vec{a}_{k-1}\|_0\,\mleft(\sum_{p=1}^{k-1} \sum_{i=1}^{s_p} a_{k-1,p}[i] x_p[i] \frac{a_{k-1,p}[i] m_p[i]^2}{x_p[i]}\mright)\\
        &= \|\vec{a}_{k-1}\|_0\,\mleft(\sum_{p=1}^{k-1} \sum_{i=1}^{s_p} a_{k-1,p}[i]^2 m_p[i]^2\mright)\\
        &\ge \mleft(\sum_{p=1}^{k-1} \sum_{i=1}^{s_p} a_{k-1,p}[i] m_p[i]\mright)^2\\
        &=\mleft(\sum_{p=1}^{k-1} \vec{a}_{k-1,p}^\top \vec{m}_p\mright)^2.
    \end{align*}
    Hence, plugging the above inequality into~\eqref{eq:hessian bound 1}, we have
    \begin{align*}
        \vec{m}^\top \nabla^2 \tilde\regu(\vec{x})\vec{m}
        &\ge \sum_{k=1}^n \mleft( \alpha_k \sum_{i=1}^{s_k} \frac{m_k[i]^2}{x_k[i]}\mright)
        - \sum_{k=2}^n \mleft(\alpha_k \|\vec{a}_{k-1}\|_0 \sum_{p=1}^{k-1}\sum_{i=1}^{s_p} \frac{a_{k-1,p}[i]\,m_p[i]^2}{x_p[i]}\mright)\\
        &= \sum_{k=1}^n \sum_{i=1}^{s_k} \mleft(\alpha_k - \sum_{p=k}^{n-1} \alpha_{p+1} \|\vec{a}_{p}\|_0 \,a_{p,k}[i]\mright) \frac{m_k[i]^2}{x_k[i]} \numberthis{eq:dge hessian diag}\\
        &\ge \sum_{k=1}^n \sum_{i=1}^{s_k} \mleft(\alpha_k - \sum_{p=k}^{n-1} \alpha_{p+1} \|\vec{a}_{p}\|_0 \, a_{p,k}[i]\mright) m_k[i]^2 \numberthis{eq:rem zero one}
    \end{align*}
    where the last inequality follows from the fact that $\cX = \Delta^{s_1}\ext^{h_1}\dots\ext^{h_{n-1}} \Delta^{s_n} \subseteq [0,1]^{s_1+\dots+s_n}$ given the assumption that $h_k(\vx_1,\dots,\vx_k) \in [0,1]$ for all $k=1,\dots,n-1$.

    In particular,~\eqref{eq:rem zero one} implies that when the coefficients $\alpha_k$ are chosen so that
    \begin{align}
        \alpha_k - \sum_{p=k}^{n-1} \alpha_{p+1} \|\vec{a}_{p}\|_0 \, a_{p,k}[i] \ge 1 \qquad\forall\,k=1,\dots,n,~~i=1,\dots,s_k, \label{eq:dge weights geq 1}
    \end{align}
    then $\tilde\regu$ is $1$-strongly convex with respect to the Euclidean norm, 
    and for the $\ell_1$ norm, we have
    \begin{align*}
        \|\vec{m}\|_1^2
        = \left(\sum_{k=1}^n \sum_{i=1}^{s_k} m_k[i]\right)^2
        &= \left(\sum_{k=1}^n \sum_{i=1}^{s_k} \frac{m_k[i]}{\sqrt{x_k[i]}} \sqrt{x_k[i]} \right)^2 \\
        &\leq \left(\sum_{k=1}^n \sum_{i=1}^{s_k} \frac{m_k[i]^2}{x_k[i]} \right) \left(\sum_{k=1}^n \sum_{i=1}^{s_k} x_k[i] \right)
        \leq M_{\cX} \vec{m}^\top \nabla^2 \tilde\regu(\vec{x})\vec{m}
    \end{align*}
    which follows by Cauchy-Schwarz, \eqref{eq:dge weights geq 1}, and \eqref{eq:dge hessian diag}.
    In other words, we have the following.

    \begin{theorem}
        Consider the dilatable global entropy DGF defined in~\eqref{eq:dgo scext} for the set $\cX$~\eqref{eq:cX} under \cref{setup:xi} and the further assumption that each $\cX_k  = \Delta^{s_k}$, where
        the coefficients $\alpha_k$ are defined recursively as
        \[
            \alpha_n = 1,\quad\text{and}\quad \alpha_k = 1 + \mleft\| \sum_{p=k}^{n-1} \alpha_{p+1}\|\vec{a}_{p}\|_0\, \vec{a}_{p,k} \mright\|_\infty \quad\forall\, k= n-1,\dots, 1.
        \] 
        Then $d$ is $1$-strongly convex with respect to the Euclidean distance, and 
        $(1/M_{\cX})$-strongly convex with respect to the $\ell_1$ norm.
    \end{theorem} 
    \section{Experiments}
\label{sec:experiments}

In this section we study the numerical performance of our DGFs. First we study the performance of the dilatable global entropy for computing Nash equilibria in zero-sum EFGs, and second we study the performance for computing correlated equilibria and team equilibria.

Our experiments will be shown on nine different games, which span a variety of poker games, other recreational games, as well as a pursuit-evasion game played on a graph.
All games are standard benchmarks in the computational game theory literature, and a full description of the games is given in the Appendix. In \cref{table:treeplex size}(a) we summarize some key dimensions of the game instances we use: the number of decision points $|\cJ_1|, |\cJ_2|$ for Player~1 and 2, respectively, the number of sequences $|\Sigma_1|,|\Sigma_2|$, and the number of terminal nodes (leaves).

\begin{table}[th]
    \centering
    \sisetup{scientific-notation=false,round-mode=places,
round-precision=2}
    \setlength{\tabcolsep}{4.0pt}\small\centering
    \scalebox{.96}{\begin{tikzpicture}
    \node[anchor=west] at (0,0) {
    \begin{tabular}{lrrrrr|||rr|||rr}
        \toprule
        \multirow{2}{*}{\bf Game instance} & \multicolumn{2}{c}{\bf Decision Points} & \multicolumn{2}{c}{\bf Sequences} & \bf Leaves & \multicolumn{2}{c|||}{\bf Weights $\beta$} & \multicolumn{2}{c}{\bf Weights $\gamma$}\\
        &$|\cJ_1|$&$|\cJ_2|$&$|\Sigma_1|$ & $|\Sigma_2|$ & $|Z|$&Avg & Max & Avg & Max\\
        \midrule
        Kuhn poker &\num{6}&\num{6}&\num{13} & \num{13} & \num{30} & \num{8.8571} & \num{38} & \num{2.28571} & \num{7}\\
        Leduc poker (3 ranks) &\num{144}&\num{144}&\num{337} & \num{337} & \num{1116} & \num{11.7655} & \num{686} & \num{2.1172} & \num{43}\\
        Leduc poker (13 ranks) &\num{2574}&\num{2574}&\num{6007} & \num{6007} & \num{98956} & \num{12.05669} & \num{12326} & \num{2.13087} & \num{703}\\
        Goofspiel & \num{17476} & \num{17476} & \num{21329} & \num{21329} & \num{13824} & \num{6.911598} & \num{23442} & \num{1.697821} & \num{917}\\
        Battleship (3 turns) &\num{18152}&\num{62875}&\num{73130} & \num{253940} & \num{552132} & \num{3.29389} & \num{2894} & \num{1.242109} & \num{99}  \\
        Battleship (4 turns) &\num{316520}&\num{734203}&\num{968234} & \num{2267924} & \num{3487428} & \num{3.753261} & \num{27470} & \num{1.330528} & \num{483} \\
        Liar's dice &\num{12288}&\num{12288}&\num{24571} & \num{24571} & \num{147420} & \num{15.5555} & \num{65546} & \num{2.043372} & \num{1399} \\
        Pursuit-evasion (4 turns) &\num{34} & \num{348}&\num{52}&\num{2029} & \num{15898} & \num{8.2857} & \num{62} & \num{1.94285} & \num{5} \\
        Pursuit-evasion (6 turns) & \num{58} & \num{11830} & \num{78} & \num{68951} & \num{118514} & \num{19.42372} & \num{254} & \num{2.50847} & \num{7}\\
        \bottomrule
    \end{tabular}};
    \draw[line width=2.3mm,white] (11.95,-2.5) -- ++(0,5.0);
    \draw[line width=2.3mm,white] (14.44,-2.5) -- ++(0,5.0);
    \node[anchor=north] at (6,-2.4) {\textbf{(a)} --- Game instances and sizes};
    \node[anchor=north] at (13.2,-2.4) {\textbf{(b)}};
    \node[anchor=north] at (15.7,-2.4) {\textbf{(c)}};
    \end{tikzpicture}}
    \caption{Column \textbf{(a)}: various measures of the size of each of the games that we test algorithms on. Columns \textbf{(b)} and \textbf{(c)}: the magnitude of the dilated entropy DGF and DGE weights.}
    \label{table:treeplex size}
\end{table}

Our experiments will show performance on three algorithms. First, we will plot the performance for both the EGT and mirror prox algorithms, with stepsizes and smoothing chosen according to the theoretical values dictated by \cref{thm:nesterovEGT,thm:mirror prox}. Second, we will also show results on a tweaked variant of EGT called EGT/AS, which implements several heuristics that typically lead to better performance in practice, as seen in \cite{Hoda10:Smoothing,Kroer20:Faster,Kroer18:Solving}. These heuristic are:
\begin{enumerate}
    \item\emph{$\mu$ balancing}: At each iteration, we take a step on the player $i$ whose smoothing parameter $\mu_i$ is larger.
    \item\emph{Aggressive Stepsizing}: The original stepsize of EGT at iteration $t$ is $\tau = 2 / (3+t)$, which is typically too conservative in practice. Instead, EGT/AS maintains some current value $\tau$, initially set at $\tau = 0.5$. EGT/AS then repeatedly attempts to take steps with the current $\tau$, and after every step checks whether the invariant condition of EGT still holds. If not, then we undo the step, decrease $\tau$, and repeat the process.
    \item\emph{Initial $\mu$ fitting}: The initial EGT values for $\mu_x,\mu_y$ are much too conservative. Instead, At the beginning of the algorithm we perform a search over initial values for $\mu_x = \mu_y$. The search starts at the candidate value $\mu = 10^{-6}$ and stops as soon as the choice of $\mu_x = \mu_y = \mu$ yields an excessive gap value above $0.1$. If the current choice does not, $\mu$ is incremented by $20\%$ and the fitting continues.
\end{enumerate}
For all parameters above, we use the same values as in \citet{Kroer18:Solving}, even though those values were tuned for the dilated entropy DGF, rather than dilatable global entropy.

In the presentation of the numerical performance, we will generally plot the number of iterations of the FOM on the x axis, rather than plot wall-clock time. Since we hold the algorithmic setup fixed in each plot, apart from the DGF, this gives a fair representation of performance, since they all use the same set of operations (in particular the same number of gradient computations, which is typically the most expensive operation).
For EGT/AS, we will instead plot the number of gradient computations on the x axis, since the number of gradient computations can vary for each DGF, depending on the amount of backtracking incurred.

\subsection{Nash Equilibrium Computation}

We will focus on comparing our new dilatable global entropy for the sequence-form polytope~(\cref{def:dilatable global entropy}) to the prior state-of-the-art dilated entropy DGF (\cref{def:kroer dgf}) from \cite{Kroer20:Faster}.

Before we study the numerical performance, we look at the size of the DGF weights $\beta$ and $\gamma$ for each of the games. \cref{table:treeplex size} column \textbf{(b)} shows the average and maximum size of the dilated entropy,  and \cref{table:treeplex size} column \textbf{(c)} shows the corresponding values for the DGE. We see that the DGE requires vastly less weight, especially in terms of the maximal weights near the root of each decision space.

\begin{figure}[]\centering
    \def\picscale{.68}
    \setlength{\tabcolsep}{0pt}
    \begin{tabular}{rrrr}
        \scalebox{\picscale}{\rotatebox{90}{Nash gap}} &    
        \raisebox{-.4\height}{\includegraphics[scale=\picscale]{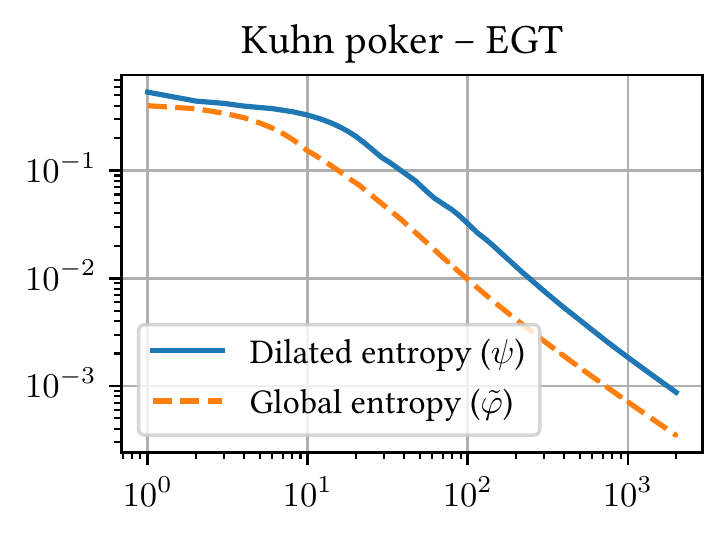}} &
        \raisebox{-.4\height}{\includegraphics[scale=\picscale]{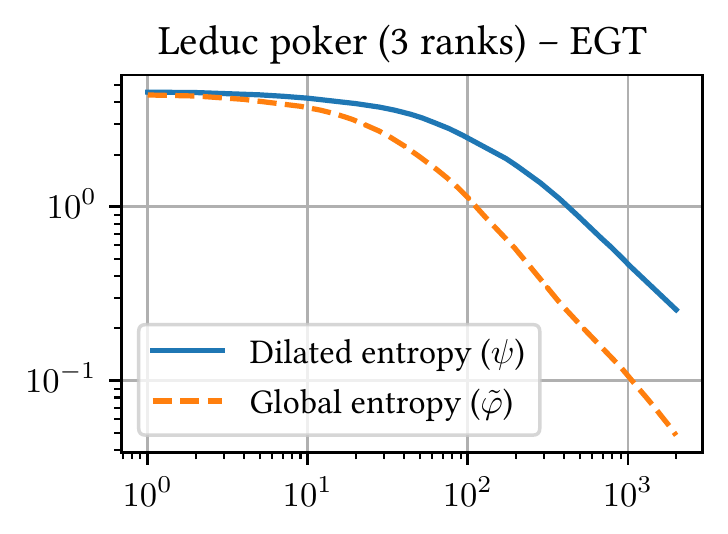}} &
        \raisebox{-.4\height}{\includegraphics[scale=\picscale]{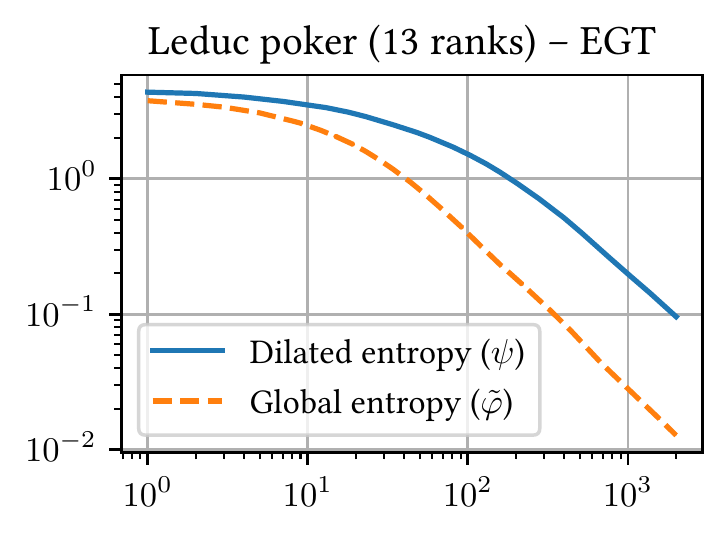}}
        \\
        \scalebox{\picscale}{\rotatebox{90}{Nash gap}} &
        \raisebox{-.4\height}{\includegraphics[scale=\picscale]{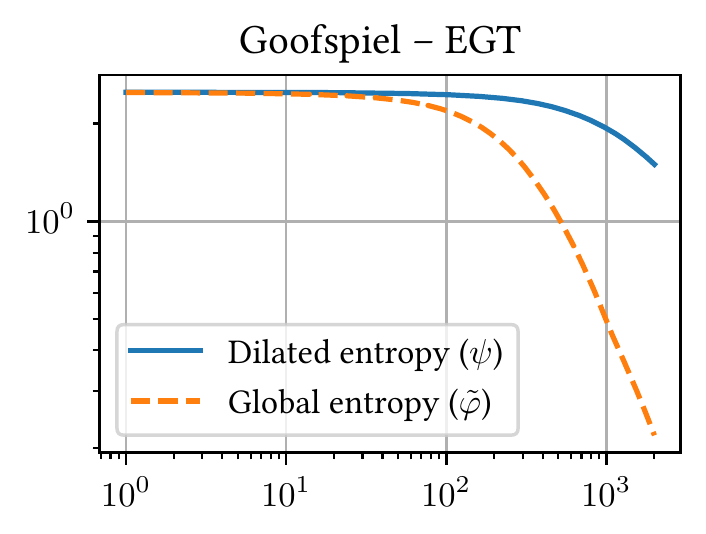}} &
        \raisebox{-.4\height}{\includegraphics[scale=\picscale]{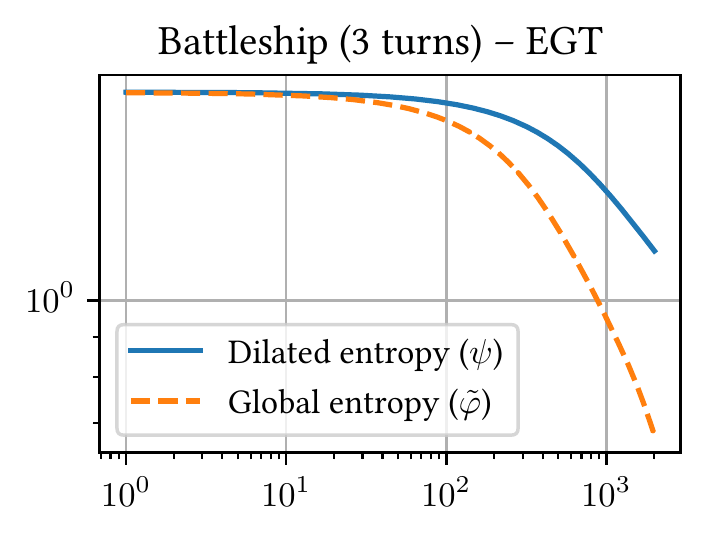}} &
        \raisebox{-.4\height}{\includegraphics[scale=\picscale]{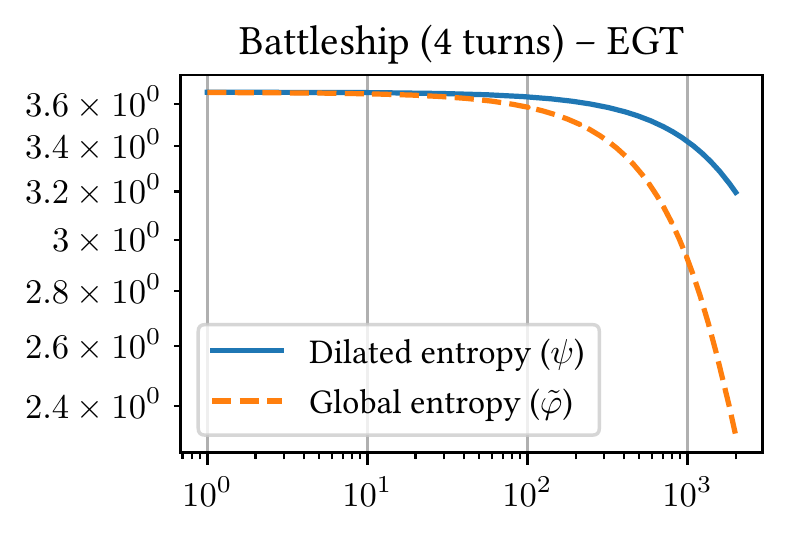}}
        \\
        \scalebox{\picscale}{\rotatebox{90}{Nash gap}} &
        \raisebox{-.4\height}{\includegraphics[scale=\picscale]{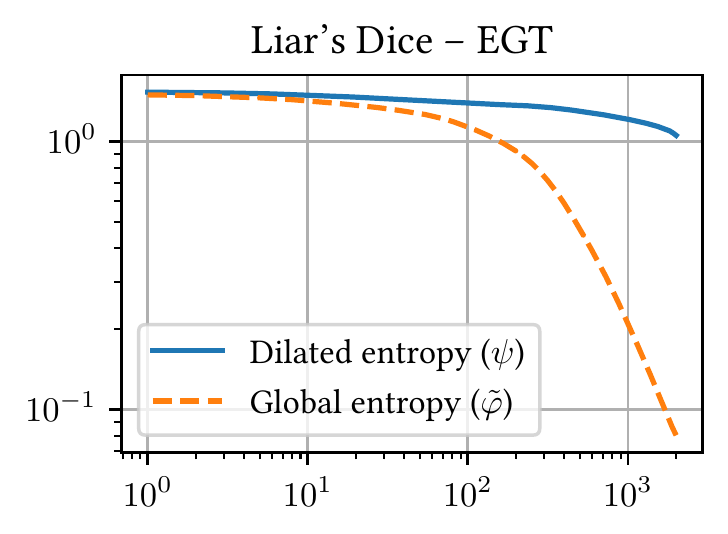}} &
        \raisebox{-.4\height}{\includegraphics[scale=\picscale]{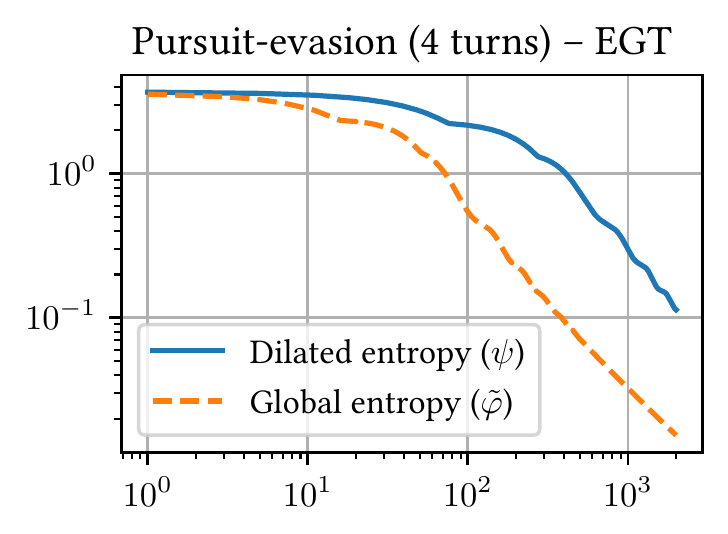}} &
        \raisebox{-.4\height}{\includegraphics[scale=\picscale]{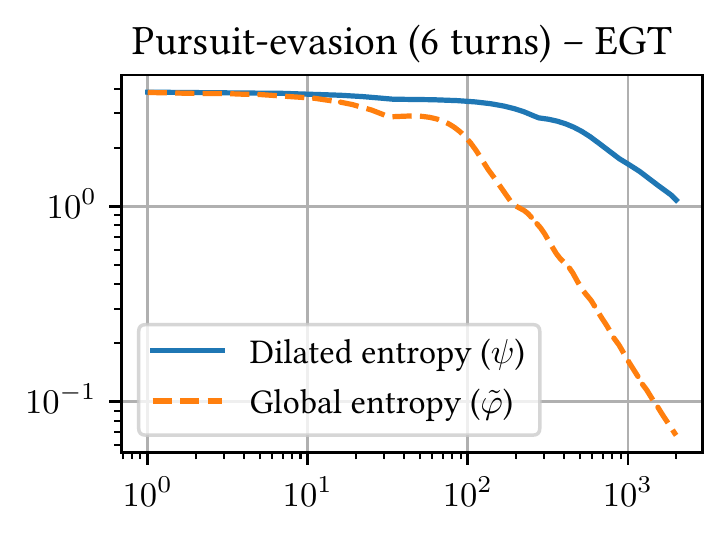}}
        \\[-2mm]
        &
        \scalebox{\picscale}{Iteration}\hspace{1.5cm} &
        \scalebox{\picscale}{Iteration}\hspace{1.5cm} &
        \scalebox{\picscale}{Iteration}\hspace{1.5cm}
        \\
    \end{tabular}
    \caption{Performance of the EGT algorithm instantiated with the two entropy DGFs across nine games. The x axis shows the number of EGT iterations, and the y axis shows the distance to Nash equilibrium.}
    \label{fig:results egt vanilla treeplex}
\end{figure}

\begin{figure}[]\centering
    \def\picscale{.68}
    \setlength{\tabcolsep}{0pt}
    \begin{tabular}{rrrr}
        \scalebox{\picscale}{\rotatebox{90}{Nash gap}} &
        \raisebox{-.4\height}{\includegraphics[scale=\picscale]{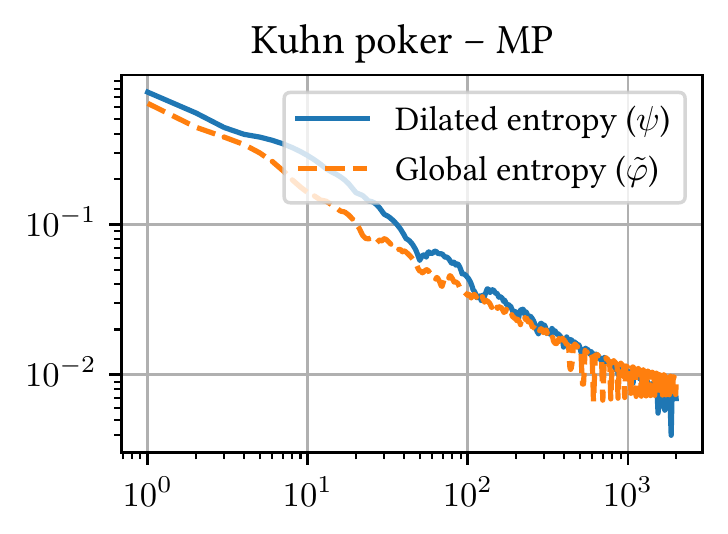}} &
        \raisebox{-.4\height}{\includegraphics[scale=\picscale]{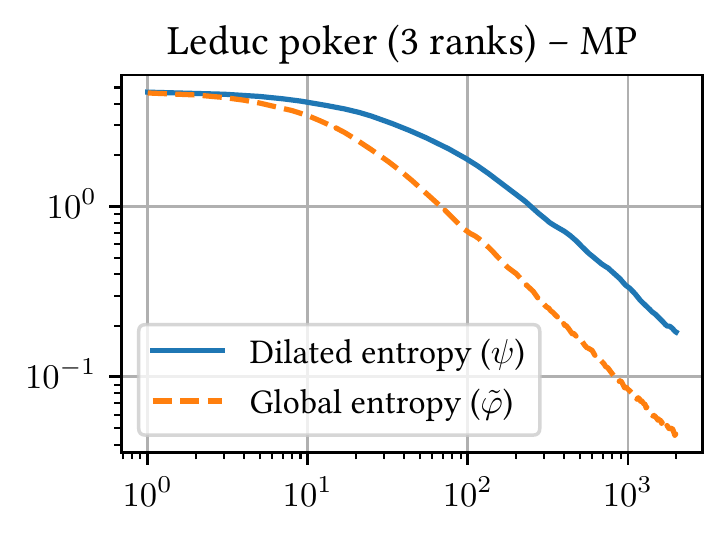}} &
        \raisebox{-.4\height}{\includegraphics[scale=\picscale]{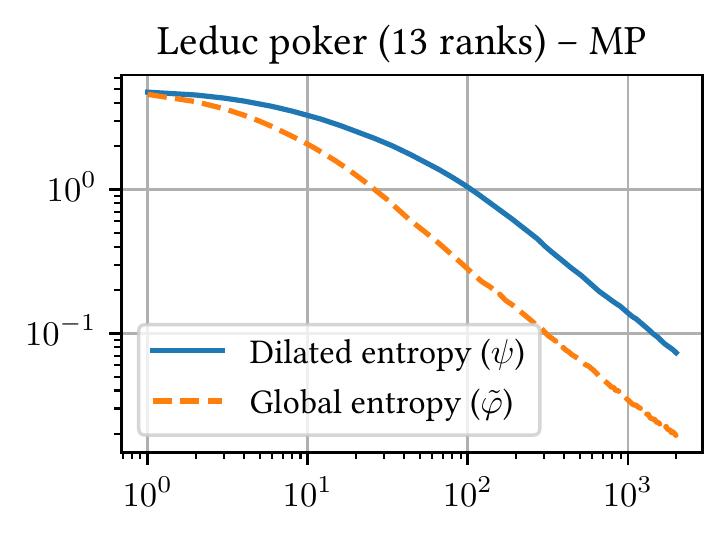}}
        \\[-2mm]
        \scalebox{\picscale}{\rotatebox{90}{Nash gap}} &
        \raisebox{-.4\height}{\includegraphics[scale=\picscale]{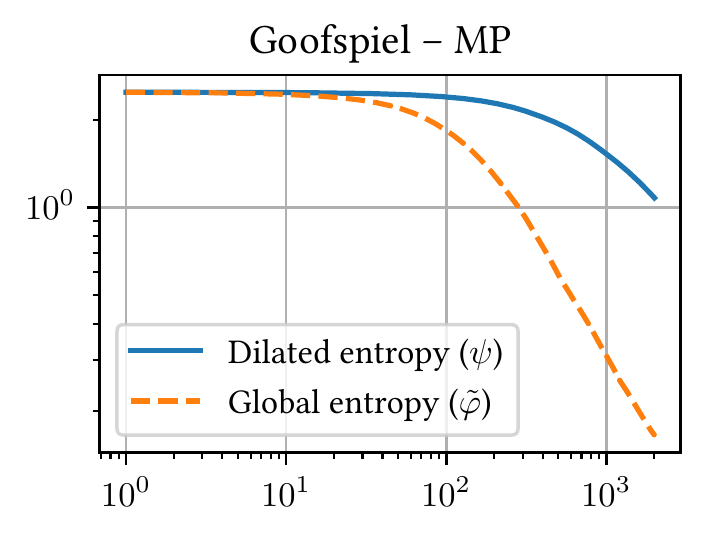}} &
        \raisebox{-.4\height}{\includegraphics[scale=\picscale]{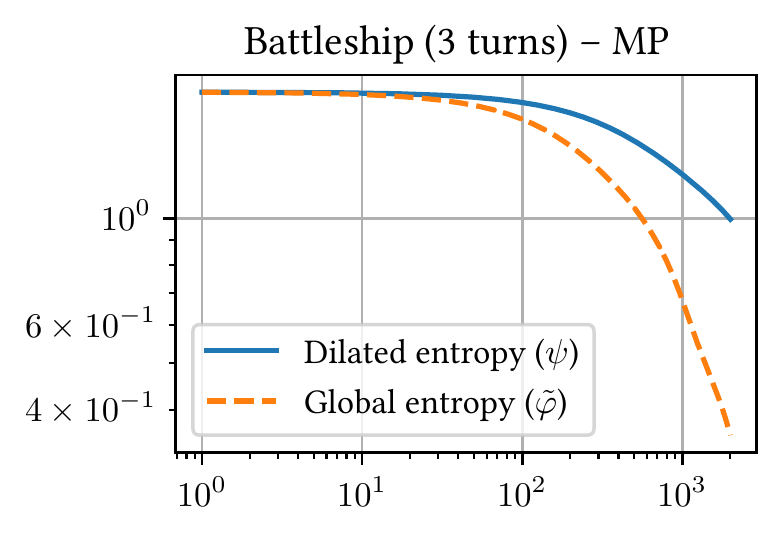}} &
        \raisebox{-.4\height}{\includegraphics[scale=\picscale]{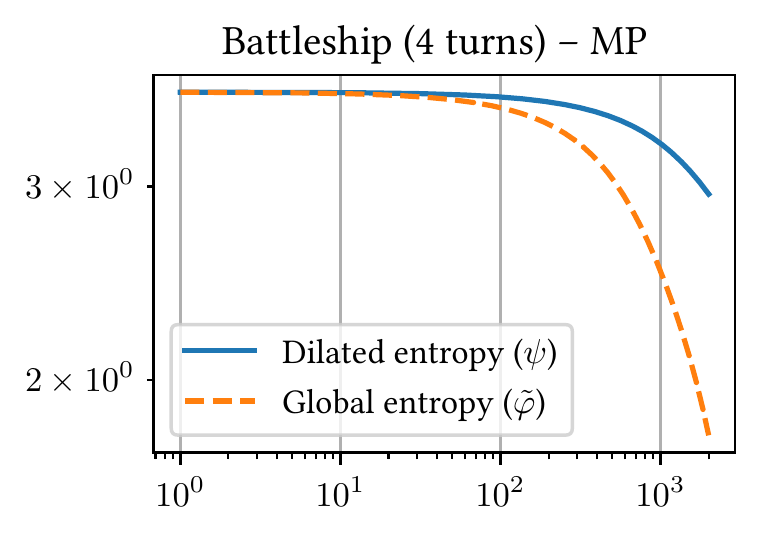}}
        \\[-2mm]
        \scalebox{\picscale}{\rotatebox{90}{Nash gap}} &
        \raisebox{-.4\height}{\includegraphics[scale=\picscale]{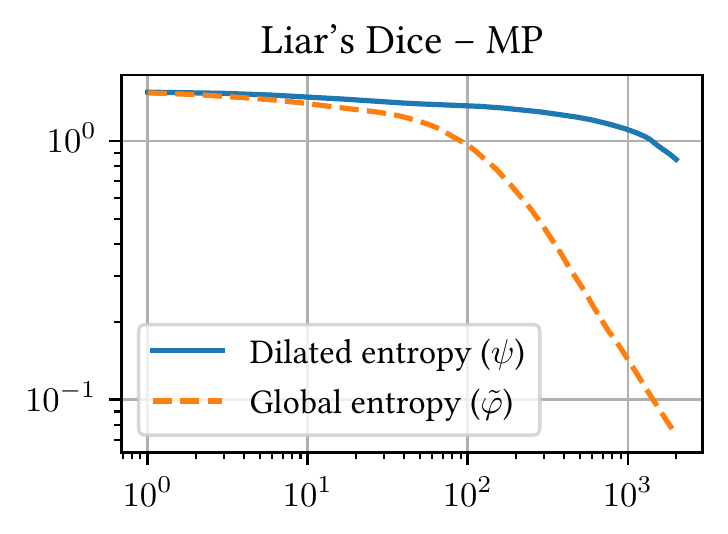}} &
        \raisebox{-.4\height}{\includegraphics[scale=\picscale]{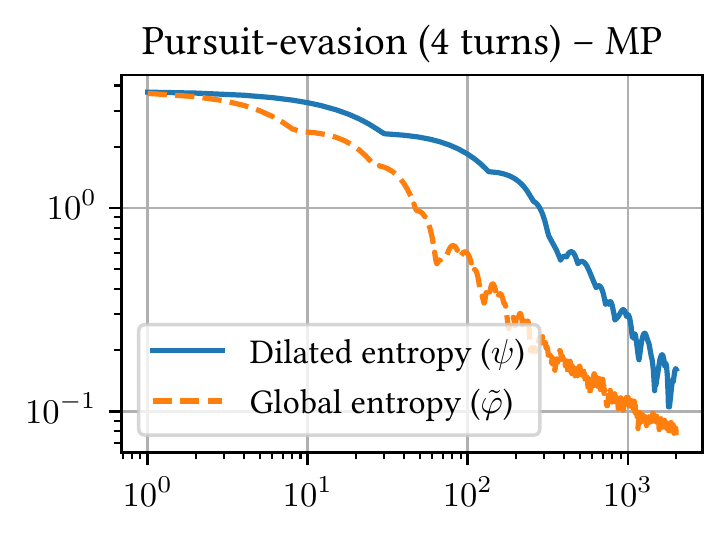}} &
        \raisebox{-.4\height}{\includegraphics[scale=\picscale]{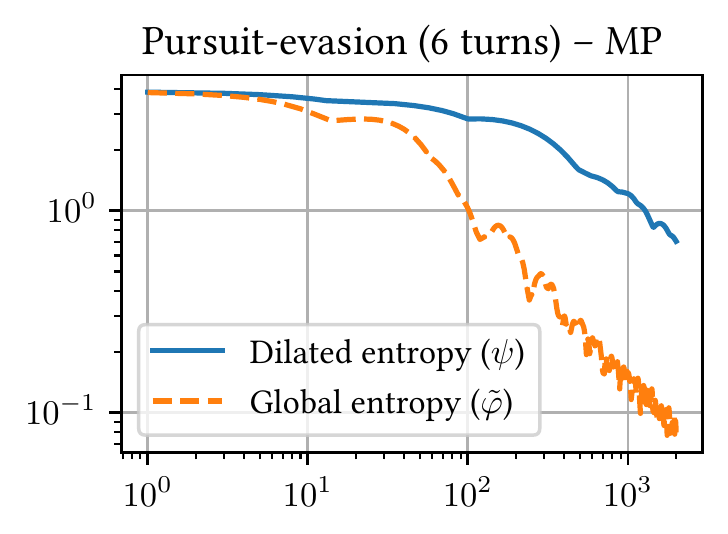}}
        \\[-2mm]
        &
        \scalebox{\picscale}{Iteration}\hspace{1.5cm} &
        \scalebox{\picscale}{Iteration}\hspace{1.5cm} &
        \scalebox{\picscale}{Iteration}\hspace{1.5cm}
        \\
    \end{tabular}
    \caption{Performance of the MP algorithm instantiated with the two entropy DGFs across nine games.
    }
    \label{fig:results mp treeplex}
\end{figure}

First, we study the theoretically-correct way to use the DGFs. In particular, we instantiate both EGT and mirror prox with the stepsizes and DGFs as specified in \cref{thm:nesterovEGT,thm:mirror prox}, for the dilatable global entropy and dilated entropy. The results for EGT are shown in \cref{fig:results egt vanilla treeplex,fig:results mp treeplex}.
Across both algorithms and all nine games, we see that our new dilatable global entropy DGF performs better, sometimes by over an order of magnitude (e.g. in Liar's dice and pursuit evasion (6 turns)).
This is in line with the fact that our new DGF has a better strong convexity modulus, which allows for a much smaller amount of smoothing, while still guaranteeing correctness. This in turns allows the algorithms to safely take larger steps, thereby progressing faster.

\begin{figure}[]\centering
    \def\picscale{.68}
    \setlength{\tabcolsep}{0pt}
    \begin{tabular}{rrrr}
        \scalebox{\picscale}{\rotatebox{90}{Nash gap}} &
        \raisebox{-.4\height}{\includegraphics[scale=\picscale]{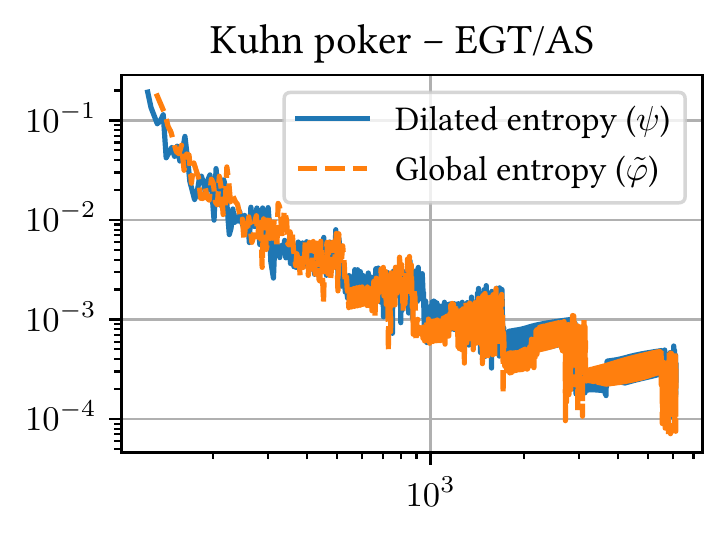}} &
        \raisebox{-.4\height}{\includegraphics[scale=\picscale]{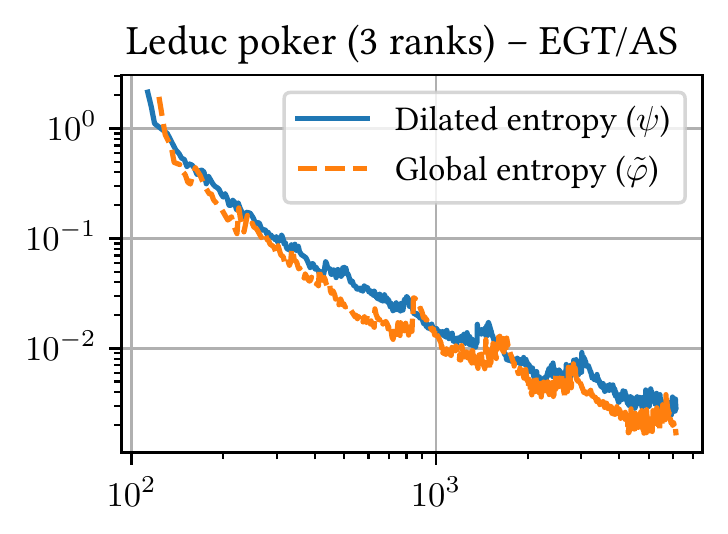}} &
        \raisebox{-.4\height}{\includegraphics[scale=\picscale]{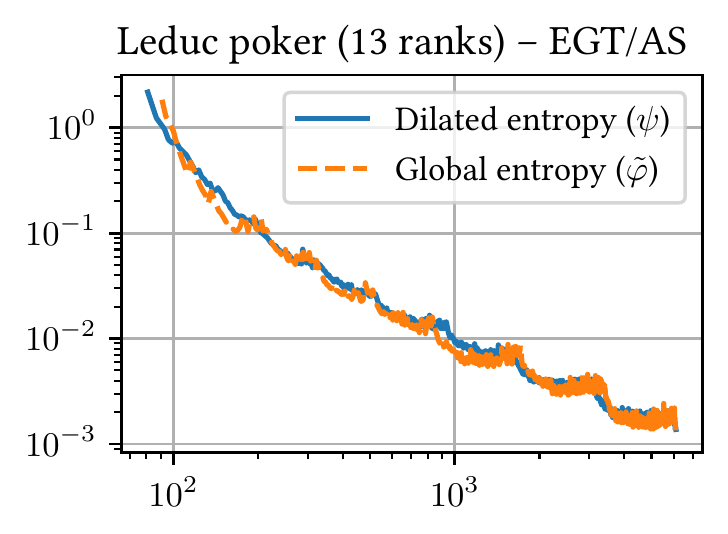}}
        \\[-2mm]
        \scalebox{\picscale}{\rotatebox{90}{Nash gap}} &
        \raisebox{-.4\height}{\includegraphics[scale=\picscale]{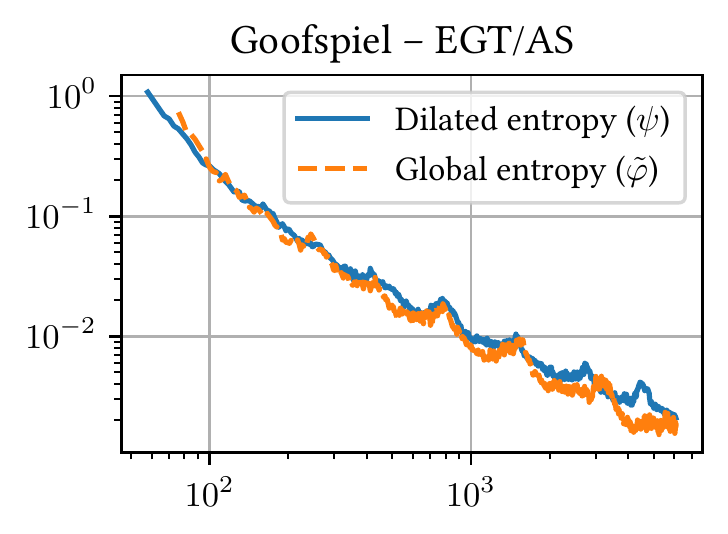}} &
        \raisebox{-.4\height}{\includegraphics[scale=\picscale]{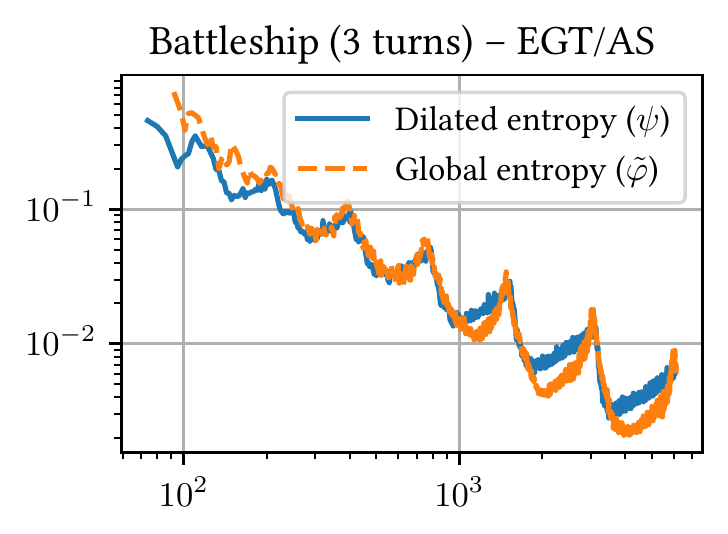}} &
        \raisebox{-.4\height}{\includegraphics[scale=\picscale]{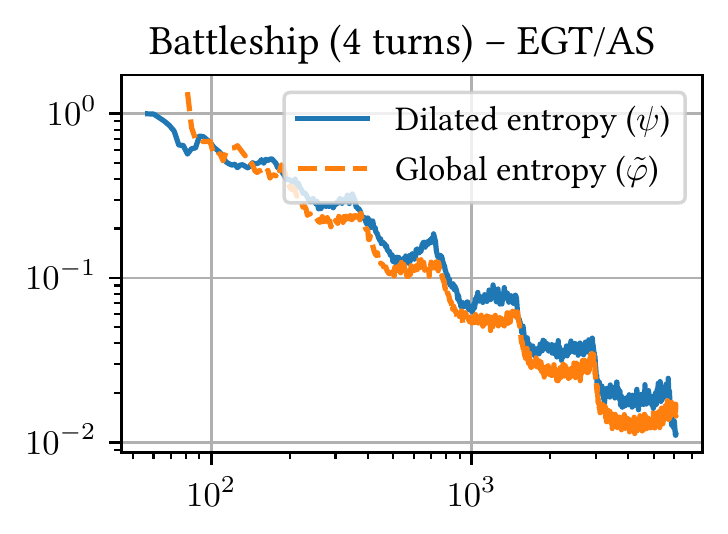}} 
        \\
        \scalebox{\picscale}{\rotatebox{90}{Nash gap}} &
        \raisebox{-.4\height}{\includegraphics[scale=\picscale]{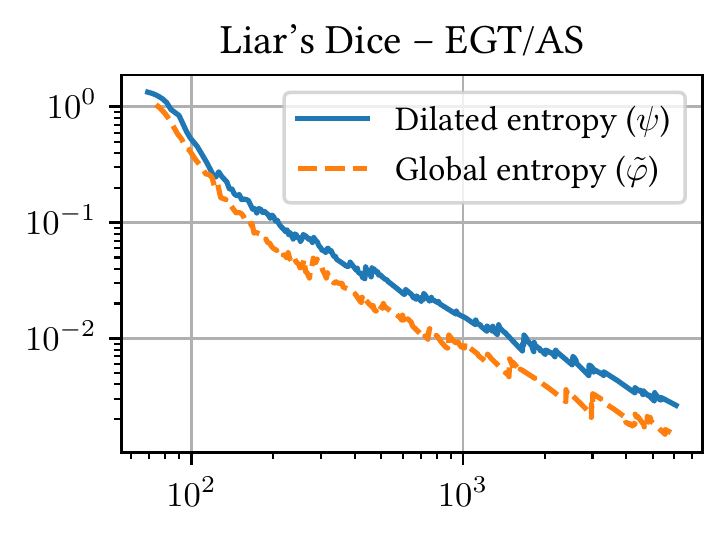}} &
        \raisebox{-.4\height}{\includegraphics[scale=\picscale]{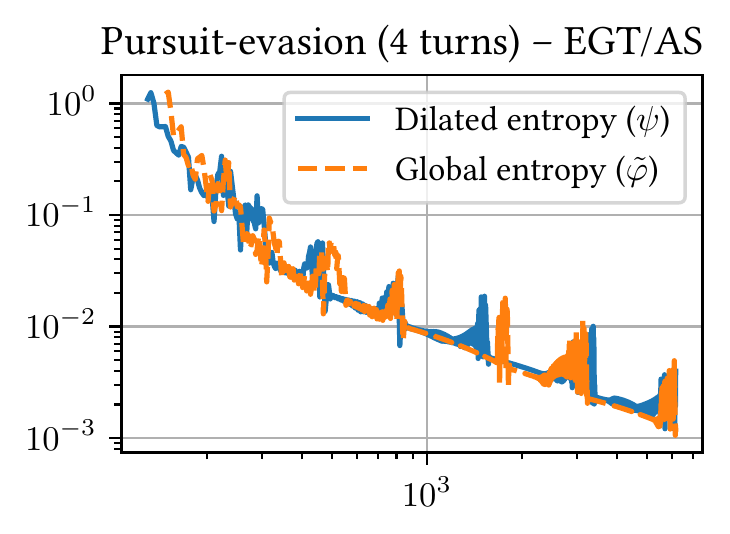}} &
        \raisebox{-.4\height}{\includegraphics[scale=\picscale]{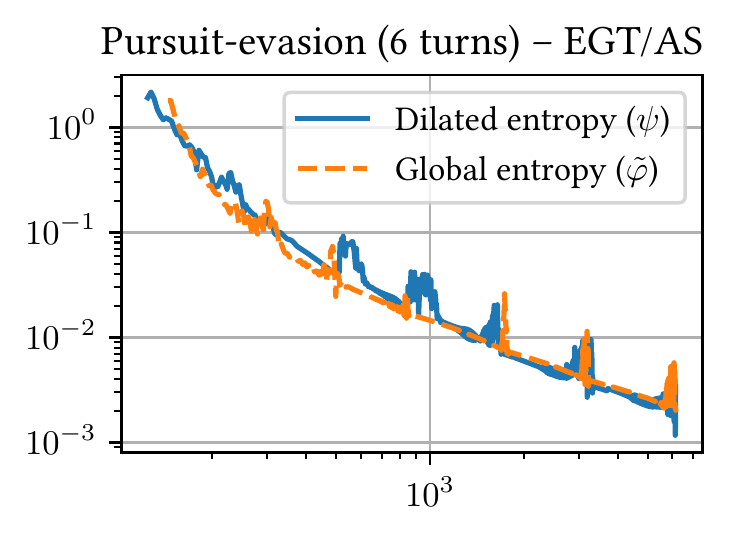}}
        \\[-2mm]
        &
        \scalebox{\picscale}{Gradient computations}\hspace{.9cm} &
        \scalebox{\picscale}{Gradient computations}\hspace{.9cm} &
        \scalebox{\picscale}{Gradient computations}\hspace{.9cm}
        \\
    \end{tabular}
    \caption{Performance of the EGT/AS algorithm instantiated with the two entropy DGFs, as well as aggressive stepsizing, $\mu$ balancing, and initial $\mu$ fitting. 
    }
    \label{fig:results egt tricks treeplex}
\end{figure}

Secondly, we investigate the numerical performance of the two entropy DGFs in the EGT/AS algorithm in \cref{fig:results egt tricks treeplex}.
Here we see a smaller performance improvement. For most of the games, we get a small factor of improvement for the first 100 or so iterations, but then the performance is similar thereafter. For Liar's Dice there is a persistent improvement to using dilatable global entropy across all iterations.

\subsection{Correlated and Team Equilibrium Computation}

Next we investigate the computational performance of our extension of both the dilated entropy DGF and the DGE DGF to the scaled extension operator. In particular, we will consider the problem of computing an NFCCE, which we saw in \cref{sec:correlation polytope} can be formulated as a BSPP via the scaled extension operator. Since the constructed polytope is the same for ex-ante team correlated equilibria, extensive form correlated, and extensive form coarse correlated, we restrict our attention to NFCCE and leave the numerical investigation on the other solution concepts for future work. We expect the takeaways to be similar.

\begin{figure}[]\centering
    \def\picscale{.68}
    \setlength{\tabcolsep}{0pt}
    \begin{tabular}{rrrr}
        \scalebox{\picscale}{\rotatebox{90}{Saddle-point gap}} &
        \raisebox{-.4\height}{\includegraphics[scale=\picscale]{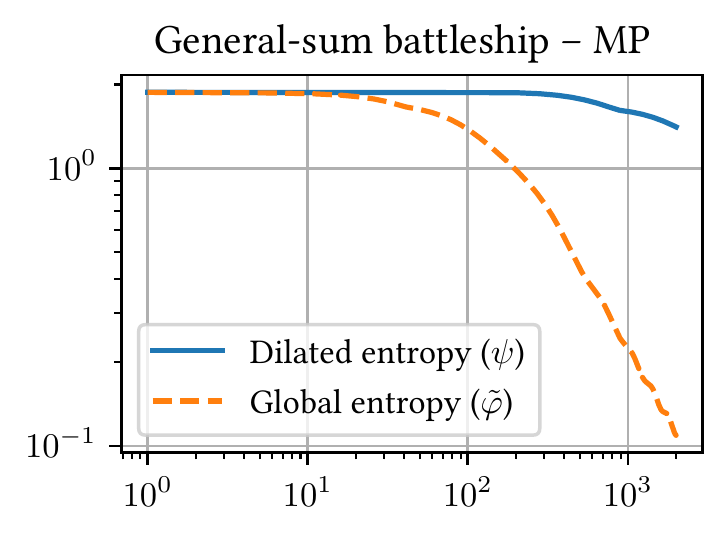}} &
        \raisebox{-.4\height}{\includegraphics[scale=\picscale]{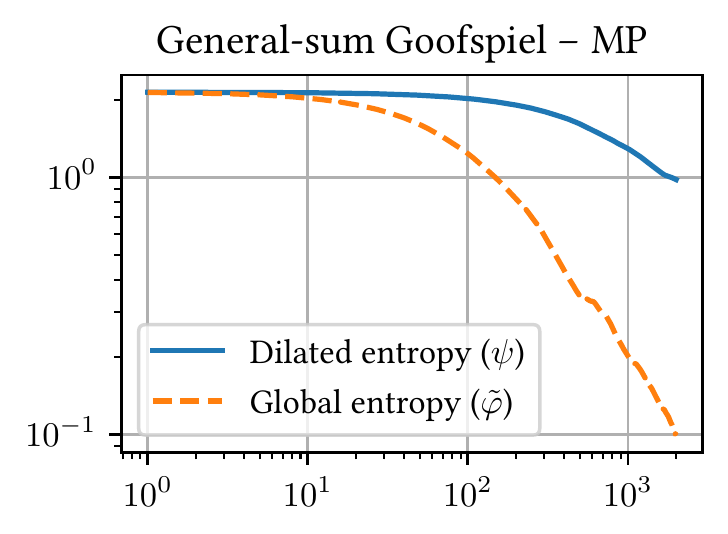}} &
        \raisebox{-.4\height}{\includegraphics[scale=\picscale]{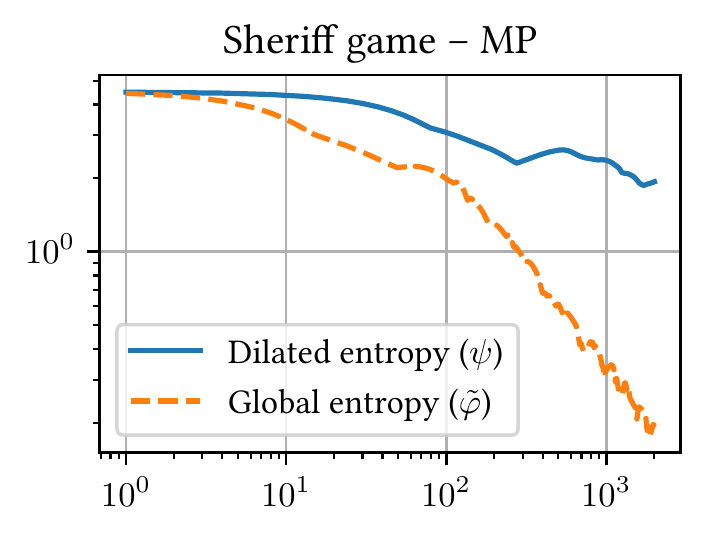}}
        \\[-2mm]
        &
        \scalebox{\picscale}{Iteration}\hspace{1.5cm} &
        \scalebox{\picscale}{Iteration}\hspace{1.5cm} &
        \scalebox{\picscale}{Iteration}\hspace{1.5cm}
        \\
    \end{tabular}
    \caption{Performance of the MP algorithm for finding normal-form coarse-correlated equilibria in three general-sum games.}
    \label{fig:results nfcce}
\end{figure}

\cref{fig:results nfcce} shows the results for instantiating the mirror prox algorithm with our two DGFs. We see that, similar to the case of zero-sum Nash equilibrium, the DGE DGF performs much better than the dilated entropy DGF, again likely due to the smaller weights needed in order to make the DGF strongly convex on the correlation-plan polytope.

    \section{Conclusions and Future Research}

We introduced the dilatable global entropy as a distance-generating function for sequential decision-making polytopes such as those encountered in sequential games. We showed that the DGE function leads to better strong-convexity properties than prior DGFs for the sequence-form polytope, and it improves the associated polytope diameter, as well as the convergence rate of FOMs, by a factor of $2^{\depth_Q}$. Experiments confirmed that this leads to a superior notion of distance. We then extended the DGE, as well as the general dilation framework, to the scaled extension operation. We thereby showed how to construct suitable DGFs for the convex polytopes encounted when computing certain correlated equilibria, as well as team equilibria. Based on these extensions, we showed the first algorithm that achieves a $1/T$ convergence rate to the set of of various correlated equilibria and ex-ante team coordinated equilibria, while requiring only  linear time (in the polytope size) for each iteration.

In future research, it would be interesting to investigate whether our new DGFs can be used to achieve numerical performance comparable to that of the currently practically-fastest algorithms, that is, new CFR variants~\cite{Brown19:Solving,Farina21:Faster,Tammelin15:Solving}, which have worse theoretical convergence rate. In particular, we think that stochastic methods could be a promising line of research for this, because it is harder to tune the stepsize in those methods, in order to account for the weights previously used in the dilated entropy DGF.

    %
    %
    %
    %
    %

    \bibliographystyle{ACM-Reference-Format}
    \bibliography{dairefs}
    
    \clearpage
    \begin{APPENDICES}
        \SingleSpacedXI
        \section{Technical Results about Scaled Extension}
        In what follows, we let $\aff S$ denote the affine hull of set $S$, that is, the set
\[
    \aff S \defeq \mleft\{\sum_{i=1}^k \alpha_i \vec{s}_i \ \Bigg|\ k \in \bbN_{>0}, \alpha_i \in \bbR, \vec{s}_i \in S, \sum_{i=1}^k \alpha_i = 1\mright\}.
\]

\begin{lemma}\label{lem:aff scext}
    Let $\cU \subset \bbR^m, \cV \subset \bbR^n$ and $h : \bbR^m \to \bbR$ be an affine function, nonnegative on $\cU$. Then,
    \[
        \aff(\cU \ext^h \cV) = (\aff \cU) \ext^h (\aff\cV).
    \]
\end{lemma}
\begin{xproof} We prove the two directions of inclusion separately.
    \begin{itemize}[leftmargin=1.1cm]
        \item[($\subseteq$)] Let $(\vu,\vw) \in \aff(\cU\ext^h \cV)$. We argue that $(\vu,\vw) \in (\aff \cU)\ext^h(\aff\cV)$.
        By definition of affine hull, there must exist $k\in\bbN_{>0}$, points $(\vu_i,\vw_i) = (\vu_i, h(\vu_i)\vv_i) \in \cU\ext^h\cV$, and affine combination coefficients $\alpha_1,\dots,\alpha_k \in \bbR, \alpha_1+\dots+\alpha_k=1$, such that
        \begin{align*}
            (\vu, \vw) &= \sum_{i=1}^k \alpha_i (\vu_i, h(\vu_i)\vv_i)
            = \mleft(\sum_{i=1}^k \alpha_i\vu_i, \sum_{i=1}^k \alpha_i h(\vu_i)\vv_i\mright).
        \end{align*}
        We break the analysis into two cases.
        \begin{itemize}
            \item[$\bullet$] If $\sum_{i=1}^k \alpha_i h(\vu_i) = 0$, that is, $\alpha_i h(\vu_i) = 0$ for all $i=1,\dots,k$, consider the point $\vu^* \defeq \sum_{i=1}^k \alpha_i \vu_i \in \aff\cU$. Since $h$ is affine, $h(\vu^*) = \sum_{i=1}^k \alpha_i h(\vu_i) = 0$. Then,
            \begin{align*}
                (\vu,\vw) &= \mleft(\sum_{i=1}^k \alpha_i\vu_i, \vec{0}\mright)
                    = (\vu^*, h(\vu^*) \vw_1) \in (\aff \cU)\ext^h (\aff \cV),
            \end{align*}
            where the inclusion holds since $\cV \subseteq \aff\cV$ (and so, in particular, $\vw_1\in\aff\cV$).
            \item[$\bullet$] Otherwise, $\sum_{i=1}^k h(\vu_i) > 0$. Then,
            \begin{align*}
                (\vu, \vw) &= \mleft(\sum_{i=1}^k \alpha_i \vu_i, \sum_{i=1}^k \alpha_i h(\vu_i)\vv_i\mright) = \mleft(\sum_{i=1}^k \alpha_i \vu_i, \mleft(\sum_{i=1}^k \alpha_i h(\vu_i)\mright)\sum_{i=1}^k \frac{\alpha_i h(\vu_i)}{\sum_{i=1}^k \alpha_i h(\vu_i)}\vv_i\mright)\\
                &= \mleft(\sum_{i=1}^k \alpha_i \vu_i, h\mleft(\sum_{i=1}^k \alpha_i \vu_i\mright)\sum_{i=1}^k \frac{\alpha_i h(\vu_i)}{\sum_{i=1}^k \alpha_i h(\vu_i)}\vv_i\mright) \in (\aff \cU)\ext^h (\aff \cV),
            \end{align*}
            where the last equality follows from the affinity of $h$, and the inclusion follows from the fact that the coefficients $\alpha_i h(\vu_i) / (\sum_{i=1}^k \alpha h(\vu_i))$ sum to $1$ (\emph{i.e.}, they are affine combination coefficients). 
        \end{itemize}
        \item[$(\supseteq)$] Let $\vu \in \aff\cU$ and $\vv \in \aff\cV$. We argue that $(\vu, h(\vu)\vv) \in \aff(\cU \ext^h \cV)$. By definition of affine hull,
        \[
            \vu = \sum_{i=1}^{k_u} \alpha_i \vu_i; \qquad \vv = \sum_{j=1}^{k_v} \beta_j \vv_j
        \]
        for appropriate $k_u,k_v\in\bbN_{>0}$, affine combination coefficients $\alpha_i, \beta_j$, and points $\vu_i\in\cU$, $\vv_j\in\cV$. Consider now the $k_u\cdot k_v$ vectors and coefficients
        \[
            \vec{x}_{ij} \defeq (\vu_i, h(\vu_i)\vv_j) \in \cU\ext^h\cV, \quad \gamma_{ij} \defeq \alpha_i\beta_j \qquad\forall\, i\in\{1,\dots,k_u\},~~j\in\{1,\dots,k_v\}.
        \]
        The coefficients $\gamma_{ij}$ are valid affine combination coefficients, since $\sum_{ij} \gamma_{ij} = \sum_{i=1}^{k_u}\sum_{j=1}^{k_v} \alpha_i\beta_j = (\sum_{i=1}^{k_u} \alpha_i)(\sum_{j=1}^{k_v}\beta_j) = 1$.
        Furthermore, the affine combination $\sum_{ij} \gamma_{ij}\vec{x}_{ij}$ is such that
        \begin{align*}
            \aff(\cU\ext^h\cV) &\ni \sum_{ij} \gamma_{ij}\vec{x}_{ij} = \sum_{i=1}^{k_u}\sum_{j=1}^{k_v} \alpha_i\beta_j (\vu_i,h(\vu_i)\vv_j)\\
            &= \mleft(\sum_{i=1}^{k_u} \alpha_i\vu_i\mleft(\sum_{j=1}^{k_v} \beta_j\mright), \mleft(\sum_{i=1}^{k_u} \alpha_i h(\vu_i) \mright)\mleft(\sum_{j=1}^{k_v}  \beta_j \vv_j \mright)\mright)\\
            &= \mleft(\sum_{i=1}^{k_u} \alpha_i\vu_i, h\mleft(\sum_{i=1}^{k_u} \alpha_i \vu_i \mright)\mleft(\sum_{j=1}^{k_v}  \beta_j \vv_j \mright)\mright) = (\vu,h(\vu)\vv).
        \end{align*}
    \end{itemize}
    The proof is now complete.
\end{xproof}

The next lemma builds on the previous results, and shows that scaled extension and relative interior commute. We recall that the the relative interior of a generic set $S \subseteq \bbR^n$ is the subset
\[
    \relint S \defeq \big\{\vec{s} \in S \mid \exists\, \epsilon > 0 : N_\epsilon(\vec{s}) \cap \aff S \subseteq S\big\} \subseteq S,
\]
where $N_\epsilon(\vec{s})$ denotes the open ball of radius $\epsilon$ centered at $\vec{s}$, that is, $N_\epsilon(\vec{s}) \defeq \{\vec{x} \in \bbR^n : \|\vec{x} - \vec{s}\|_2 < \epsilon\}$. 

\lemrelintscext*
\begin{xproof}
    Since $\cU$ and $\cV$ are bounded, there exists constants
    \[
        M_u \defeq \sup_{\vu\in\cU} \|\vu\|_2 < \infty, \qquad M_v \defeq \sup_{\vv\in\cV} \|\vv\|_2 < \infty.
    \]
    Furthermore, note that
    \[
        0 < \zeta \defeq \sup_{\vu\in\cU} h(\vu) \le M_u \|\vec{a}\|_2 + b < \infty.
    \]
    
    We prove the statement by proving the two directions of inclusion separately.
    \begin{itemize}[leftmargin=1.1cm]
        \item[($\subseteq$)] Let $\vx \defeq (\vu,\vw) \in \relint(\cU\ext^h \cV)$. We argue that $(\vu,\vw) \in (\relint \cU)\ext^h(\relint\cV)$. Since $\relint S \subseteq S$, then $\vw = h(\vu) \vv$ for some appropriate $\vv\in\cV$. Furthemore, by definition of relative interior, there exists a real number $\epsilon > 0$ such that 
        \begin{equation}\label{eq:relint sub}
            N_\epsilon(\vx) \cap \aff(\cU\ext^h \cV) \subseteq \cU\ext^h \cV. 
        \end{equation}
        Let $\epsilon' \defeq \epsilon / \sqrt{1 + \|\vec{a}\|_2^2 M_v^2} > 0$, consider any $\vu'\in B_{\epsilon'}(\vu) \cap \aff\cU$, and let $\vec{x}' \defeq (\vu', h(\vu')\vv)$. The point $\vec{x}'$ satisfies two properties:
        \begin{itemize}
            \item[$\bullet$] First, since $\vu' \in \aff\cU$ by hypothesis, and $\cV \in \cV \subseteq \aff\cV$, it follows that
            \[
                \vec{x}' = (\vu', h(\vu')\vv) \in (\aff \cU)\ext^h (\aff\cV) = \aff(\cU\ext^h\cV),
            \]
            where the last equality follows from \cref{lem:aff scext}.
            \item[$\bullet$] Second, 
                \begin{align*}
                    \|\vec{x}' - \vec{x} \|^2_2 &= \|(\vu,h(\vu)\vv) - (\vu',h(\vu')\vv)\|_2^2 = \|\vu - \vu'\|^2_2 + (h(\vu) - h(\vu'))^2 \|\vv\|_2^2 \\
                        &\le \|\vu - \vu'\|^2_2 + (\vec{a}^\top (\vu - \vu'))^2 M_v^2\\ & \le (1 + \|\vec{a}\|_2^2 M_v^2)\|\vu-\vu'\|_2^2 \hspace{2cm}\text{(apply the Cauchy-Schwarz inequality)}\\& < \epsilon^2,
                \end{align*}
                that is, $\vec{x}' \in N_{\epsilon}(\vec{x})$.
            \end{itemize}
            Combining the two properties above, we have that $\vec{x}' \in N_\epsilon(\vx) \cap \aff(\cU\ext^h\cV)$. So, using~\eqref{eq:relint sub}, we obtain that $\vx' \in (\cU \ext^h \cV)$, which implies $\vu' \in \cU$. Now, since $\vu'\in\aff\cU$ was an arbitrary point such that $\|\vu'-\vu\|_2 < \epsilon'$, we obtain that 
            \[
                N_{\epsilon'}(\vu) \cap \aff\cU \subseteq \cU, \quad\text{which implies that}\quad \vu\in\relint\cU.\numberthis{eq:vu}
            \]
            We prove that $\vv \in \relint\cV$ in a similar fashion. Let $\epsilon'' \defeq \epsilon / \zeta > 0$ and consider any $\vv'' \in B_{\epsilon''}(\vv) \cap \aff\cV$. The point $\vec{x}'' \defeq (\vu, h(\vu)\vv'')$ belongs to $(\aff\cU)\ext^h(\aff\cV) = \aff(\cU\ext^h\cV)$, and furthermore 
            \begin{align*}
                \|\vec{x}'' - \vec{x} \|^2_2 &= \|(\vu,h(\vu)\vv) - (\vu,h(\vu)\vv'')\|_2^2 = h(\vu)^2 \|\vv - \vv'\|_2^2 \le \zeta^2 \|\vv-\vv''\|_2^2 \\
                &< \epsilon^2.
            \end{align*}
            This shows that
            \[
                \vx'' \in N_{\epsilon}(\vx) \cap \aff(\cU\ext^h\cV) \subseteq \cU\ext^h\cV.
            \]
            Since $\vx'' = (\vu, h(\vu)\vv'')$, this implies that $\vv'' \in \cV$. Since $\vv''$ was arbitrary, we have proved that
            \[
                N_{\epsilon''}(\vv) \cap \aff\cV \subseteq \cV,\quad\text{which implies that}\quad \vv\in\relint\cV.\numberthis{eq:vv}
            \]
            Putting~\eqref{eq:vu} and~\eqref{eq:vv} together, we have that
            \[
                \vx = (\vu,h(\vu)\vv) \in (\relint \cU)\ext^h (\relint \cV),
            \]
            as we wanted to show.
        \item[$(\supseteq)$] Let $\vu \in \relint\cU$ and $\vv \in \relint\cV$. We argue that $(\vu, h(\vu)\vv) \in \relint(\cU \ext^h \cV)$. By definition of relevative interior, there exist constants $\epsilon_u, \epsilon_v > 0$ such that
        \[
            N_{\epsilon_u}(\vu) \cap \aff\cU \subseteq \cU, \qquad\text{and}\qquad N_{\epsilon_v}(\vv) \cap \aff\cV \subseteq \cV.
            \numberthis{eq:defn sups}
        \]
        Let
        \[
            \epsilon' \defeq \frac{\min\{\epsilon_u,\epsilon_v\}}
            {\sqrt{\max\mleft\{\frac{2}{h(\vu)^2}, 1+\frac{2\|\vec{a}\|_2^2 M_v^2}{h(\vu)^2}\mright\}}} > 0
            \numberthis{eq:eps prime}
        \]
        and let $\vx' \in B_{\epsilon'}(\vx) \cap \aff(\cU\ext^h\cV)$ be arbitrary. Since from \cref{lem:aff scext} $\aff(\cU\ext^h\cV) = (\aff \cU)\ext^h (\aff\cV)$, then $\vx' = (\vu', h(\vu')\vv')$ for appropriate $\vu' \in \aff\cU, \vv' \in \aff\cV$. Now, since $\vu\in\relint\cU$ and $h$ is strictly positive in the relative interior of $\cU$ by hypothesis, we can write 
        \begin{align*}
            \|\vu-\vu'\|_2^2 + \|\vv-\vv'\|_2^2
            &= \|\vu-\vu'\|_2^2 + \frac{1}{h(\vu)^2}\|(h(\vu)\vv - h(\vu')\vv') + (h(\vu')-h(\vu))\vv'\|_2^2\\
            &\le \|\vu-\vu'\|_2^2 + \frac{2}{h(\vu)^2}\|h(\vu)\vv - h(\vu')\vv'\|_2^2 + \frac{2}{h(\vu)^2}(h(\vu)-h(\vu'))^2\|\vv'\|_2^2\\
            &\le \|\vu-\vu'\|_2^2 + \frac{2}{h(\vu)^2}\|h(\vu)\vv - h(\vu')\vv'\|_2^2 + \frac{2 \|\vec{a}\|^2_2 M_v^2}{h(\vu)^2}\|\vu-\vu'\|_2^2 \numberthis{eq:used cs}\\
            &\le \max\mleft\{\frac{2}{h(\vu)^2}, 1+\frac{2\|\vec{a}\|_2^2 M_v^2}{h(\vu)^2}\mright\}\Big(\|\vu-\vu'\|_2^2 + \|h(\vu)\vv - h(\vu')\vv'\|_2^2\Big)\\
            &= \max\mleft\{\frac{2}{h(\vu)^2}, 1+\frac{2\|\vec{a}\|_2^2 M_v^2}{h(\vu)^2}\mright\}\|(\vu,h(\vu)\vv) - (\vu',h(\vu')\vv')\|_2^2\\
            &= \max\mleft\{\frac{2}{h(\vu)^2}, 1+\frac{2\|\vec{a}\|_2^2 M_v^2}{h(\vu)^2}\mright\} \|\vx - \vx'\|_2^2 < \min\{\epsilon_u,\epsilon_v\}^2,\numberthis{eq:core}
        \end{align*}
        where we applied the Cauchy-Schwarz inequality in~\eqref{eq:used cs}, and the last inequality follows from the hypothesis that $\vx' \in B_{\epsilon'}(\vx)$ together with the definition of $\epsilon'$ given in~\eqref{eq:eps prime}. Now, since $\|\vu-\vu'\|_2^2$ and $\|\vv-\vv'\|_2^2$ are nonnegative quantities,~\eqref{eq:core} implies that
        \[
            \|\vu-\vu'\|_2 < \min\{\epsilon_u,\epsilon_v\},\qquad\text{and}\qquad
            \|\vv-\vv'\|_2 < \min\{\epsilon_u,\epsilon_v\}.
        \]
        Furthermore, since $\vu'\in\aff\cU,\vv'\in\aff\cV$ by construction, we have that
        \[
            \vu' \in (B_{\min\{\epsilon_u,\epsilon_v\}}(\vu)\cap\aff\cU) \subseteq (B_{\epsilon_u}(\vu)\cap\aff\cU), \qquad
            \vv' \in (B_{\min\{\epsilon_u,\epsilon_v\}}(\vv)\cap\aff\cV) \subseteq (B_{\epsilon_v}(\vv)\cap\aff\cV).
        \]
        Hence, using the hypothesis~\eqref{eq:defn sups}, we obtain that $\vu' \in \cU$ and $\vv'\in \cV$, which implies that
        \[
            \vx' = (\vu', h(\vu')\vv') \in \cU\ext^h\cV.
        \]
        Now, since $\vx' \in B_{\epsilon'}(\vx) \cap \aff(\cU\ext^h\cV)$ was arbitrary, this implies
        \[
            B_{\epsilon'}(\vx) \cap \aff(\cU\ext^h\cV) \subseteq \cU\ext^h\cV,
        \]
        and so $\vx \in \relint (\cU\ext^h\cV)$, which is what we wanted to show.
    \end{itemize}
\end{xproof}
        \section{Further Details}\label{app:details}
        \lemhandysc*
\begin{xproof}
Using the expression~\eqref{eq:gradient dz} for the gradient of $d_z$, for all $(\vu,\vv),(\vu',\vv')\in\relint\cZ$ we have
\begin{align*}
    \nabla d_z(\vu,\vv)^\top\begin{pmatrix}\vu'\\\vv'\end{pmatrix} &=
        \mleft(\nabla d_u(\vu)  + \alpha_v \mleft(d_v\mleft(\frac{\vv}{h(\vu)}\mright)- \nabla d_v\mleft(\frac{\vv}{h(\vu)}\mright)^{\!\!\top}\!\! \frac{\vv}{h(\vu)}\mright)\vec{a}\mright)^{\!\!\top}\!\! \vu'+
    \mleft(\alpha_v \nabla d_v\mleft(\frac{\vv}{h(\vu)}\mright)\mright)^{\!\!\top}\!\! \vv'\\
    &= \nabla d_u(\vu)^\top \vu' + \alpha_v d_v\mleft(\frac{\vv}{h(\vu)}\mright)h(\vu') - \alpha_v \nabla d_v\mleft(\frac{\vv}{h(\vu)}\mright)^{\!\!\top}\!\!\frac{\vv}{h(\vu)}h(\vu') +
    \alpha_v \nabla d_v\mleft(\frac{\vv}{h(\vu)}\mright)^{\!\!\top}\!\! \vv'\\
    &= \nabla d_u(\vu)^\top \vu' - \alpha_v h(\vu')\mleft[-d_v\mleft(\frac{\vv}{h(\vu)}\mright)-\nabla d_v\mleft(\frac{\vv}{h(\vu)}\mright)^{\!\!\top}\!\!\mleft(\frac{\vv'}{h(\vu')}-\frac{\vv}{h(\vu)}\mright)
    \mright],
    \numberthis{eq:simple gradient1}
\end{align*}
where the second equality uses the fact that $h( \vu' ) = \vec{a}^\top\vu'$ by hypothesis. In particular, when $(\vu',\vv') = (\vu,\vv)$ we have
\begin{align*}
    \nabla d_z(\vu,\vv)^\top\begin{pmatrix}\vu\\\vv\end{pmatrix} 
    &=\nabla d_u(\vu)^\top\vu + \alpha_v h(\vu) d_v\mleft(\frac{\vv}{h(\vu)}\mright),\numberthis{eq:simple gradient2}
\end{align*}
and therefore
\begin{align*}
    \nabla d_z(\vu,\vv)^\top\begin{pmatrix}\vu-\vu'\\\vv-\vv'\end{pmatrix} 
    &= \nabla d_u(\vu)^\top (\vu-\vu') + \alpha_v h(\vu)d_v\mleft(\frac{\vv}{h(\vu)}\mright)\\&\hspace{3cm} + \alpha_v h(\vu')\mleft[-d_v\mleft(\frac{\vv}{h(\vu)}\mright)-\nabla d_v\mleft(\frac{\vv}{h(\vu)}\mright)^{\!\!\top}\!\!\mleft(\frac{\vv'}{h(\vu')}-\frac{\vv}{h(\vu)}\mright)
    \mright]\\
    &= \nabla d_u(\vu)^\top (\vu-\vu') + \alpha_v h(\vu)d_v\mleft(\frac{\vv}{h(\vu)}\mright) - \alpha_v h(\vu')d_v\mleft(\frac{\vv'}{h(\vu')}\mright)\\&\hspace{1cm} + \alpha_v h(\vu')\mleft[d_v\mleft(\frac{\vv'}{h(\vu')}\mright)-d_v\mleft(\frac{\vv}{h(\vu)}\mright)-\nabla d_v\mleft(\frac{\vv}{h(\vu)}\mright)^{\!\!\top}\!\!\mleft(\frac{\vv'}{h(\vu')}-\frac{\vv}{h(\vu)}\mright)
    \mright].
\end{align*}
Since $\alpha_v \ge 0$ and $h$ is a nonnegative function by hypothesis, using the strong convexity hypothesis on $d_v$ yields
\begin{align*}
    \nabla d_z(\vu,\vv)^\top\begin{pmatrix}\vu-\vu'\\\vv-\vv'\end{pmatrix} 
    &\ge \nabla d_u(\vu)^\top (\vu-\vu') + \alpha_v h(\vu)d_v\mleft(\frac{\vv}{h(\vu)}\mright) - \alpha_v h(\vu')d_v\mleft(\frac{\vv'}{h(\vu')}\mright)\\&\hspace{7cm} + \alpha_v\frac{h(\vu')}{2}\mleft\|\frac{\vv'}{h(\vu')} - \frac{\vv}{h(\vu)}\mright\|^2.
            \numberthis{eq:simple gradient3}
\end{align*}
Symmetrically,
\begin{align*}
    \nabla d_z(\vu',\vv')^\top\begin{pmatrix}\vu'-\vu\\\vv'-\vv\end{pmatrix} 
    &\ge \nabla d_u(\vu')^\top (\vu'-\vu) + \alpha_v h(\vu')d_v\mleft(\frac{\vv'}{h(\vu')}\mright) - \alpha_v h(\vu)d_v\mleft(\frac{\vv}{h(\vu)}\mright)\\&\hspace{7cm} + \alpha_v\frac{h(\vu)}{2}\mleft\|\frac{\vv'}{h(\vu')} - \frac{\vv}{h(\vu)}\mright\|^2.\numberthis{eq:simple gradient4}
\end{align*}
Adding~\eqref{eq:simple gradient4} and~\eqref{eq:simple gradient3} together we obtain
\begin{align*}
    (\nabla d_z(\vu,\vv) - \nabla d_z(\vu',\vv'))^\top \begin{pmatrix}\vu - \vu'\\\vv - \vv'\end{pmatrix}
    &\ge
        \big(\nabla d_u(\vu) - \nabla d_u(\vu')\big)^\top(\vu-\vu')\\&\hspace{3.5cm} + \alpha_v \mleft(\frac{h(\vu)}{2} + \frac{h(\vu')}{2}\mright)\mleft\|\frac{\vv}{h(\vu)} - \frac{\vv'}{h(\vu)}\mright\|^2\\
    &=\big(\nabla d_u(\vu) - \nabla d_u(\vu')\big)^\top(\vu-\vu')\\&\hspace{3.5cm} + \alpha_v h\mleft(\frac{\vu+\vu'}{2}\mright)\mleft\|\frac{\vv}{h(\vu)} - \frac{\vv'}{h(\vu)}\mright\|^2
,\end{align*}
which is what we wanted to prove.
\end{xproof}

\propellonesc*
\begin{xproof}
    Strong convexity with modulus $1$ with respect to the Euclidean norm follows directly from \cref{cor:alpha scext dilated}. So, we only need to establish $(1/M_\cX)$-strong convexity with respect to the $\ell_1$ norm. 
    
    Let $\psi$ be the dilated entropy DGF for $\cX$, as defined in \cref{def:dilent scext}. As a first step, we will show by induction on $n$ that for all $\vec{m}=(\vec{m}_1,\dots,\vec{m}_n)\in\bbR^{s_1}\times\dots\times\bbR^{s_n}$ and $\vec{x}=(\vec{x}_1,\dots,\vec{x}_n)\in\relint \cX = (\relint \cX_1)\ext^{h_1}\dots \ext^{h_{n-1}} \cX_n$,
    \[
        \vec{m}^\top \nabla^2 \psi(\vec{x}) \vec{m} \ge \sum_{k=1}^n \sum_{i=1}^{s_k} \mleft(\frac{\alpha_k}{2} - \sum_{p=k}^{n-1} \alpha_{p+1}\|\vec{a}_p\|_0\, a_{p,k}[i]\mright) \frac{m_k[i]^2}{x_k[i]}.
        \numberthis{eq:dilent scext bound}
    \]
    \begin{description}
        \item[Base case] The base case corresponds to the case where $n=1$ an $\cX = \Delta^{s_1}$, that is, no scaled extension is performed. In that case,
        \[
            d: \cX \ni \vx_1 \to \alpha_1 d_1(\vx_1) = \alpha_1 \log(s_1) + \alpha_1\sum_{i=1}^{s_1} x_1[i]\log x_1[i].
        \]
        Since $\nabla^2 d(\vx_1) = \textrm{diag}\mleft(\frac{1}{x_1[i]} : i = 1,\dots,s_1\mright)$, we have that for all $\vec{m}\in\bbR^{s_1}$ and all $\vec{x}\in\relint\cX$,
        \[
            \vec{m}^\top \nabla^2 d(\vx) \vec{m} = \sum_{i=1}^{s_1} \alpha_1 \frac{m[i]^2}{x[i]} \ge \sum_{i=1}^{s_1} \frac{\alpha_1}{2} \frac{m[i]^2}{x[i]},
        \]
        which satisfies the inductive statement.
        \item[Inductive step] Let $\cU \defeq \cX_1 \ext^{h_1} \dots \ext^{h_{n-2}} \cX_{n-1}$, and let $\tilde d$ be the dilated entropy DGF constructed for $\cU$. We will show that the inductive statement continues to hold after one further application of the inductive DGF construction is performed, that is, for the dilated entropy DGF
        \[
            d: (\cU \ext^{h_{n-1}} \Delta^{s_n}) \ni (\vec{u}, \vec{w}) \mapsto \tilde d(\vec{u}) + \alpha_n\mleft(\log(s_n) + \sum_{i=1}^{s_n} w[i] \log \frac{w[i]}{h_{n-1}(\vec{u})}\mright).
        \]
        Note that for all $(\vec{u},\vec{w}) \in (\relint \cU) \ext^{h_{n-1}} (\relint \Delta^{s_n})$,
        \[
            \nabla_{\vec{u}}^2 d(\vec{u}, \vec{w}) = \nabla^2 \tilde d(\vec{u}) + \alpha_n \frac{\vec{a}_{n-1}\vec{a}_{n-1}^\top}{h_{n-1}(\vec{u})}, \qquad
            \nabla^2_{\vec{w}} d(\vu, \vw) = \alpha_n\textrm{diag}\mleft(\mleft\{\frac{1}{w[i]} : i = 1,\dots,s_n\mright\}\mright),
        \]
        and
        \[
            \frac{\partial^2}{\partial u[i] \partial w[j]} d(\vu, \vw) = \frac{\partial^2}{\partial w[j] \partial u[i]} d(\vu, \vw) =
            -\alpha_n\frac{ a_{n-1}[i]}{h_{n-1}(\vu)} \qquad\begin{array}{ll}\forall& i=1,\dots,s_1+\dots+s_{n-1},\\& j = 1,\dots, s_{n}.\end{array}
        \]
        Let $\vm=(\vec{m}_1,\dots,\vec{m}_n) \in \bbR^{s_1}\times\dots\times\bbR^{s_n}$ and $\vx=(\vec{x}_1,\dots,\vec{x}_n) \in \relint \cX = (\relint \cU) \ext^{h_{n-1}} (\relint\Delta^{s_n})$ be arbitrary, and introduce the vector $\tilde{\vx} \defeq (\vx_1,\dots,\vx_n) \in \relint\cU$ and $\tilde{\vm}\defeq (\vm_1,\dots,\vm_{n-1})$. Then, 
        \begin{align*}
            &\vec{m}^\top \nabla^2 d(\vec{x}) \vm \\
            &\hspace{1cm}= \tilde{\vm}^\top \nabla^2 \tilde d(\tilde{\vx})\tilde{\vm} + \alpha_n \frac{(\vec{a}_{n-1}^\top \tilde{\vm})^2}{h_{n-1}(\tilde{\vx})} - 2\alpha_n \frac{ \vec{a}_{n-1}^\top \tilde{\vm}}{h_{n-1}(\tilde{\vx})}(\vec{1}^\top \vm_n) + \alpha_n\sum_{i=1}^{s_n} \frac{m_n[i]^2}{x_n[i]}\\
            &\hspace{1cm}= \tilde{\vm}^\top \nabla^2 \tilde d(\tilde{\vx})\tilde{\vm} + \alpha_n\underbrace{\mleft[ \frac{(\vec{a}_{n-1}^\top \tilde{\vm})^2}{h_{n-1}(\tilde{\vx})} - 2\frac{ \vec{a}_{n-1}^\top \tilde{\vm}}{h_{n-1}(\tilde{\vx})}(\vec{1}^\top \vm_n) + \frac{1}{2}\sum_{i=1}^{s_n} \frac{m_n[i]^2}{x_n[i]}\mright]}_{\eqqcolon\ \Lambda(\vm_n)} + \frac{\alpha_n}{2}\sum_{i=1}^{s_n} \frac{m_n[i]^2}{x_n[i]}.\numberthis{eq:mhessianm}
        \end{align*}
        We now lower bound the expression $\Lambda$ in square brackets by explicitly computing a minimizer for $\vm_n$. The term is strongly convex in $\vm_n$, and therefore the minimizer is the only point in $\bbR^{s_n}$ for which the gradient is the $\vec{0}$ vector. Specifically, given the partial derivatives
        \[
        \frac{\partial}{\partial m_n[i]}\mleft[ \frac{(\vec{a}_{n-1}^\top \tilde{\vm})^2}{h_{n-1}(\tilde{\vx})} - 2\frac{ \vec{a}_{n-1}^\top \tilde{\vm}}{h_{n-1}(\tilde{\vx})}(\vec{1}^\top \vm_n) + \frac{1}{2}\sum_{i=1}^{s_n} \frac{m_n[i]^2}{x_n[i]}\mright] = -2\frac{\va_{n-1}^\top \tilde{\vm}}{h_{n-1}(\tilde{\vx})} + \frac{m_n[i]}{x_n[i]},
        \]
        we find that the minimizer of $\Lambda$ is the vector $\vm_n^*$ whose coordinates are
        \[
            m_n^*[i] = 2 x_n[i] \frac{\va_{n-1}^\top \tilde{\vm}}{h_{n-1}(\tilde{\vx})}.
        \]
        Evaluating $\Lambda$ in $\vm_n^*$ in particular yields
        \begin{align*}
            \Lambda(\vm_n) &\ge \Lambda(\vm_n^*) \\&= \frac{(\vec{a}_{n-1}^\top \tilde{\vm})^2}{h_{n-1}(\tilde{\vx})} - 2\frac{ \vec{a}_{n-1}^\top \tilde{\vm}}{h_{n-1}(\tilde{\vx})}\mleft(\sum_{i=1}^{s_n} 2 x_n[i] \frac{\va_{n-1}^\top \tilde{\vm}}{h_{n-1}(\tilde{\vx})}\mright) + \frac{1}{2}\sum_{i=1}^{s_n} \frac{1}{x_n[i]}\mleft(2 x_n[i] \frac{\va_{n-1}^\top \tilde{\vm}}{h_{n-1}(\tilde{\vx})}\mright)^2\\
            &=\frac{(\vec{a}_{n-1}^\top \tilde{\vm})^2}{h_{n-1}(\tilde{\vx})} - 4\frac{ (\vec{a}_{n-1}^\top \tilde{\vm})^2}{h^2_{n-1}(\tilde{\vx})}\mleft(\sum_{i=1}^{s_n} x_n[i] \mright) + 2\frac{ (\vec{a}_{n-1}^\top \tilde{\vm})^2}{h^2_{n-1}(\tilde{\vx})}\mleft(\sum_{i=1}^{s_n} x_n[i]\mright).
        \end{align*}
        Finally, note that by definition of scaled extension, $\sum_{i=1}^{s_n} x_n[i] = h_{n-1}(\tilde{\vx})$. So,
        \[
            \Lambda(\vm_n) \ge \Lambda(\vm_n^*) = -\frac{(\vec{a}_{n-1}^\top \tilde{\vm})^2}{h_{n-1}(\tilde{\vx})}.
            \numberthis{eq:lambda min}
        \]
        Plugging~\eqref{eq:lambda min} into~\eqref{eq:mhessianm} yields
        \begin{align*}
            \vec{m}^\top \nabla^2 d(\vec{x}) \vm &\ge \tilde{\vm}^\top \nabla^2 \tilde d(\tilde{\vx})\tilde{\vm} - \alpha_n\frac{(\vec{a}_{n-1}^\top \tilde{\vm})^2}{h_{n-1}(\tilde{\vx})}+\frac{\alpha_n}{2}\sum_{i=1}^{s_n} \frac{m_n[i]^2}{x_n[i]}\\
            &= \tilde{\vm}^\top \nabla^2 \tilde d(\tilde{\vx})\tilde{\vm} - \alpha_n\frac{(\vec{a}_{n-1}^\top \tilde{\vm})^2}{\vec{a}_{n-1}^\top \tilde{\vx}}+\frac{\alpha_n}{2}\sum_{i=1}^{s_n} \frac{m_n[i]^2}{x_n[i]},
            \numberthis{eq:mhessianm2}
        \end{align*}
        where the equality follows from expanding the definition of $h_{n-1}$. At this point we upper bound the fraction $\frac{(\vec{a}_{n-1}^\top \tilde{\vm})^2}{\vec{a}_{n-1}^\top \tilde{\vx}}$. First, using the Cauchy-Schwarz inequality,
        \begin{align*}
            (\vec{a}_{n-1}^\top \tilde{\vm})^2 &\le
            \|\vec{a}_{n-1}\|_0 \sum_{q=1}^{n-1} \sum_{i=1}^{s_q} a^2_{n-1, q}[i]\, m_q[i]^2.
        \end{align*}
        So,
        \begin{align*}
            \frac{(\vec{a}_{n-1}^\top \tilde{\vm})^2}{\vec{a}_{n-1}^\top \tilde{\vx}} &=
            \frac{(\vec{a}_{n-1}^\top \tilde{\vm})^2}{\sum_{q=1}^{n-1} \sum_{i=1}^{s_q} a_{n-1, q}[i]\, x_q[i]}\\ & \le \|\vec{a}_{n-1}\|_0\frac{ \sum_{q=1}^{n-1} \sum_{i=1}^{s_q} a^2_{n-1, q}[i]\, m_q[i]^2}{\sum_{q=1}^{n-1} \sum_{i=1}^{s_q} a_{n-1, q}[i]\, x_q[i]}.
        \end{align*}
        Since the denominator is positive by the assumption that $h_{n-1}$ is a nonnegative function, we can further upper bound a fraction of sums of the form $(\sum c_i)/(\sum d_i)$ with the sum of fractions $\sum (c_i/d_i)$, and obtain
        \begin{align*}
            \frac{(\vec{a}_{n-1}^\top \tilde{\vm})^2}{\vec{a}_{n-1}^\top \tilde{\vx}} & \le \|\vec{a}_{n-1}\|_0 \sum_{q=1}^{n-1} \sum_{i=1}^{s_q} \frac{a_{n-1,q}[i]\,m_q[i]^2}{x_q[i]}.
            \numberthis{eq:cancellation terms}
        \end{align*}
        Finally, plugging~\eqref{eq:cancellation terms} into~\eqref{eq:mhessianm2}, we obtain
        \begin{align*}
            \vec{m}^\top \nabla^2 d(\vec{x}) \vm &\ge \tilde{\vm}^\top \nabla^2 \tilde d(\tilde{\vx})\tilde{\vm} - \mleft(\alpha_n\|\vec{a}_{n-1}\|_0 \sum_{q=1}^{n-1} \sum_{i=1}^{s_q} \frac{a_{n-1,q}[i]\,m_q[i]^2}{x_q[i]}\mright) + \frac{\alpha_n}{2}\sum_{i=1}^{s_n} \frac{m_n[i]^2}{x_n[i]}.
        \end{align*}
        Substituting the inductive hypothesis, we find
        \[
        \tilde{\vm}^\top \nabla^2 \tilde d(\tilde{\vx})\tilde{\vm} \ge  \sum_{k=1}^{n-1} \sum_{i=1}^{s_k} \mleft(\frac{\alpha_k}{2} - \sum_{p=k}^{n-2} \alpha_{p+1}\|\vec{a}_p\|_0\, a_{p,k}[i]\mright) \frac{m_k[i]^2}{x_k[i]}
        \]
        and consolidating terms completes the inductive proof.
    \end{description}
    Plugging in the definition of the weights $\alpha_k$ defined in \cref{cor:alpha scext dilated} into \eqref{eq:dilent scext bound} yields
    \[
        \vec{m}^\top \nabla^2 \psi(\vec{x}) \vec{m} \ge \sum_{k=1}^n\sum_{i=1} \frac{m_k[i]^2}{x_k[i]} \qquad\forall\, \vec{m} \in \mathbb{R}^{s_1+\dots+s_n}, \vec{x}\in\relint\cX. 
    \]
    Hence, using the Cauchy-Schwarz inequality,
    \begin{align*}
        \|\vec{m}\|_1^2
        = \left(\sum_{k=1}^n \sum_{i=1}^{s_k} m_k[i]\right)^2
        &= \left(\sum_{k=1}^n \sum_{i=1}^{s_k} \frac{m_k[i]}{\sqrt{x_k[i]}} \sqrt{x_k[i]} \right)^2 \\
        &\leq \left(\sum_{k=1}^n \sum_{i=1}^{s_k} \frac{m_k[i]^2}{x_k[i]} \right) \left(\sum_{k=1}^n \sum_{i=1}^{s_k} x_k[i] \right)
        \leq M_{\cX} \vec{m}^\top \nabla^2 \psi(\vec{x})\vec{m},
    \end{align*}
    which shows that $\psi$ is $(1/M_\cX)$-strongly convex on $\relint\cX$ with respect to the $\ell_1$ norm.
\end{xproof}
        \section{Detailed Description of Game Instances Used in Numerical Experiments}

Here we describe each of the games that we consider in the experimental section of the paper.

\begin{description}
  \item[Kuhn poker] is a standard benchmark in the EFG-solving community \citep{Kuhn50:Simplified}. 
  In Kuhn poker, each player puts an ante worth $1$ into the pot. Each player is then privately dealt one card from a deck that contains $3$ unique cards (Jack, Queen, King). Then, a single round of betting then occurs, with the following dynamics. First, Player $1$ decides to either check or bet $1$. Then,
  \begin{itemize}[nolistsep]
  \item If Player 1 checks Player 2 can check or raise $1$.
    \begin{itemize}[nolistsep]
      \item If Player 2 checks a showdown occurs; if Player 2 raises Player 1 can fold or call.
        \begin{itemize}
          \item If Player 1 folds Player 2 takes the pot; if Player 1 calls a showdown occurs.
          \end{itemize}
        \end{itemize}
      \item If Player 1 raises Player 2 can fold or call.
        \begin{itemize}[nolistsep]
        \item If Player 2 folds Player 1 takes the pot; if Player 2 calls a showdown occurs.
        \end{itemize}
      \end{itemize}
      When a showdown occurs, the player with the higher card wins the pot and the game immediately ends.


    \item[Leduc poker] is another standard benchmark in the EFG-solving
community~\cite{Southey05:Bayes}. The game is played with a deck of $R$ unique
cards, each of which appears exactly twice in the deck. The game is composed of two rounds. In the
first round, each player places an ante of $1$ in the pot and is dealt a single private card. A round of betting then takes place, with Player 1 acting first. At most two bets are allowed per player. Then, a card is is revealed face up and another
round of betting takes place, with the same dynamics described above. After the two betting round, if one of the players has a pair with the public card, that player
wins the pot. Otherwise, the player with the higher card wins the pot. All bets in the first
round are worth $1$, while all bets in the second round are $2$.

    \item[Goofspiel] is another popular benchmark game, originally proposed by \citet{Ross71:Goofspiel}. It is a two-player card game, employing three identical decks
of $k$ cards each whose values range from $1$ to $k$. At the beginning of the game, each player gets dealt a full deck as their hand, and the third deck (the ``prize'' deck) is shuffled and put face down on the board. In each turn, the topmost card from the prize deck is revealed. Then, each player privately picks a card from their hand. This card acts as a bid to win the card that was just revealed from the prize deck. The selected cards are simultaneously revealed, and the highest one
wins the prize card.

    In the zero-sum version of the game, if the players' played cards are equal, the prize card is split. In the general-sum version of the game, denoted ``General-sum Goofspiel'' and used in the experiments on NFCCE, the prize card is thrown out on tie. Either way, the players' scores are computed as the sum of the values of the prize cards they have won.


    \item[Pursuit-evasion] is a security-inspired pursuit-evasion game played on the graph shown in Figure~\ref{fig:search_game}. It is a zero-sum variant of the one used by \citet{Kroer18:Robust}, and a similar search game  has been considered by \citet{Bosansky14:Exact} and \citet{Bosansky15:Sequence}.

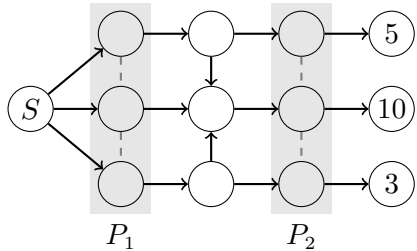
\begin{figure}[ht]
  \centering
  \begin{tikzpicture}
        \path[fill=black!10!white] (.8, -1.4) rectangle (1.6, 1.4);
        \path[fill=black!10!white] (3.2, -1.4) rectangle (4.0, 1.4);
        \node at (1.2, -1.7) {$P_1$};
        \node at (3.6, -1.7) {$P_2$};

          \node[draw, circle, minimum width=.6cm, inner sep=0] (A) at (0, 0) {$S$};
          \node[draw, circle, minimum width=.6cm] (B) at (1.2, 1) {};
          \node[draw, circle, minimum width=.6cm] (C) at (1.2, 0) {};
          \node[draw, circle, minimum width=.6cm] (D) at (1.2, -1) {};
            \node[draw, circle, minimum width=.6cm] (E) at (2.4, 1) {};
            \node[draw, circle, minimum width=.6cm] (F) at (2.4, 0) {};
            \node[draw, circle, minimum width=.6cm] (G) at (2.4, -1) {};
              \node[draw, circle, minimum width=.6cm] (H) at (3.6, 1) {};
              \node[draw, circle, minimum width=.6cm] (I) at (3.6, 0) {};
              \node[draw, circle, minimum width=.6cm] (J) at (3.6, -1) {};
                    \node[draw, circle, minimum width=.6cm,inner sep=0] (K) at (4.8, 1) {$5$};
                    \node[draw, circle, minimum width=.6cm,inner sep=0] (L) at (4.8, 0) {$10$};
                    \node[draw, circle, minimum width=.6cm,inner sep=0] (M) at (4.8, -1) {$3$};

        \draw[thick,->] (A) edge (B);
        \draw[thick,->] (A) edge (C);
        \draw[thick,->] (A) edge (D);
        \draw[thick,->] (B) edge (E);
        \draw[thick,->] (C) edge (F);
        \draw[thick,->] (D) edge (G);
        \draw[thick,->] (E) edge (F);
        \draw[thick,->] (G) edge (F);
        \draw[thick,->] (E) edge (H);
        \draw[thick,->] (F) edge (I);
        \draw[thick,->] (G) edge (J);
        \draw[thick,->] (H) edge (K);
        \draw[thick,->] (I) edge (L);
        \draw[thick,->] (J) edge (M);

        \draw[thick,gray,dashed] (B) edge (C);
        \draw[thick,gray,dashed] (C) edge (D);

        \draw[thick,gray,dashed] (H) edge (I);
        \draw[thick,gray,dashed] (I) edge (J);
    \end{tikzpicture}
  \caption{The graph on which the search game is played.}
  \label{fig:search_game}
\end{figure}

In each turn, the attacker and the defender act simultaneously. The defender controls two patrols, one per each
respective patrol areas labeled $P_1$ and $P_2$. Each patrol can move by one step along the grey dashed lines, or stay in place. The attacker starts from the leftmost node (labeled $S$) and at each turn can move to
any node adjacent to its current position by following the black directed edges. The attacker can also choose to wait
in place for a time step in order to hide all their traces. If a patrol visits a
node that was previously visited by the attacker, and the attacker did not wait
to clean up their traces, they will see that the attacker was there. The goal of the attacker is to reach any of the rightmost nodes, whose corresponding payoffs are $5$, $10$, or $3$, respectively, as indicated in \cref{fig:search_game}. If at any time the attacker and any patrol
meet at the same node, the attacker is loses the game, which leads to
a payoff of $-1$ for the attacker and of $1$ for the defender. The game times out after $m$ simultaneous moves, in which case both players defender
receive payoffs $0$. 

    \item[Battleship] is a parametric version of a classic board game, where two
competing fleets take turns shooting at each other~\citep{Farina19:Correlation}.
At the beginning of the game, the players take turns at secretly placing a set
of ships on separate grids (one for each player) of size $3\times 2$. Each ship
has size 2 (measured in terms of contiguous grid cells) and a value of $4$, and
must be placed so that all the cells that make up the ship are fully contained
within each player’s grids and do not overlap with any other ship that the
player has already positioned on the grid. After all ships have been placed. the
players take turns at firing at their opponent. Ships that have been hit at all
their cells are considered sunk. The game continues until either one player has
sunk all of the opponent’s ships, or each player has completed $R$ shots. At the
end of the game, each player’s payoff is calculated as the sum of the values of
the opponent’s ships that were sunk, minus the sum of the values of ships which
that player has lost.

In the general-sum variant we consider in the NFCCE experiments, we set $R=3$, and furthermore we set 
each player’s payoff is calculated as the sum of the values of
the opponent’s ships that were sunk, minus the sum of the values of ships which
that player has lost \emph{times two}. This modification makes the game general-sum, and makes the players
more risk-averse. Because of that, it was observed by \citet{Farina19:Correlation} that the
introduction of a mediator in the game (through the correlated solution concept)
enables to players to reach equilibrium states with significantly larger social welfare.

  \item[Liar's dice] is another standard benchmark in the
      EFG-solving community~\citep{Lisy15:Online}. In our instantiation, each
      of the two players initially privately rolls an unbiased $6$-face die.
      The first player begins bidding, announcing any face value up to $6$ and
      the minimum number of dice that the player believes are showing that value
      among the dice of both players. Then, each player has two choices during
      their turn: to make a higher bid, or to challenge the previous bid by
      declaring the previous bidder a ``liar''. A bid is higher than the
      previous one if either the face value is higher, or the number of dice is
      higher. If the current player challenges the previous bid, all dice are
      revealed. If the bid is valid, the last bidder wins and obtains a reward
      of $+1$ while the challenger obtains a negative payoff of $-1$. Otherwise,
      the challenger wins and gets reward $+1$, and the last bidder obtains reward of $-1$.
      
\item[Sheriff] The Sheriff game is inspired by the Sheriff of Nottingham board game and 
was introduced by \citet{Farina19:Correlation} as a benchmark game for correlated
solution concepts in extensive-form game. Player 1 (the "smuggler") selects the number of illegal items to be placed in the cargo (in our case, between 0 and 3). The selected number is unknown to
Player 2 (the "sheriff").

Then, the game proceeds for 3 bargaining rounds. In each round,
the following happens:
\begin{itemize}
    \item The smuggler selects an integer bribe amount, in the range 0 to 3
   (inclusive). The selected amount is public information. However, the
   smuggler does not actually give money to the sheriff, unless this is the
   final round.
   \item Then, the sheriff tells the smuggler whether he is planning to inspect the
   cargo. However, no cargo is actually inspected other than in the final
   round. The sheriff can change his mind in later rounds, except for the
   final round.
\end{itemize}


\end{description}
    \end{APPENDICES}
\end{document}